\documentclass[aps,physrev,preprint,groupedaddress]{revtex4-2}

\usepackage{amsmath}
\usepackage{amsfonts}
\usepackage{amssymb}
\usepackage{mathtools}
\usepackage[margin=0.75in]{geometry}
\usepackage{graphicx}
\usepackage{hyperref}
\hypersetup{colorlinks, 
        linkcolor=blue,citecolor=blue
}
\usepackage{cleveref}
\usepackage{bm}
\usepackage{ifpdf}
\usepackage{slashed}
\usepackage{tabularx}
\ifpdf
\fi
\usepackage{feynmp}
\usepackage[outdir=./]{epstopdf}
\newcommand{\fphi}{\widetilde{\phi}}

\newcommand{\rvec}{\mathbf{r}}

\newcommand{\kvec}{\mathbf{k}}
\newcommand{\qvec}{\mathbf{q}}

\renewcommand{\[}{\begin{equation}}
\renewcommand{\]}{\end{equation}}

\begin{document}

\title{Pattern formation in a coupled driven diffusive system}
\author{Guilherme E. Freire Oliveira}
\email{guilhermeefoliveira@gmail.com}
\affiliation{Departamento de F\'isica and National Institute of Science and Technology for Complex Systems, ICEx, Universidade Federal de Minas Gerais, C. P. 702, 30123-970 Belo Horizonte, Minas Gerais, Brazil}
\author{Ronald Dickman}
\email{dickman@fisica.ufmg.br }
\affiliation{Departamento de F\'isica and National Institute of Science and Technology for Complex Systems, ICEx, Universidade Federal de Minas Gerais, C. P. 702, 30123-970 Belo Horizonte, Minas Gerais, Brazil}
\author{Maxim O. Lavrentovich}
\email{lavrentm@gmail.com}
\affiliation{Department of Earth, Environment, and Physics, Worcester State University, Worcester, Massachusetts 01602, USA}
\affiliation{Department of Physics \& Astronomy, University of Tennessee, Knoxville, Tennessee 37996, USA}
\author{R. K. P. Zia}
\email{rkpzia@vt.edu}
\affiliation{Center for Soft Matter and Biological Physics, Department of Physics, Virginia Polytechnic Institute \& State University, Blacksburg, Virginia 24061, USA}
\affiliation{Physics Department, University of Houston, Houston, Texas 77204, USA}

\date{\today}

\begin{abstract}
We investigate pattern formation in a driven mixture of two mutually repulsive particle species via a  Field-based Lattice Model (FLM), a hybrid model that combines aspects of the driven Widom-Rowlison lattice gas (DWRLG) and its statistical field theory \cite{DWRLG1,DWRLG2}. We find that the FLM effectively captures the bulk behavior of the DWRLG in both low- and high-density phases, suggesting that phase transitions in these models may share a common universality class. Under the effect of a drive, the FLM additionally reveals an intermediate regime, not reported in previous DWRLG studies, characterized by irregular stripes with widely fluctuating widths, contrasting with the regular, well-ordered stripes found at higher densities. In this intermediate phase, the system exhibits long-range order predominantly perpendicular to the drive direction. To construct a  continuum description, we derive two coupled partial differential equations for the particle densities via a gradient expansion of the FLM mean mass-transfer equations, supplemented with additive noise. Designing a numerical solver using the pseudospectral method with dealiasing and stochastic time differencing, we reproduce the low-density microemulsion phase (characterized by a nonzero characteristic wavenumber $q^*$) and stripes perpendicular to the drive at high density. We identify the nonzero difference in the characteristic velocities of the sum and difference of the particle densities as a necessary condition for perpendicular stripe formation in the high-density phase. The continuum model exhibits novel behaviors not observed in the FLM, such as stripes aligned parallel to the drive, and chaotic patterns. This work highlights how the interplay of external drive, particle interactions, and noise can lead to a rich phenomenology in strongly driven binary mixtures.
\end{abstract}
\maketitle

\newpage
\section{Introduction\label{sec:introduction}}

While statistical mechanics has excelled in describing systems at equilibrium, a significant portion of the world operates far from equilibrium. Among the most fascinating properties of nonequilibrium systems are the complex spatiotemporal patterns they exhibit \cite{CrossHohenberg, HohenbergKrekhov, Cross2}, including those associated with phase transitions and critical phenomena \cite{HohenbergHalperin}. There has been growing interest in understanding how such patterns emerge and how they can be used to infer important properties of physical systems, whether active or driven externally.

In the study of systems driven out of equilibrium, significant open questions remain, including how to define phase coexistence \cite{Ron2016} or a thermodynamic quantity of the same usefulness as entropy at equilibrium \cite{Calazans2019}. A driven system is subject to an external force (or ``drive", for simplicity) that maintains it out of equilibrium. Examples include \textit{driven diffusive systems} in which the drive biases particle diffusion along an axis with periodic boundaries, as in the pioneering studies of driven Ising, Potts, and Blume-Emery-Griffths models (in their lattice-gas representations) \cite{KLS2, KLS1, Garrido1987, LSZ89, VLZ89, GLMS90, CGLV91, SZ91, SHZ92, Bassler1994, KSZ97}. Driven diffusive systems exhibit a range of intriguing behaviors \cite{DDSbook, MarroDickman99}, including generic long-range correlations at all temperatures \cite{KLS2, KLS1} and critical properties governed by a nonequilibrium fixed point \cite{JS86, LC86}. At low temperatures, these models typically exhibit phase-separation into stripes which form along the drive direction \cite{KLS2, KLS1}. Recently, some of the present authors have considered a driven version \cite{DWRLG1, DWRLG2} of the Widom-Rowlinson lattice gas \cite{eqRWLG1} (WRLG). The WRLG is a lattice gas model in which the two species of particles ($A$ and $B$, say) diffuse freely, apart from the constraint of strict exclusion of nearest neighbor (NN) $A$-$B$ pairs. With no interactions except volume exclusion, the only control parameters of the equilibrium WRLG are the overall densities of the two species. Governed by entropy alone, this equilibrium system transitions, as the densities increase, from a homogeneous disordered state to one with phase separation, much like the Ising lattice gas \cite{eqRWLG1}. When both species are \textit{driven} in the \textit{same} direction, striking features emerge \cite{DWRLG1, DWRLG2}: In the disordered phase, the $A$-$B$ correlation function develops an intrinsic, preferred wavenumber ($\mathbf{q}^*\propto\hat{\mathbf{x}}$, the drive direction)! Correlations grow as the overall density increases, until long-range order appears -- in the form of alternating $A$- and $B$-rich stripes, characterized by $\mathbf{q}^*$. In stark contrast with the driven Ising case \cite{KLS1}, these stripes are \textit{always} perpendicular to the drive. Further, since $\mathbf{q}^*$ is independent of $L$, the system is \textit{not} scale invariant.

Notably, similar nonequilibrium phenomena, such as pattern formation and phase separation, are found in active matter, where scalar active mixtures have highlighted the importance of nonreciprocal couplings and characteristic length scales far from equilibrium \cite{Marchetti2024, Fausti2021, Fruchart2021, Golestanian2023, Golestanian2025}. Although the active matter literature is filled with interesting examples of pattern formation, many such phenomena already appear in simpler externally driven systems. The driven Widom-Rowlinson lattice gas (DWRLG) provides an especially interesting and simple setting for exploring patterns. The model was originally introduced and studied in detail in Ref.~\cite{DWRLG1}, where its phase behavior and stripe formation were characterized through simulations. More recently, a coarse-grained field-theoretic description of its disordered phase was formulated using the Doi-Peliti formalism \cite{Doi, Peliti} in Ref.~\cite{DWRLG2}. In the latter work, the field equations describe the nonequilibrium dynamics as two coupled driven diffusive systems: the sum and difference of $A$ and $B$ particle densities (referred to as the ``density'' and ``charge'' fields in the following), with different characteristic velocities. The field theory is capable of capturing key features observed in the simulations at low particle densities, including the development of a characteristic wavenumber $\mathbf{q}^*$, and further highlights the importance of noise in generating this characteristic wavenumber in the disordered phase.

The present study builds upon Refs.~\cite{DWRLG1,DWRLG2}, bridging the gap between them. More specifically, we seek a minimal continuum description, inspired by the Doi-Peliti formalism, of the ordered (heterogeneous) phase, where open questions remain regarding the structure of the equations of motion and whether they are capable of capturing the stripe orientation observed in lattice simulations. We also revisit the properties of the low-density phase. To this end, we first introduce a new Field-based Lattice Model (FLM) that preserves all the symmetries of the original DWRLG. We find excellent agreement with the phenomenology reported in Refs.~\cite{DWRLG1, DWRLG2}, strongly suggesting that both models belong to the same universality class. While discrete field theories are commonly employed in the quantum domain \cite{Skopenkov2023}, their stochastic counterparts are far less explored; our lattice model provides a simple and straightforward example of a stochastic discrete field theory. Next, we perform a gradient expansion (GE) of the FLM master equation, leading to a pair of coupled field equations that, we shall argue, satisfactorily describe both low- and high-density regimes.

The equations we derive are consistent with the Doi-Peliti and Martin-Siggia-Rose-Janssen-De Dominicis (MSRJD) field equations developed for the DWRLG in Ref.~\cite{DWRLG2}, but include additional terms necessary for describing a stable ordered phase. In the absence of an external drive, our equations of motion follow the structure of a two-component Cahn-Hilliard-like equation \cite{Mao2019}. But a drive breaks rotational symmetry, leading to stripe formation, or traveling waves, similar to phenomena observed in scalar active matter \cite{Marchetti2024, Golestanian2020}. We demonstrate here that aspects of the phenomenology reported in \cite{Marchetti2024, Golestanian2020} are already evident in a driven matter context, without requiring non-reciprocal particle interactions or particle activity. 

The remainder of the paper is organized as follows: In the next section, we introduce the FLM, providing a brief characterization of the model. In Sec.~\ref{sec:Continuum}, we derive the field equations via a GE, including explicit expressions for the linear coefficients, with additional details outlined in Appendices~\ref{appx:gradient-expansion} and \ref{appx:coefficients}. In Sec.~\ref{sec:numer-int}, we integrate these equations numerically using the pseudospectral method combined with stochastic second-order Heun exponential time-differencing (PSSETD2H) \cite{Cox, Toral} and smooth dealiasing techniques \cite{Hou}, with further numerical details provided in Appendices~\ref{appx:pseudospectral} and~\ref{appx:dealiasing}. Both deterministic and stochastic versions of the field equations are analyzed. One of the key numerical findings is that the difference in the characteristic velocity of the charge and density fields plays a crucial role in the nonequilibrium behavior of the model, supporting the conclusions of \cite{DWRLG2}, but in the ordered phase as well. Finally, Sec.~\ref{sec:conclusion} presents our conclusions and outlines potential future directions.

 \section{Field-based Lattice Model \label{sec:Model}}
  
Early studies of Widom-Rowlinson systems were based on two distinct microscopic models, the original one of interacting particles in the continuum \cite{eqRWLG1}, and a lattice-gas model \cite{WR70}. In more recent studies \cite{DWRLG1, DWRLG2}, a stochastic field theoretic version was also considered, with the bridge between the two established either phenomenologically or using the Doi-Peliti formalism. All exhibit two distinct regimes: a disordered low-density phase and a high-density ordered phase.  Here, we consider a hybrid model \footnote{The rudiments of this model were introduced in \cite{DWRLG1}, where it played the role of an intermediary between simulations of the lattice gas and studies of a phenomenological field theory. In the present study, we investigate a full stochastic version of it.}, with continuous variables (densities, $\rho_{A,B}\in\left[0,1\right]$) defined on a $L\times L$ lattice \footnote{Generalizations to lattices in higher dimensions are straightforward.}. This approach follows the spirit of coarse graining a lattice gas and will be referred to as the Field-based Lattice Model (FLM). Expecting it to be qualitatively similar to the WRLG, we perform simulations to explore its behavior. As we will show, the driven FLM has a richer phenomenology than the DWRLG, in that a third regime appears as the overall density is varied between the homogeneous and striped phases. In this regime,  which we call the irregular stripe regime in the following, there is long-range order only in the direction \textit{perpendicular} to the drive, with well-defined stripes, but of \textit{markedly different widths}. In the next sections, we present a straightforward gradient expansion of the FLM to arrive at a field-theoretic description with equations of motion for the densities, and compare our results with the field theories studied earlier \cite{DWRLG1, DWRLG2}. We begin with a review of the driven WRLG.

\subsection{Model specification}

The driven WRLG \cite{DWRLG1} consists of two species of particles, $A$ and $B$, diffusing on a square lattice of $L^{2}$ sites with periodic boundary conditions. Each site $\mathbf{s}=\left(i,j\right)$ is either empty or occupied by a single particle. In addition, opposite species repel each other: nearest neighbor (NN) $A$-$B$ pairs are forbidden. In the stochastic dynamics, a random particle is chosen and moved to a randomly chosen NN or next NN (NNN) site, provided the destination site is empty and its NN sites are not occupied by a particle of the opposite species. When an external drive $\delta$ is imposed, particle moves in the $+x$ (increasing $i$) direction are favored over moves in the $-x$ (decreasing $i$) direction. Following previous simulation studies, we set the numbers of $A$ and $B$ particles equal, so that we have only two control parameters: the total particle density, $\rho$, and the drive, $\delta$. For $\delta=0$ (the WRLG), the dynamics satisfies detailed balance and the properties of its stationary state can be predicted by considering entropy alone, as every allowed configuration is equally probable. Phase separation occurs when $\rho$ is increased beyond a critical density, $\rho_{c}^{\left(0\right)}\simeq0.618(1)$, with critical behavior in the Ising universality class \cite{eqRWLG1}. For $\delta>0$ (the DWRLG), detailed balance is violated, the stationary distribution is unknown a priori, while novel phenomena emerge in the nonequilibrium steady-state, for all densities (below, near, and above a $\delta$-dependent $\rho_{c}^{\left(\delta\right)}$, with $\rho_{c}^{\left(\delta\right)}>\rho_{c}^{\left(0\right)}$, in general) \cite{DWRLG1, DWRLG2}. 

The FLM differs from the DWRLG mainly in having \textit{continuous} (instead of discrete) variables at each site of the lattice, i.e., $\rho_{A,B}\in\left[0,1\right]$ instead of $0,\pm1$. Both densities are conserved; here we consider only the case of equal overall densities:%
\begin{equation}
\sum_{\mathbf{s}}\rho_{A}(\mathbf{s})=\sum_{\mathbf{s}}\rho_{B}(\mathbf{s})=\frac{\rho
L^{2}}{2}. \label{eq:dens-cons}%
\end{equation}
The evolution is defined through the following set of rules, which are motivated, in part, by the excellent agreement between simulation results for the current-density relation, $J\left(  \rho\right)$, and a simple phenomenological form, discussed in more detail in Ref.~\cite{DWRLG1}:
\begin{enumerate}
    \item At each iteration, select a site $\mathbf{s}=(i,j)$ at random and one of the eight NN or NNN sites $\mathbf{s}'=(i',j')$, according to the following drive-dependent probabilities
    \begin{equation} \label{eq:trates}
    w_{\mathbf{s} \to \mathbf{s}'} =\frac{1}{8} \times
        \begin{dcases}
             1+\delta &  \mathrm{if} \ \Delta i = 1\\
              1 &   \mathrm{if} \ \Delta i = 0\\
             1-\delta  & \mathrm{if} \ \Delta i = -1
        \end{dcases}
    \end{equation}       
    where $\Delta i = i' -i$ and $\delta \in \left[0,1\right]$ is the drive. In other words, hops along (opposite) the direction of the drive are enhanced (suppressed).
    
    \item Next, select one of the species at random. In case $A$ is chosen, then transfer the following amount of $A$-density from $\mathbf{s}$ to $\mathbf{s}'$:
    \begin{equation}\label{eq:amount_moved}
        \Delta\rho_A(\mathbf{s}) = \epsilon\,\rho_A(\mathbf{s})h(\mathbf{s}')\slashed{\rho}_B(\mathbf{s}'),
    \end{equation}
     where $\epsilon$ is a random variable \footnote{In our simulations, $\epsilon$ is chosen from a uniform distribution. From the results, $\epsilon$ serves only to redefine a diffusion length. Note that this multiplicative aspect of the noise may become relevant near the boundary values: $\rho_{A,B}=0,1$.} in $\left[0,1\right]$, $\rho_A(\mathbf{s})$ is the density of $A$-particles at site $\mathbf{s}$, $h(\mathbf{s}')=1-\rho_A(\mathbf{s}')-\rho_B(\mathbf{s}')$ is the vacancy density at $\mathbf{s}'$, and $\slashed{\rho}_B(\mathbf{s}')$ is defined as
\begin{equation}\label{eq:neighbor-function}
     \slashed{\rho}_B(\mathbf{s}') \equiv  \prod_{\mathbf{s}'' }[1-\rho_B(\mathbf{s}'')],
 \end{equation}%
where the product is over all nearest neighbors $\mathbf{s}''$ of $\mathbf{s}'$, excluding the original site $\mathbf{s}$.  Note that this product contains three factors for NN hops, and four in the case of hopping to a NNN. The factor $\slashed{\rho}_B$ models $A$-$B$ repulsion through an estimate of the probability that the nearest neighbors of the target site are free of $B$'s \footnote{The form of Eq.~\eqref{eq:amount_moved} and these suppression factors are motivated in part by the remarkably good agreement between a conjectured formula for the current density and the observed one in DWRLG: Eq. (8) and Fig. 5 in \cite{DWRLG1}.}. If instead of $A$, species $B$ is chosen, the same rules apply, simply exchanging $A$ and $B$ in all expressions. 
\end{enumerate}

These steps are iterated to generate the evolution. We emphasize that $\Delta\rho_{A,B}(\mathbf{s})\leq h(\mathbf{s}')$, so that it is impossible for a site to be \textquotedblleft overfilled\textquotedblright. As usual, a Monte Carlo Step (MCS) is defined as $2L^{2}$ iterations, so that, on average, each species
at each site is given one attempt to transfer some of its contents.

\subsection{Results from simulations and comparisons with the lattice gas}

\begin{figure*}[ht]
    \centering
    \includegraphics[width=3.5in]{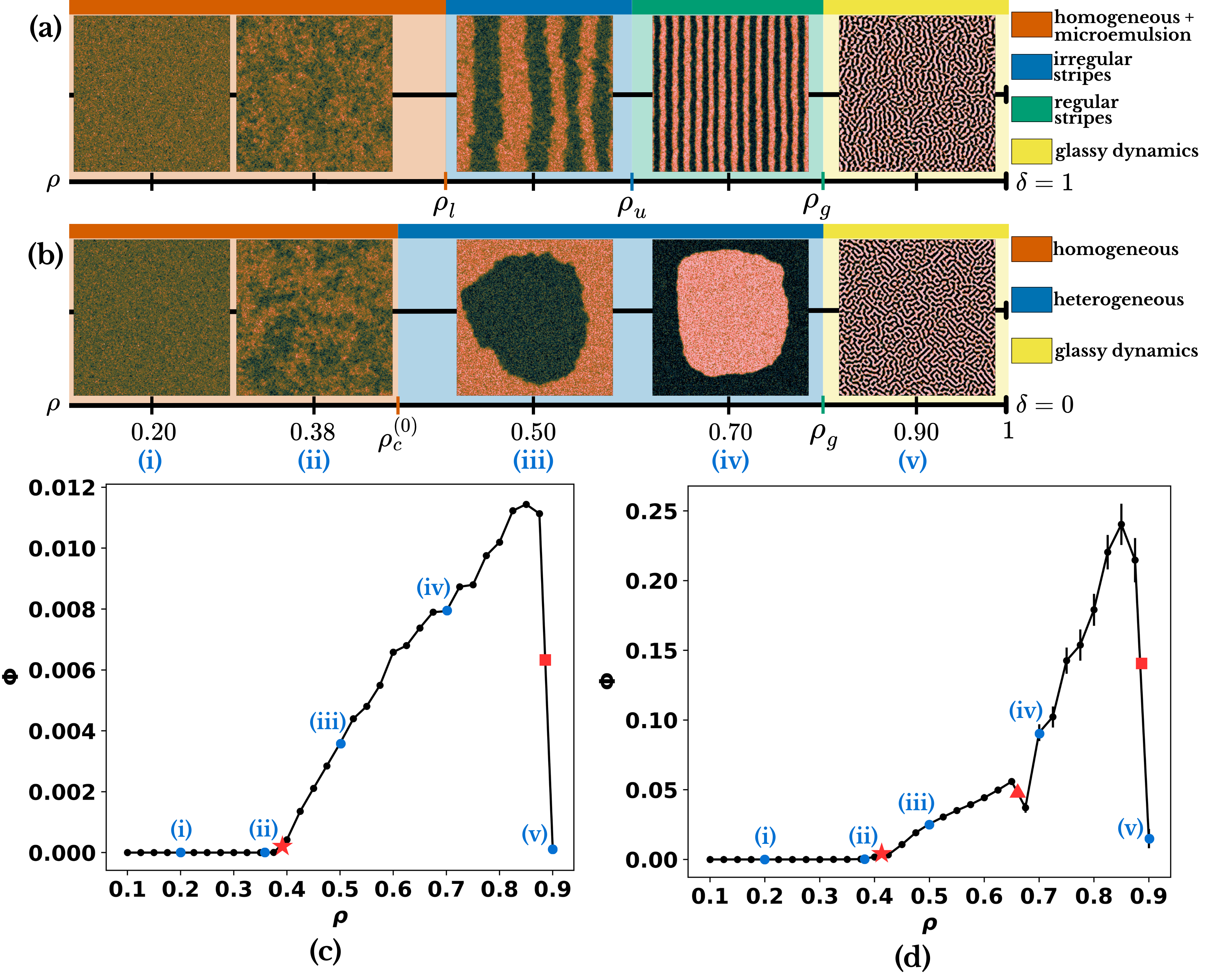} 
    \caption{Summary of steady-states in the FLM. See text for details. Upper panels (a)-(b): Typical charge fluctuation field configurations in a $L = 192$ system, also observed in the other sizes studied. (a): $\delta = 1$. For $\rho < \rho_\ell $ (orange background), the system is well mixed and homogeneous, with the structure factors exhibiting a discontinuity at the origin and a peak at $(q_x,q_y)=(q^*,0)$. The intermediate regime (blue background), $\rho_\ell < \rho < \rho_u $, is marked by \textit{irregular stripes} which suffer long-wavelength instabilities and display a range of widths. For $\rho>\rho_u$ (green background), the  stripes are fully developed and long-wavelength instabilities are not sufficient to destroy long-range order in the $x$(drive)-direction. This regime is referred to  as the \textit{regular stripe} regime. (b): $\delta =0$. For $\rho< \rho_c^{(0)} $ (orange background) the system is well mixed and homogeneous, with the structure factors displaying the  Ornstein-Zernike form. For $\rho> \rho_c^{(0)}$ (blue background) the system is heterogeneous, exhibiting coarsening dynamics that follows the Lifshitz-Slyozov law. For both  $\delta=0$ and $1$, the system undergoes a glassy transition when $\rho>\rho_g$ (yellow background). Lower panels (c)-(d): order parameter plots [Eq.~\eqref{eq:op}] for $\delta = 0 $ (c) and $\delta = 1$ (d); using $L=64$ and averaged over $100$ realizations. The blue Roman numerals (i-v) near $\rho=0.2, 0.4, 0.5, 0.7, 0.9$ correspond to the typical configurations shown in (a) and (b).  Red stars are our crude estimates of the critical density; $\rho_c^{(0)} \simeq 0.39$ and $\rho_\ell \simeq 0.40$.  Red squares are estimates of the glassy transition density, $\rho_g \simeq 0.90$. For $\delta =1$, the red triangle marks the transition to regular stripes at the density $\rho_u \simeq 0.65$. A movie related to this figure is available in \cite{movies}.}
    \label{fig:phase-diagram}
\end{figure*}

To explore the behavior of this model, we carry out simulations on lattices of size $L$ ranging from $32$ to $256$, varying the overall density, $\rho$, from $0.10$ to $0.95$. Here, we report only results from the extreme cases of $\delta=0$ and $1$, which exhibit all the key phenomenology of the model at intermediate drive values. The main set of studies employs a uniform initial state with $\rho_{A,B}\left(  \mathbf{s}\right)  = \rho/2$ and, for measurements in the steady-state, up to $100$ realizations per run, for up to $2\cdot10^{7}$ MCS. With no drive ($\delta = 0$), the behavior is essentially the same as the WRLG, displaying a disordered (phase-separated) state at low (high) densities, with the critical density estimated to be $\rho_{c}^{\left(0\right)}\simeq0.39$. For $\rho$'s below this value, the structure factors are isotropic and display the standard Ornstein-Zernike form. Near criticality, we expect the properties to be in the Ising class \cite{eqRWLG1} and did not pursue this issue. At densities above $\rho_{c}^{\left(0\right)}$, the usual coarsening dynamics is observed, and we verify \cite{thesisgui} that our system follows the Lifshitz-Slyozov law, i.e., characteristic length scales associated with the peaks of structure factors grow initially as $t^{1/3}$. For densities closer to unity, the system exhibits arrested coarsening, with the lack of vacancies and $A$-$B$ repulsion leading to considerable frustration. The system begins to lose ergodicity \footnote{Indeed, for $\rho$ as low as just $0.5$, certain configurations are \textquotedblleft frozen\textquotedblright. Examples include the checkerboard pattern (all white squares vacant, alternate black ones filled with $A$ or $B$) and similarly alternate rows or columns. Note that such configurations are also present in the WRLG, as $\rho=0,\pm1$ appears simply as $n=0,\pm1$. As a result, the particular density $\rho_g$ at which this transition occurs may depend on the initial condition.} and exhibits glassy behavior. Systems starting with $\rho_{A,B}\left( \mathbf{s}\right)  =\rho/2$ quickly freeze into labyrinthine patterns that resemble those of the Turing-Swift-Hohenberg model \cite{Alar-Clerc}. A rough estimate of the \textquotedblleft glass transition\textquotedblright\ density is $\rho_{g}\sim0.9$. Typical configurations from these two (expected) phases and the final glassy regime of the $\delta=0$ system are schematically depicted in Fig.~\ref{fig:phase-diagram}(b).

The behavior of the \textit{driven} FLM (at $\delta=1$ here) is more surprising. Though many of its properties are much the same as those in the DWRLG (i.e., disordered at low densities and ordering into regular stripes at high densities), there are some major differences. In particular, it may be argued that there is an extra regime (as $\rho$ is varied) in the FLM, consisting of a phase in between the analogs of the ordered (high-density) and disordered (low-density) phases of the DWRLG. 

At low densities, both the DWRLG and the FLM display a homogeneous, \textquotedblleft microemulsion\textquotedblright\ phase, in which the correlation functions show a periodic structure in the drive direction. For larger $\rho$, both display inhomogeneities. While the DWRLG develops well defined $A$- and $B$-rich stripes when $\rho>\rho_{c}^{\left(0\right)} \sim0.6$, stripes begin to emerge in the FLM at a lower value, $\rho_{\ell}\sim0.4$. In addition, the widths of the stripes in the DWRLG are relatively small and (mostly) independent of $\rho$ for maximal drive ($\delta=1$) \cite{DWRLG1}. By contrast, the stripes in the FLM that first emerge for $\rho>\rho_{\ell}$  appear to be  ``irregular'' in that their widths fluctuate widely. This   regime is apparently not accessible to the DWRLG. Such behavior persists up to an ``upper'' density, $\rho_{u} \sim 0.65$.   

As we increase $\rho$ beyond $\rho_{u}$ and all the way up to the glass transition density $\rho_g$, we find properties that more closely resemble those of the DWRLG high-density phase: starting from a uniform initial state, the system quickly develops many stripes with thin, regular widths  with a well defined characteristic wavenumber $q^*$. We will label this the regular stripe regime, $\rho \in \left(\rho_{u},\rho_{g}\right)$. Although investigations with some other initial conditions suggest that there are many long-lived metastable states (i.e., not relaxing to regular stripes within the time limits of our runs), a definitive study of them is challenging and beyond the scope of this article \cite{thesisgui}. Instead, our main focus here will be the regular and irregular stripe regimes of the FLM. Although the various regimes encountered in our simulations and numerical integration appear to represent distinct phases, we caution the reader that the term is used somewhat informally. Deferring more systematic studies (including finite-size scaling analyses) required to establish the existence of well defined phases (and associated phase transitions) to future work, our goal is to present qualitative observations in the next subsection.

At the highest densities, excluded-volume considerations mark the behavior of the DWRLG and the driven FLM. In the DWRLG, exclusion of $A$-$B$ NN pairs prohibits the use of random initial configurations. In \cite{DWRLG1}, such densities were studied using single-stripe initial configurations consisting of two blocks (A particles plus vacancies, B particles plus vacancies) separated by empty columns. For the FLM with $\rho>\rho_{g}$, we encounter extremely slow dynamics reminiscent of glass, just as in the $\delta=0$ case. Comparing panels (c) and (d) of Fig.~\ref{fig:phase-diagram}, we note that the precipitous drop in $\Phi$, marking the onset of the glassy regime, occurs at nearly the same density in the undriven and maximally-driven cases, suggesting that this kinetic phenomenon is independent of the drive, and occurs well before ordering can develop. To summarize, these different behaviors and regimes for the driven FLM are depicted in Fig.~\ref{fig:phase-diagram}(a).

\subsection{Brief analysis of various regimes}

\begin{figure}[th]
\centering\includegraphics[width=0.6\columnwidth]{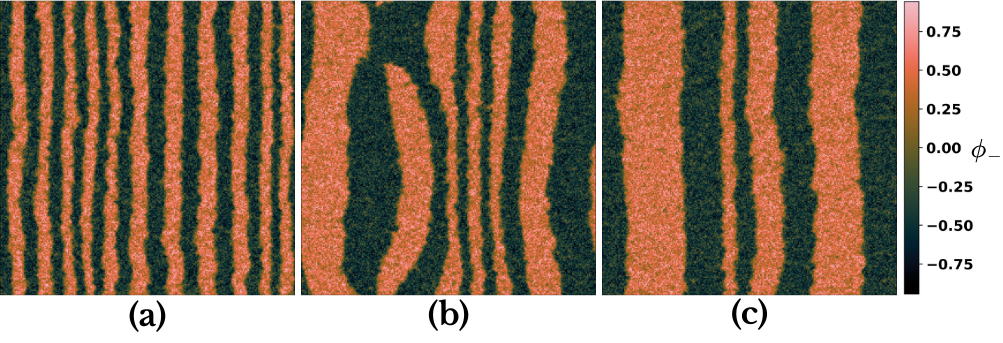}
\caption{Charge fluctuation field configurations of the driven FLM for $\rho=0.60,\ L=256$, at times $5.40 \times 10^{5}$ (a), $2.26 \times 10^{6}$ (b), and $1.02 \times 10^{7}$ (c) MCS, after starting from a uniform initial state. Panel (a) shows quite regular, thin stripes. Breakup and merging take place in panel (b). Panel (c) shows a typical configuration at late times, with few stripes of widely differing widths.}%
\label{fig:600movie}%
\end{figure}

To highlight the novelty of the irregular stripe regime in the driven FLM, we will present a number of approaches to analyzing the $\delta=1$ system. To best identify the various regimes accessible to this model, we focus here on systems with $L=32, 64, 128, 256$ and three densities, $\rho=0.39, 0.60, 0.80$,  corresponding to the homogeneous, irregular stripe and regular stripe regimes, respectively. As mentioned above, the typical configurations in the first and last regimes are similar to the \textquotedblleft microemulsions\textquotedblright\ and the striped phases of the DWRLG \cite{DWRLG2}. By contrast, starting with uniform $\rho_{A,B}$ with overall $\rho\in\left(\rho_{\ell},\rho_{u}\right) $, the system quickly (e.g., $\sim 10^5 $ MCS for $\rho=0.60$ and $L=256$) forms a relatively large number of thin stripes (of regular width, mainly), as if it were settling into regular stripes. Soon thereafter, these stripes suffer a long-wavelength instability, reminiscent of the Eckhaus instability of convection rolls \cite{Eckhaus}, and break up. They subsequently recombine and, by the end of the runs ($\sim 10^7$ MCS), settle into \textit{relatively few} well-defined (irregular) stripes with a \textit{wide range} of widths. We illustrate this unusual behavior in Fig.~\ref{fig:600movie} with three snapshots of a $\rho =0.60, \ L=256$ system  at about $ 10^{5},10^{6}$, and $10^{7}$ MCS.

There are a number of ways to characterize quantitatively the atypical behavior in the irregular stripe regime. We focus only on the steady-state, deferring an analysis of the interesting evolution process to the future. A standard quantity is the (two-point) correlation function of the fluctuating densities. Instead of deviations of $\rho_{A,B}$ from $\rho/2$, let us consider their sum and difference:%
\begin{align} \label{eq:charge-dens}
    & \phi_\pm(\mathbf{s},t) = \left(\rho_A(\mathbf{s},t)-\frac{\rho}{2}\right) \pm \left(\rho_B(\mathbf{s},t)-\frac{\rho}{2}\right),
\end{align}
labeled as the density (DF) and charge (CF) fluctuations: $\phi_{+}$ and $\phi_{-}$, respectively. Translational invariance of the steady-state allows us to exploit their Fourier transforms
\begin{align} \label{eq:ft-charge-dens}
    & \fphi_\pm(\qvec,t) = \sum_{\mathbf{s}} \phi_{\pm}(\mathbf{s},t)e^{i\qvec\cdot\mathbf{s}} 
\end{align}
and study the transform of the correlations, i.e., the structure factors (SF) \footnote{Due to the symmetries of the systems we are considering, it can be shown that the cross correlation (associated with $S_{+-}$) vanishes, so that we investigated on these two SFs.}
\begin{align} \label{eq:sf-charge-dens}
    & S_\pm(\mathbf{q},t) = \frac{1}{L^2}\left\langle\left|\fphi_\pm(\qvec,t)\right|^2\right\rangle.
\end{align}
The average $\left\langle \cdots\right\rangle $ is over the stationary-state ensemble. In practice, we perform both an ensemble average (with up to $100$ independent runs) and time average (of each run, measuring $1000$ times in the last $10^{7}$ MCS).

As in the DWRLG, $S_{-}(\mathbf{q})$ is found to peak at a nonzero wave-vector: $\mathbf{q}^{\ast}\equiv\left( q^{\ast},0\right) $, even in the homogeneous phase, while $S_{+}(\mathbf{q})$ peaks at $2\mathbf{q}^{\ast}$ at high densities (when stripes become more well-established). In fact, the doubling of the frequency for the DF can be explained in simple terms: since the density field marks the interfaces between A- and B-rich regions, a modulation for the CF must be accompanied by a modulation of twice the frequency in the density. Let us denote the value of $S_{-}$ at the peak as $S_{-}^{\ast}\equiv S_{-}(\mathbf{q}^{\ast})$. As in ordinary equilibrium systems (in $d$ dimensions) with transitions to long-range order, we expect this maximum to be $O\left( 1\right)$ in the disordered phase, but rise to $O\left(L^{d}\right)$ in the ordered one. Thus, we follow Ref.~\cite{DWRLG1} and define the order parameter as
\begin{equation}\label{eq:op}
    \Phi = \frac{S^*_-}{L^2}.
\end{equation}
Associated with the SFs are sum rules:%
\begin{equation}\label{eq:sigma-def}%
\Sigma_{\pm}\equiv L^2\sum_{\mathbf{q}}  S_{\pm}(\mathbf{q}).
\end{equation}
These would be constants in the DWRLG, but, for the FLM, only bounds can be placed on them:
\[
 \rho^2 \leq \Sigma_{+}\leq \rho  \quad \mbox{ and }\quad 0 \leq \Sigma_{-}\leq \rho.%
\]
The two extremes come from a state of complete disorder [$\rho_{A,B}\left(\mathbf{s}\right) =\rho/2$, such as those used as initial states in simulations] and one of complete order [$\rho_{A,B}\left(\mathbf{s}\right)=0$ or $1$ only]. As the crudest measure of order, it is interesting to report that, e.g., $\Sigma_{-}$ is approximately $0.75\%$ of being maximal for $\rho=0.39$, rising to approximately $15\%$ for the irregular stripe regime case of $\rho = 0.60$, while reaching about $21\%$ in the high-density $\rho = 0.80$ case with regular stripes.

\begin{figure}[ht]
    \centering
 \includegraphics[width=0.7\columnwidth]{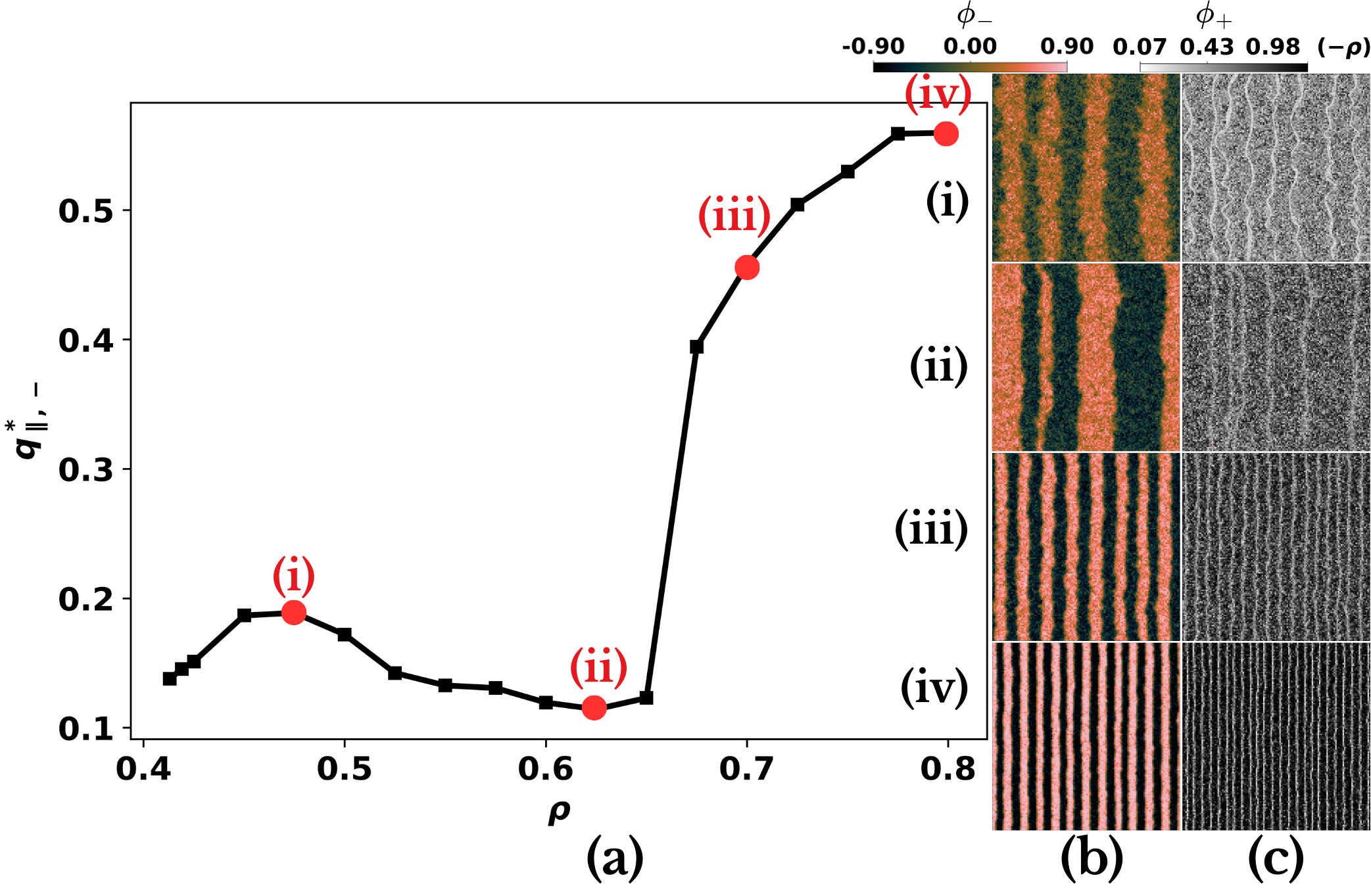}
    \caption{(a): Peak position of the charge fluctuation structure factor (CF SF) along the drive, $q^*$, versus the density $\rho$, for $\delta =1 $ and $L=128$, averaged over $100$ realizations starting from a uniform initial condition. Color plots: typical field configurations of the CF (b) and the density fluctuation (DF) (c), for: (i) $\rho= 0.475$, (ii) $\rho = 0.625$, (iii) $\rho=0.7$ and (iv) $\rho = 0.8$. Observe that the DF has minima at the interfaces of the CF. The color map scale for the DF is shifted by $\rho$. Since the DF signal for $\rho = 0.475$ is very weak, we mark the interfaces (regions where $\phi_+ \sim-\rho$) in white.}
    \label{fig:selectedmode}
\end{figure}

We also demonstrate the different regimes by plotting the $\rho$-dependence of the order parameter, $\Phi\left(\rho\right)$, over a wide range of $\rho$: $\left(0.1,0.9\right) $. In the lower panels of Fig.~\ref{fig:phase-diagram}, we illustrate for both $\delta=0$ and $\delta=1$ [Figs.~\ref{fig:phase-diagram}(c) and (d), respectively],  with $L=64$. In the former, there are clearly two (or three, if the the glassy regime at the highest density is included) regions with very different behavior: $\rho$ below or above $\rho \sim 0.4$. From the associated typical configurations  with $L=256$ in (i) and (ii), we identify these regions with the disordered and ordered phases of the $\delta=0$ system, leading to the estimate $\rho_{c}^{\left(0\right)}\simeq 0.39$. By contrast, in Fig.~\ref{fig:phase-diagram}(d), there are arguably \textit{three} (four, including the glassy) regimes for the $\delta=1$ case, from which we estimate $\rho_{\ell}\simeq 0.40$ and $\rho_{u}\simeq 0.65$. We note that the order parameter in the ordered regimes of the driven ($\delta=1$) system is over an order of magnitude larger than in the corresponding regimes of the undriven case. Fig.~\ref{fig:selectedmode}(a) further supports these observations by displaying the peak position of the charge fluctuation structure factor, $q^*$, as a function of $\rho$ for $\delta = 1$ and $L = 128$. A clear transition is observed around $\rho \sim 0.65$, where $q^*$ exhibits a sharp jump, indicating a qualitative change in behavior within the ordered phase. Figs.~\ref{fig:selectedmode}(b) and (c) show representative steady-state configurations for four selected densities: $\rho = 0.475$ (i), $0.625$ (ii), $0.700$ (iii), and $0.800$ (iv),  demonstrating that long-range order has developed along the drive direction.

\begin{figure}[th]
\centering\includegraphics[width=0.5\columnwidth]{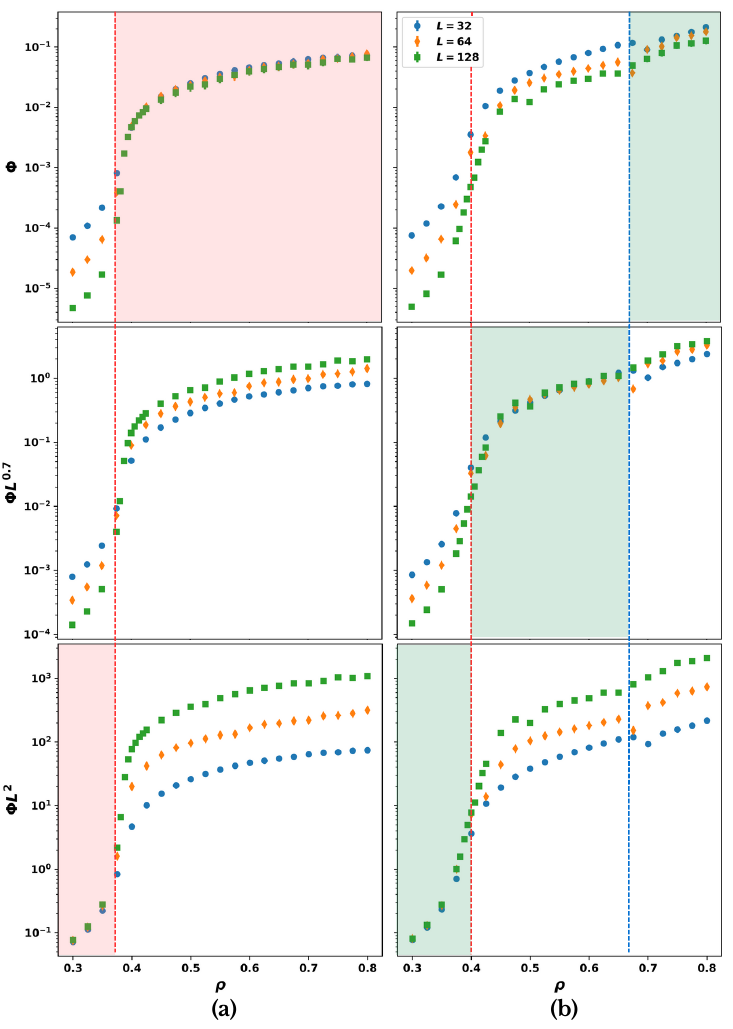}
\caption{Data collapse for the order parameter in the different density regimes, when scaled by different powers of $L$. (a): $\delta = 0$. (b): $\delta = 1$. Dashed lines mark the transition densities. In each case, the different regimes, where data best collapse, are highlighted in red (green) for $\delta =0(1)$. Error bars are mainly smaller than the size of the symbols. Observe the absence of a third regime in the undriven $\delta=0$  case (middle-(a) panel); see text for details.}%
\label{fig:ScaledOP}%
\end{figure}

Another indication of the presence of a new regime between the disordered and regular-stripe regimes is that $\Phi\left(\rho,L\right) $ scales differently with $L$ in three regions. As noted above, $\Phi\sim O\left(L^{-d}\right) $ in the disordered phase, rising to $O\left(1\right)$ in the ordered one for standard statistical systems. For example, for a non-conserved Ising model, the magnetization is nonzero only in the ordered phase (in the thermodynamic limit). In the driven FLM, it can be argued that we encounter \textit{anomalous} scaling in the intermediate, irregular stripe regime. To illustrate this phenomenon, we display plots of $\Phi\left(\rho,L\right)$ for $\delta=0$ [panels (a)] and $\delta = 1$ [panels (b)], scaled by various powers of $L$, in Fig.~\ref{fig:ScaledOP}. Note that these are semi-log plots, so that data of a wide range of magnitudes can be shown together. We can discern different scaling properties in the three regimes even for a limited range of $L$'s ($32,64,128$). In the lower panels, $\Phi L^{2}$ (i.e., $S_{-}^{\ast}$ itself) is shown and, noting the data collapse in the low density regime, $\rho<\rho_{\ell}$, for $\delta=1$ (or $\rho<\rho_{c}^{(0)}$ for $\delta=0$), we may conclude that $\Phi$ has the typical scaling behavior of a disordered phase. In the upper panels, we show that the $\Phi$ values collapse  best in the high density regime $\rho>\rho_{u}$ for $\delta =1$ (or $\rho> \rho_c^{(0)}$ for $\delta=0$), though considerable finite-size corrections may be at play. By contrast, for $\delta = 1$, the middle-right panel, corresponding to intermediate densities $\rho_{\ell}<\rho<\rho_u$, displays good data collapse for $\Phi L^{p}$, where $p$ is estimated as $p\sim 0.7$ for the best data collapse. Note that no such collapse occurs for $\delta = 0$ (or for any other value of $p$ between $0$ and $2$), in the middle-left panel, highlighting the absence of a third regime in the equilibrium FLM. The conclusion is that, for the driven FLM, there is a regime where $\Phi$ scales with a power of $L$ that is \textquotedblleft intermediate\textquotedblright\ between the standard values of $-2$ and $0$, corresponding to the usual disordered and ordered phases, respectively.
This is our  irregular stripe regime.

\begin{figure*}[th]
\centering\includegraphics[width=6in]{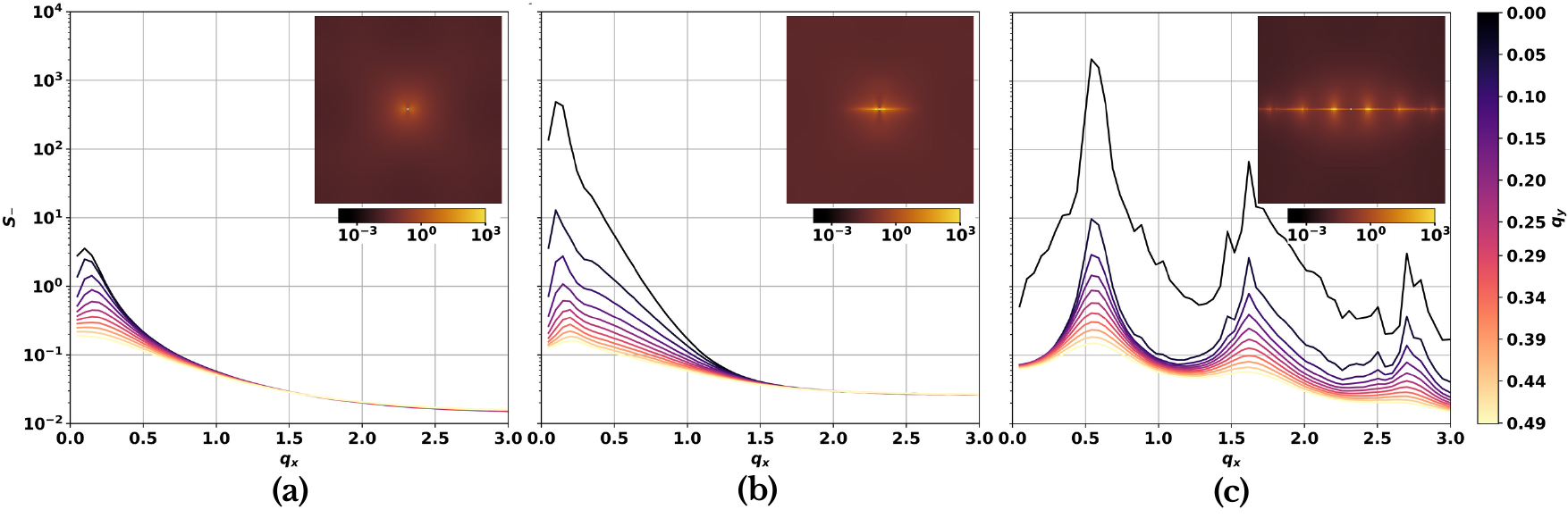}
\caption{$\delta = 1$: The charge fluctuation structure factor, as a function of $q_{x}$ for $q_{y}\,L/2\pi=0,...,10$ (using $L=128$), in the three regimes: microemulsion ($\rho=0.39$) (a), irregular stripes ($\rho=0.60$) (b), and regular stripes ($\rho=0.80$) (c). Note that the $S_{-}$ axis is logarithmic, so that, for the ordered case, there is an extremely sharp \textquotedblleft ridge\textquotedblright\ at $q_{y}=0$. For the irregular stripe regime in  (b), this ridge is quite sharp also, but not as extreme.}%
\label{fig:S_(qx,qy)}%
\end{figure*}

A further argument for the irregular stripe regime comes from more detailed data  involving   sections of $S_{-}\left(q_{x},q_{y}\right)$. In Fig.~\ref{fig:S_(qx,qy)}(a-c), we present an illustration ($\delta=1,L=128$) of how $S_{-}$ varies with $q_{x}$ for the first decade of $q_{y}\, L/2\pi=0,...,10$ in the three regimes. Deep in the homogeneous (microemulsion) regimes, $S_{-}$ \textit{roughly} follows the Orstein-Zernike form [with two major differences: a discontinuity singularity at $\mathbf{q}\equiv (q_x,q_y)=(0,0)$ and a peak at a nonzero, $L$-independent, $\mathbf{q}^{\ast}=\left( q^{\ast},0\right)$]. In other words, the values of $S_{-}\left(q_{x},q_{y}\right) $ are comparable to $S^{\ast}$ (i.e., the peak is quite broad), as seen in Fig.~\ref{fig:S_(qx,qy)}(a). In the opposite extreme, deep in the ordered regime, the peak is sharp and narrow  (with a value increasing with the system size), as seen in Fig.~\ref{fig:S_(qx,qy)}(c). In the irregular stripe regime [Fig.~\ref{fig:S_(qx,qy)}(b)], the peak value at $\mathbf{q}^*=(q^*,0)$  falls between the two phases and also falls off more slowly as we increase $q_y$.cTo quantitatively appreciate how rapidly $S_{-}$ falls from its maximum values $S_{-}^{\ast}$ as we move away from the ridge, we display the numerical values in Table~\ref{table:S_values}.

\begin{table}[htbp]
    \centering
    \begin{tabular}{|c|c|c|c|c|}
\hline
$\rho$ & $q^* $ & $S_{-}\left(q^{\ast},0\right)$  & $S_{-}\left(  q^{\ast}, 0.05\right)$  & $S_{-}\left(q^{\ast},0.1\right)$ \\
\hline
\hline
$0.39$ & $0.1$  & $3.55$    & $2.47$ & $1.27$\\
$0.60$ & $0.1$  & $486.55$  & $12.90$ & $2.26$\\
$0.80$ & $0.54$ & $2083.81$ & $9.68$  & $2.93$\\
\hline
    \end{tabular}
    \caption{Comparison of the charge fluctuation structure factor values (using $L=128$), $S_{-}\left(q^{\ast},q_y\right)$,  at the peak position occurring at $q_y=0$  (third column) and at locations slightly displaced from the peak, $q_{y} \, L/2\pi=1,2$ (fourth and fifth columns, respectively) for different values of $\rho$ corresponding to the three regimes: microemulsion (first row), irregular stripes (second row), and regular stripes (third row).}
    \label{table:S_values}
\end{table}

\begin{figure}[ht]
    \centering
    \includegraphics[width=0.6\columnwidth]{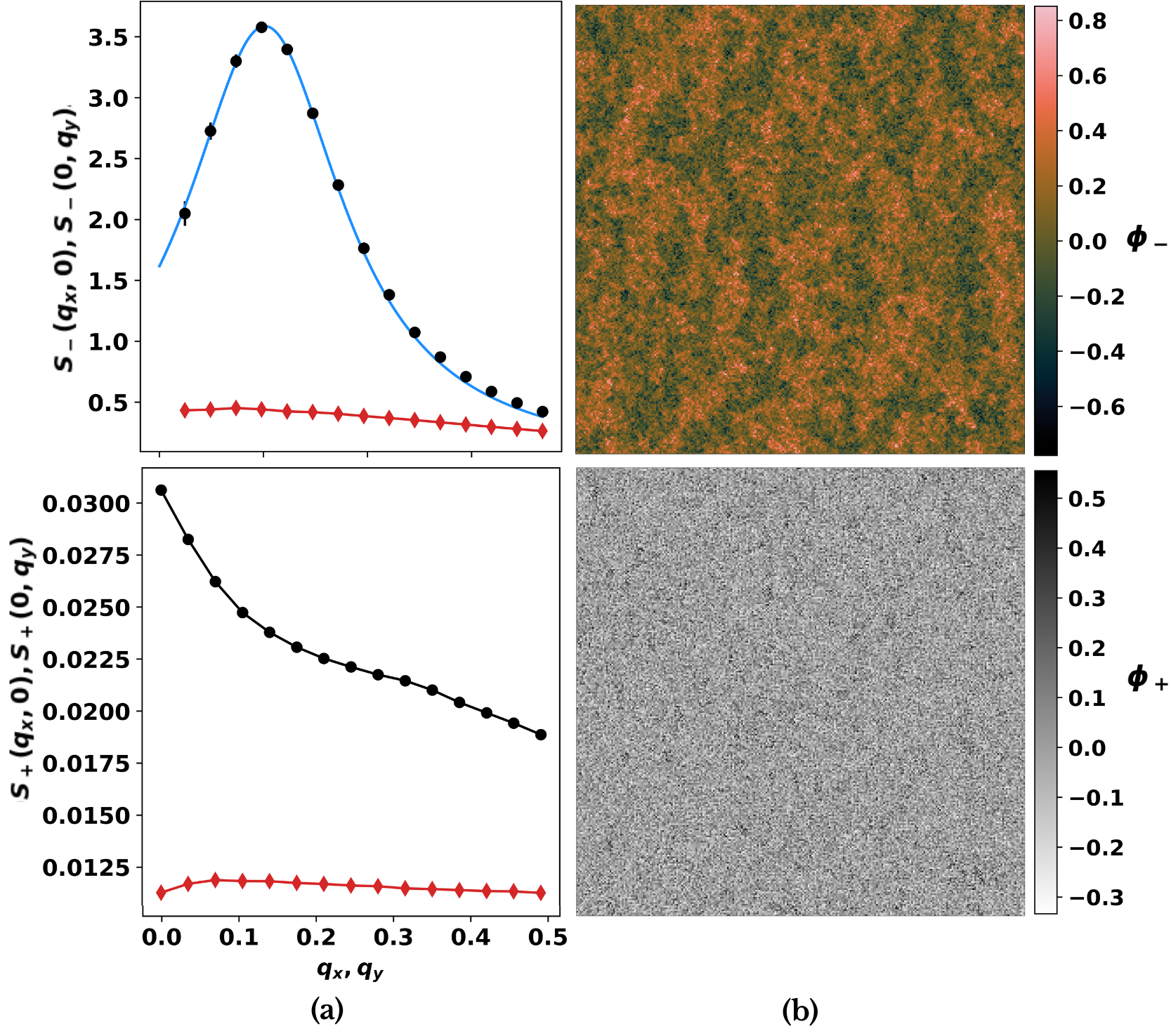}
    \caption{Steady-state structure factor (SF) in the low-density regime using $L=256$ for $\rho=0.39$ and $\delta = 1$, averaged over $25$ realizations. Upper panels: (a) Charge fluctuation (CF) SF. (b) CF typical field configuration. Lower panels: (a) Density fluctuation structure factor (DF SF) (b) DF typical field configuration. The black circles (red diamonds) are the SF parallel (perpendicular) to the drive direction, $S_{\pm}(q_x,0)$ and $S_{\pm}(0,q_y)$, respectively. Error bars are smaller than the symbols. The blue line is the fit $A_1\, q^2[A_2\, q^2 + q^2(q-q^*)^2]^{-1}$ using $20$ points around the peak. Analogous behavior is found for $L=128, 192$. We find $q^* = 0.101(1)$ for $L=128,192$ and $256$. Note the shoulder in the black line at $2q^*_-$, indicating the emergence of the coupling of the density and charge fields.}
    \label{fig:lowd-figs-largeL}
\end{figure}

The peak positions of the CF SF, which occur at $\left(q^{\ast},0\right)$, are listed in the second column, while the next three columns are associated with the $S_-$ values at $\left(q^{\ast}, q_y\right)$, with $q_{y}\, L/2\pi=0, 1$ and $2$. Note that the ratios of the entries in the third and fourth columns are dramatically different in the three cases: $\sim 1.4$, $38$, and $215$ for the microemulsion phase, the irregular stripes, and the regular stripes, respectively. Remarkably, once $q_{y}$ reaches $3 \cdot2\pi/L$, there is hardly any difference between the three rows! We also call attention to the $q^{\ast}$ of the irregular stripe regime, which is the same (for this $L$) as that of the disordered phase, but considerably smaller than that in the regular stripe regime, which may be also observed in Fig.~\ref{fig:selectedmode} where the stripes at high density are visibly narrower. In Fig.~\ref{fig:lowd-figs-largeL}, we plot the structure factors for $L=256$. The peak in the low density, disordered phase is well fit by a function of the form $A_1 q^2 \left[A_2 q^2 + q^2(q - q^*)^2\right]^{-1}$, consistent with the behavior observed in the DWRLG \cite{DWRLG1, DWRLG2}. Upon examination of other system sizes, e.g., $L=128$ and $192$, we consistently find $q^* = 0.101 \pm 0.001$ for the disordered  regime.

We conjecture that as $\rho$ is increased, the system first has no order, then long-range order just along $\hat{\mathbf{j}}$ (\textit{transverse} to the drive), and lastly, transitions to a state of full order. A final demonstration of this can be done by binning the  column-averaged particle densities:
\begin{equation} \label{eq:avg-over-columns-rhos}
\bar{\rho}_{A,B}\left(i\right)  \equiv\frac{1}{L}\sum_{j}\rho_{A,B}\left(i,j\right),
\end{equation}
which lie in $\left[ 0,1\right]$. Histograms of these densities  are sensitive to ordering only in columns, i.e., a measure of the presence of well-defined stripes (along the direction perpendicular to the drive), but not their widths. Exploiting the equivalents in the DF and CF description, we define%
\begin{align} 
&\bar{\phi}_{+}\left(  i\right)  \equiv\bar{\rho}_{A}\left(i\right) +\bar{\rho}_{B}\left(  i\right)  -\rho, \nonumber \\
&\bar{\phi}_{-}\left(  i\right)\equiv\bar{\rho}_{A}\left(  i\right) -\bar{\rho}_{B}\left(i\right). \label{eq:avg-over-columns-phis}
\end{align}
Since $\bar{\phi}_+(i)$ is related to the hole density in a column, it carries crucial information on the location of interfaces between the two species. Our goal is to study histograms, $H\left(\bar{\phi}_{\pm}\right)$, constructed from $1000$ time points sampled every $10^4$ MCS in the steady-state, each of which provides $L$ values of $\bar{\phi}_{\pm}$.

\begin{figure*}[th]
\centering\includegraphics[width=5in]{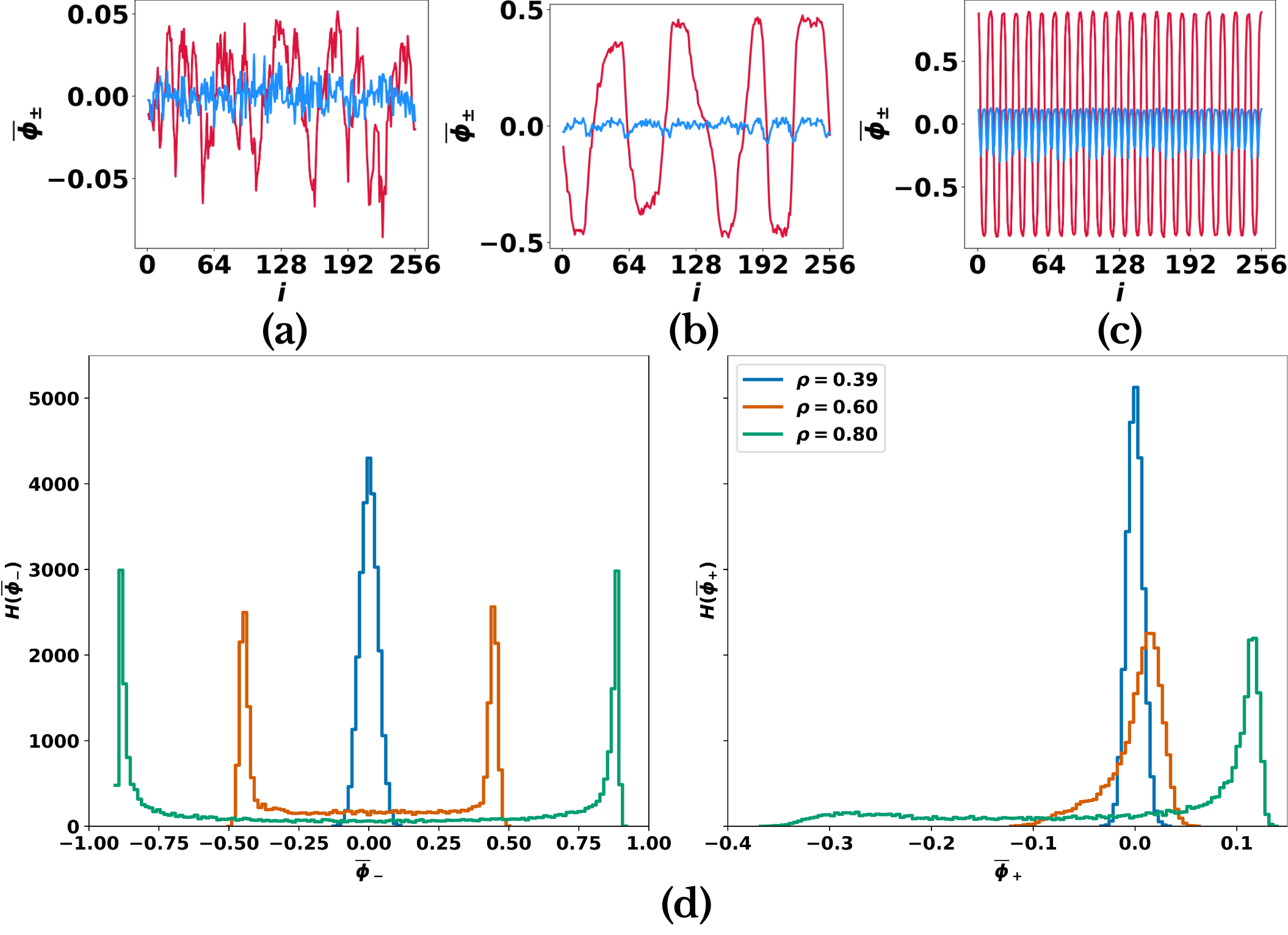}
\caption{Column-averaged fields $\bar{\phi}_{\pm}$ [Eq.~\eqref{eq:avg-over-columns-phis}]. Panels (a)-(c): examples of $\bar{\phi}_{\pm}$ at a single representative time point for overall densities: $\rho = 0.39$ (a), $0.60$ (b), and $0.80$ (c). Panel (d) Histograms $H\!\left(\bar{\phi}_{\pm}\right)$ of the densities presented in panels (a-c), compiled in a single plot for comparison. Curves for the three densities are labled by blue, red, green, respectively.}%
\label{fig:Hist}%
\end{figure*}

When the system is in the homogeneous phase, we expect the histograms to have a simple Gaussian-like peak around $0$. As ordering builds in each column, the peak in $H\left(\bar{\phi}_{-}\right)$ should split symmetrically, much like a histogram of the magnetization in an Ising model when $T$ drops below $T_{c}$. This phenomenon indeed occurs, as illustrated by the blue and orange lines on the left panels of Fig.~\ref{fig:Hist}(d) for $\rho=0.39$ and $\rho=0.60$, which correspond to densities in the microemulsion phase (just below stripe formation) and the irregular stripe regime, respectively. In this context, we may roughly estimate $\rho_{\ell}\simeq 0.40$ from the peak split. With further increases of $\rho$, we see that the twin peaks move further apart, but with no obvious signal corresponding to the upper transition density $\rho_u$. As shown by the green line on the left panel of Fig.~\ref{fig:Hist}(d), the peaks are located close to the extremal values ($\bar{\phi}_{-}=\pm 0.9$) for the case of $\rho=0.80$. On the other hand, the density fluctuation histograms, $H\left(\bar{\phi}_{+}\right)$, seem to be sensitive to the upper transition density $\rho_u$ but not to the lower one, $\rho_{\ell}$, as the histograms seen on the right panels of Figs.~\ref{fig:Hist}(d) indicate. For $\rho=0.39$ (blue curve), $H\left(\bar{\phi}_{+}\right)$ is still sharply peaked around the origin. Then, we see that $H\left(\bar{\phi}_{+}\right)$ begins to develop considerable asymmetry for $\rho=0.60$ (orange curve), indicating the development of an (asymmetric) bimodal distribution. The histogram is then clearly bimodal at $\rho=0.80$ (green curve), where we see a sharp peak (near $\bar{\phi}_+ \sim 0.1 $) accompanied by a broader distribution of hole-rich columns ($\bar{\phi}_+ \lesssim -0.2$). The transition in the shape of $H\left(\bar{\phi}_{+}\right)$ can be used to estimate $\rho_{u}$, yielding a consistent result with our previous analyses, $\rho_u\simeq 0.65$. Figs.~\ref{fig:Hist}(a-c) are the typical configurations of $\bar{\phi}_\pm(i)$, with the average DF(CF) in blue(red).

To summarize, we introduced a novel driven diffusive system: a hybrid between the driven Widom-Rowlinson lattice gas \cite{DWRLG1} and its associated statistical field theory \cite{DWRLG2}. Following the spirit of coarse graining a lattice gas, this \textquotedblleft field-based lattice model\textquotedblright\ consists of continuous densities defined on a discrete lattice. Though the properties of the \textit{undriven} version are, as expected, those of the WRLG, unexpected phenomena arise under a drive. While the two are comparable for $\rho\lesssim 0.40$ and share some similarities for $\rho\gtrsim0.70$, conspicuous differences appear at intermediate densities. We propose the presence of an irregular stripe regime in which, qualitatively, there is  long-range order  perpendicular to the drive and disorder along the drive direction. To arrive at quantitative conclusions would require extensive simulations beyond the scope of the present study.

\section{\label{sec:Continuum}Continuum Description}

The out-of-equilibrium transition to a striped regime from a structured disordered regime (microemulsion) of both the FLM and the DWRLG for $\delta>0$ raises the possibility that similar behavior characterizes a broader class of models. Whether a continuum coarse-grained field-theoretic approach can describe such class is unclear, especially for high particle densities. In a previous study \cite{DWRLG1}, the simplest continuum field theory, a linear stochastic partial differential equation (SPDE)  based on a FLM-like model, was proposed, yielding a successful phenomenological picture of the low-density phase of the DWLRG. A more systematic derivation of a stochastic field theory, exploiting the Doi-Peliti formalism, was undertaken in Ref.~\cite{DWRLG2}, providing good approximations for the noise terms and the drive-dependent quadratic terms in the SPDE. Treating fluctuations perturbatively, the study recovered the chief novel feature of the disordered phase: a charge fluctuation structure factor peak at a nonzero wavelength ($\mathbf{q}^{\ast }\neq 0$), characteristic of a microemulsion. However, approximations necessary within the latter scheme render higher-order terms in the field theory questionable, motivating a different approach for finding a continuum description of the stripe formation at high densities and a full picture of the transition to stripes. 

Here, we build on the success of the FLM, developed in Sec. \ref{sec:Model}, examining its continuum limit [integer-valued $\mathbf{s}=\left( i,j\right)$ to continuous $\mathbf{r}=\left( x ,y \right)$]. Key details of the calculation are given in Appendices~\ref{appx:gradient-expansion} and \ref{appx:coefficients}. Specifically, we seek a minimal coarse-grained model of a two-component, coupled diffusive system exhibiting a transition at $\rho_c$, such that:
\begin{itemize}
    \item the system is phase-separated for $\rho>\rho^{\mathrm{FT}}_c$, exhibiting usual coarsening for $\delta = 0$ and stripes perpendicular to the drive for $\delta > 0$.
    \item the system is well mixed for $\rho<\rho^{\mathrm{FT}}_c$, exhibiting a homogeneous disordered phase for $\delta=0$ and a ``microemulsion'' with a characteristic length scale along the drive direction for $\delta >0$. We also expect a $\delta$-dependent discontinuity of the SFs at the origin.
\end{itemize}

Let us begin with a short overview of the procedure: we start with the rules introduced in Sec.~\ref{sec:Model} for mass transfer for $\rho _{A,B}(\mathbf{s},t)$ associated with all possible NN and NNN hops to/from site $\mathbf{s}$. Then, we set $\mathbf{r}=\mathbf{s}\ell $, where $\ell $ is a lattice spacing, and write the corresponding difference equations for the updated densities $\rho_{A,B}(\mathbf{s},t +\Delta t)$ due to a single hop occurring over a time interval $\Delta t$. Taking the $\ell,\Delta t \rightarrow 0$ limits and applying a gradient expansion (GE), we cast the difference equations in the form of driven-diffusion continuity equations for the coupled pair of charge and density fluctuations, $\phi_{\pm }(\mathbf{r},t)$:
\begin{equation}\label{eq:continuity}
    \partial_t\phi_\pm(\mathbf{r},t) + \nabla\cdot \mathbf{J}_\pm + \partial_xJ^\delta_\pm = 0 ,
\end{equation}
where $\mathbf{J}$ and $J^\delta$ are the drive-independent and drive-dependent currents, respectively (with $\ell $ and $\Delta t$ absorbed into the coefficients of the various terms in the currents). Seeking minimal expressions for the $J$'s via a GE, we will arrive at a continuum description [see Eq.~\eqref{eq:eom-final} below] which we study in detail in the next section. Note that the noise terms cannot be derived systematically in this approach. Instead, we will add the simplest form of noise (zero mean, conserved-Gaussian), keeping the variance as a free parameter. To arrive at a minimal description, we limit ourselves to the lowest possible order, in both the fields and gradients, maintaining, in the spirit of Ginzburg-Landau theory, high-order terms only as needed for stability. 

Deferring the details of the derivation to Appendix~\ref{appx:gradient-expansion}, we arrive at the following SPDEs,
\begin{align}\label{eq:eom-final}
\partial _{t}\phi _{+}& =D_{+}\nabla ^{2}\phi _{+} - \Gamma_+\nabla^4\phi_+\nonumber \\ 
&\qquad \quad   - \lambda \nabla ^{2}\phi_{-}^{2} -g_{-}\partial _{x}\phi _{-}^{2} + g_{+}\partial _{x}\phi _{+}^{2}+\xi _{+},  \nonumber \\
\partial _{t}\phi _{-}& =D_{-}\nabla ^{2}\phi _{-}-\Gamma _{-}\nabla^{4}\phi _{-}-\Delta v\,\partial _{x}\phi _{-} \nonumber \\
&\qquad \quad  +\lambda \nabla (\phi_{-}\nabla \phi _{+})+g_{0}\partial _{x}(\phi _{+}\phi _{-})+\xi _{-},
\end{align}%
where $\xi_{\pm}$ are conserved-Gaussian noises with zero mean, zero cross-correlation, and autocorrelation
\begin{align}\label{eq:noise-in-eq}
&\langle\xi_\pm(\mathbf{r},t)\xi_\pm(\mathbf{r}',t')\rangle = -\sigma^2\nabla^2\delta(\mathbf{r}-\mathbf{r}')\delta(t-t'),
\end{align}
with $\sigma$ an effective noise intensity \footnote{Note that, more generally, the noises acting on $\phi_{\pm}(\rvec,t)$ originate from the noises in the underlying densities $\rho_{A,B}(\rvec,t)$, which are inherently multiplicative: i.e., $\xi_A$ vanishes in regions devoid of A particles, and analogously for $\xi_B$. If we consider small perturbations from a state with $\phi_+ \sim \rho$ and $\phi_- \sim 0$, the fluctuations are additive to a first approximation (as assumed here)}. 

In the equation above, $D_{\pm}$, $\Gamma_{\pm}$ and $\lambda$ are $\rho$--dependent coefficients, while $g_{\pm,0}$ and $\Delta v$ are proportional to $\delta$. The diffusion constants for the DF and CF are given by $D_+$ and $D_-$, respectively. In equilibrium, the free energy cost per unit area to form an interface for each corresponding field is $\propto \Gamma_{+,-}$. Moreover, in analogy to Model B dynamics, the charge diffusion coefficient $D_-$ changes sign from positive to negative at the transition point and, as such, the fourth-derivative terms are necessary to ensure the stability of the theory as well. We have moved to the co-moving frame of the DF, so that the fields' characteristic velocities are encoded in a single drive-dependent linear parameter, $\Delta v$. As pointed out in Ref.~\cite{DWRLG2}, and as will be shown in our numerical studies, $\Delta v$ plays the \textit{key} role in generating both the periodic correlations in the low-density phase and stripes in the high-density regime \footnote{A similar feature occurs in the analysis of kinematic waves in drifting crystals \cite{driftingcrystal1}, where different characteristic velocities also play a central role in creating those periodic structures.}. The nonlinear parameters $g_{\pm,0}$ and $\lambda$ account for the drive-independent and drive-dependent contributions to the effective $A$-$B$ interaction, respectively. 

The choice of signs in Eq.~\eqref{eq:eom-final} follows naturally from the GE of the FLM \cite{thesisgui}, so that the equation is manifestly stable by taking all couplings positive. The linear coefficients are fixed through the GE, but the functional dependence on $\rho$ of the nonlinear parameters is challenging to establish, which is why we shall keep them free. From the functional dependence of $D_-$ on $\rho$, we find that this quantity changes sign at $\rho_c^{\mathrm{FT}}= 0.37\dots$, which is remarkably close to the $\rho _{c}^{\left( 0\right) }\simeq 0.39$ result for the critical density estimated from simulations. The exact expression of the linear coefficients obtained via the GE, and key details of the calculation, are outlined in Appendix~\ref{appx:coefficients}. In summary, our formulation in terms of Eq.~\eqref{eq:eom-final} depends on five free parameters: $\lambda$, $g_\pm$, $g_0$, and $\sigma$. As noted in Appendix~\ref{appx:coefficients}, the term proportional to $g_+$ can be neglected for $\rho > \rho_c^{\mathrm{FT}}$. For $\delta = 0$, we have $\Delta v = g_{\pm,0} = 0$, leaving only the terms proportional to $\lambda$.

Finally, it is interesting to observe that the large-scale properties of the $\delta =0$ system are isotropic, so that there is just one $D, \Gamma, \sigma$ for each field (instead of $D_{x},\,D_{y}$, etc.). Moreover, the fluctuation-dissipation relation constrains the noise correlations to be isotropic, as well. However, it is also known that this relation is \textit{violated} in driven-diffusive systems and various anisotropies emerge, with the most significant manifestation being a \textit{discontinuity singularity} of $S\left( \mathbf{q}\right) $ at the origin \cite{DDSbook}. Although it is easy to argue qualitatively how the (anisotropic) drive induces such features, it is difficult to obtain quantitative predictions from the microscopic models. Thus, previous studies of the DWRLG \cite{DWRLG1,DWRLG2} tend to rely on phenomenological approaches to arrive at, say, the difference $D_{x}/\sigma _{x}-D_{y}/\sigma_{y}$. In Sec.~\ref{sec:numer-int}, we show that, \textit{despite} starting with isotropic choices for parameters like $D$, $\Gamma $, $\lambda $, and $\sigma$, the many surprising phenomena of our driven system \footnote{As a reminder, a linear SPDE can be solved analytically. Not surprisingly, as they correspond to non-interacting fields, having isotropic parameters, the structure factors display only isotropic Ornstein-Zernike forms, despite the presence of $\Delta v$}, e.g., the discontinuity singularity, emergence of a peak at $\mathbf{q}\neq \mathbf{0}$, and formation of stripes arise in our numerical solutions due to the interplay of the drive (via $\Delta v$), the interactions (via non-linearities), and the noise. 

\section{Numerical integration of the Stochastic Field Theory\label{sec:numer-int}}

In this section, we integrate Eq.~\eqref{eq:eom-final} numerically to study how the possible steady states depend on the various parameters of the model. Our goal in this section is to assess whether the continuum stochastic field theory we propose in Eq.~\eqref{eq:eom-final} is capable of reproducing the different regimes encountered both in the DWRLG and the FLM, namely the microemulsion structure in the low-density phase, and stripes aligned perpendicular to the drive in the high-density phase. Additionally, we check whether the continuum model also exhibits the irregular stripe regime present in the FLM.

To integrate Eq.~\eqref{eq:eom-final} numerically in 2D with periodic boundary conditions, we design a numerical SPDE solver using the PSSETD2H and smooth dealiasing. To fix the timestep $\Delta t$, we run simulations at several values of $\Delta t$ and choose the largest value such that there are no appreciable changes in the results; in practice we use $\Delta t= 0.5$. Unless otherwise specified, simulations are performed on a domain of size $L = 128$, using grid resolutions of $N = 200$ points per direction \footnote{This mesh size is chosen to balance numerical accuracy and computational cost. We have verified that further refining the mesh does not produce any appreciable changes in the results.}. We also adopt the following initial conditions for the fields $\phi_\pm$:

\begin{equation}
\begin{dcases}
     \phi_-(\rvec,0) = \sigma_0 \, \xi(\rvec) \\
     \phi_+(\rvec,0) = 0
\end{dcases}
\end{equation}
where $\sigma_0=10^{-2}$ and $\xi(\rvec)$ is a conserved Gaussian noise with zero mean and unit variance.

\begin{figure}[ht]
    \centering  \includegraphics[width=0.6\columnwidth]{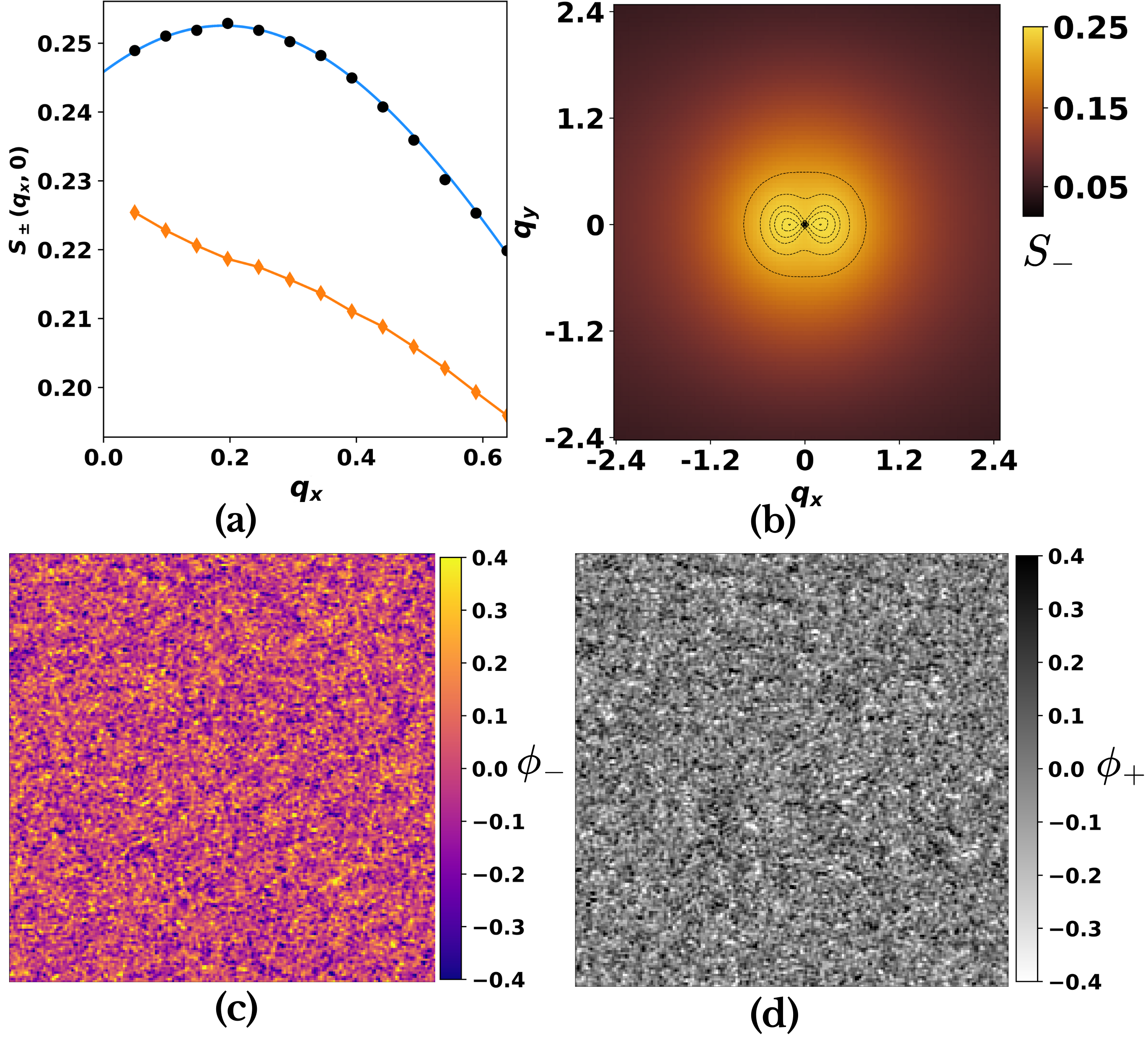}
    \caption{Microemulsion phase for the stochastic equations of motion; Eq.~\eqref{eq:eom-final}. Linear coefficients are fixed by Eq.~\eqref{eq:linear-coefs}, using $\rho = 0.2,$ and $\delta =1$. The noise intensities are fixed to $\sigma = 0.27$. The other parameters are retrieved from \cite{DWRLG2} using the same $\rho$ and $\delta$: $\lambda=0,\, \, g_0 = 4.243,\, g_-=0.707$ and $g_+=2.828$. (a) Charge fluctuation (CF, black circles) and density fluctuation  (DF, orange diamonds) structure factors (SFs) along the drive direction. The error bars are smaller than the symbols. Blue line is the best fit $A_1\, q^2\left[A_2\, q^2 + q^2(q-q^*)^2\right]^{-1}$, using $10$ points around the peak. We find $q^* = 0.190\pm 0.009$. Note the shoulder in the DF SF at $2q^*$. (b) CF SF. Since the peak near the origin is difficult to discern from just the color map, we also mark the contour of the levels at $0.8, 0.9, 0.95, 0.97, 0.99$ and $0.999\cdot S_-(q^*,0) \ (\simeq 0.253)$ by black dashed lines to guide the eyes. (c) CF and (d) DF typical steady-state field configurations. This figure is analogous to Fig.~\ref{fig:lowd-figs-largeL} for the FLM (and Fig. 10 in \cite{DWRLG2} for the DWRLG). A movie related to this figure is available in \cite{movies}.}
    \label{fig:ft-microemulsion}
    \end{figure}

In the low-density equilibrium regime, $\rho < \rho_c^{\mathrm{FT}} = 0.37\dots$ and $\delta =0$, for $\lambda \geq 0 $, both the CF and DF steady-state structure factors are isotropic and follow the usual Ornstein-Zernike form. For a fixed drive $\delta$ in this regime, exploring the full phenomenology of the model requires scanning a four-dimensional parameter space (or five, if the noise amplitudes are included), which is computationally demanding. Fortunately, we can simplify the task since our main goal is to find the various regimes of the lattice models. Thus, first, at low densities $\rho$,  the non-linearities facilitating coarsening dynamics should be irrelevant (as $D_->0$), so we  set $\lambda=0$. Then, taking advantage of the perturbative analysis  in Ref.~\cite{DWRLG2}, we set $g_0 = 4.243$, $g_- = 0.707$, $g_+ = 2.828$, and $\sigma = 0.27$, which should correspond to the microemulsion regime. Fig.~\ref{fig:ft-microemulsion} shows the steady-state charge fluctuation SF, $S_-(\mathbf{q})$, in the resulting microemulsion phase, averaged over $5000$ time points evenly spaced across the final $10^6$ time steps, and using $150$ independent realizations. As expected, $S_-(\mathbf{q})$ exhibits both a discontinuity at the origin and the emergence of a peak at a nonzero wavevector, $q^* = 0.190 \pm 0.009$, along the drive, closely matching the behavior of the FLM.  The peak is well fit by the same fitting form used for the FLM,  $A_1\, q^2\left[A_2\, q^2 + q^2(q-q^*)^2\right]^{-1}$.  These results are consistent with previous studies \cite{DWRLG2}, where it was also noted that the field theory smoothes the peak region, making it noticeably less sharp than in the lattice-based models.

\begin{figure}[ht]
    \centering
\includegraphics[width=0.65\columnwidth]{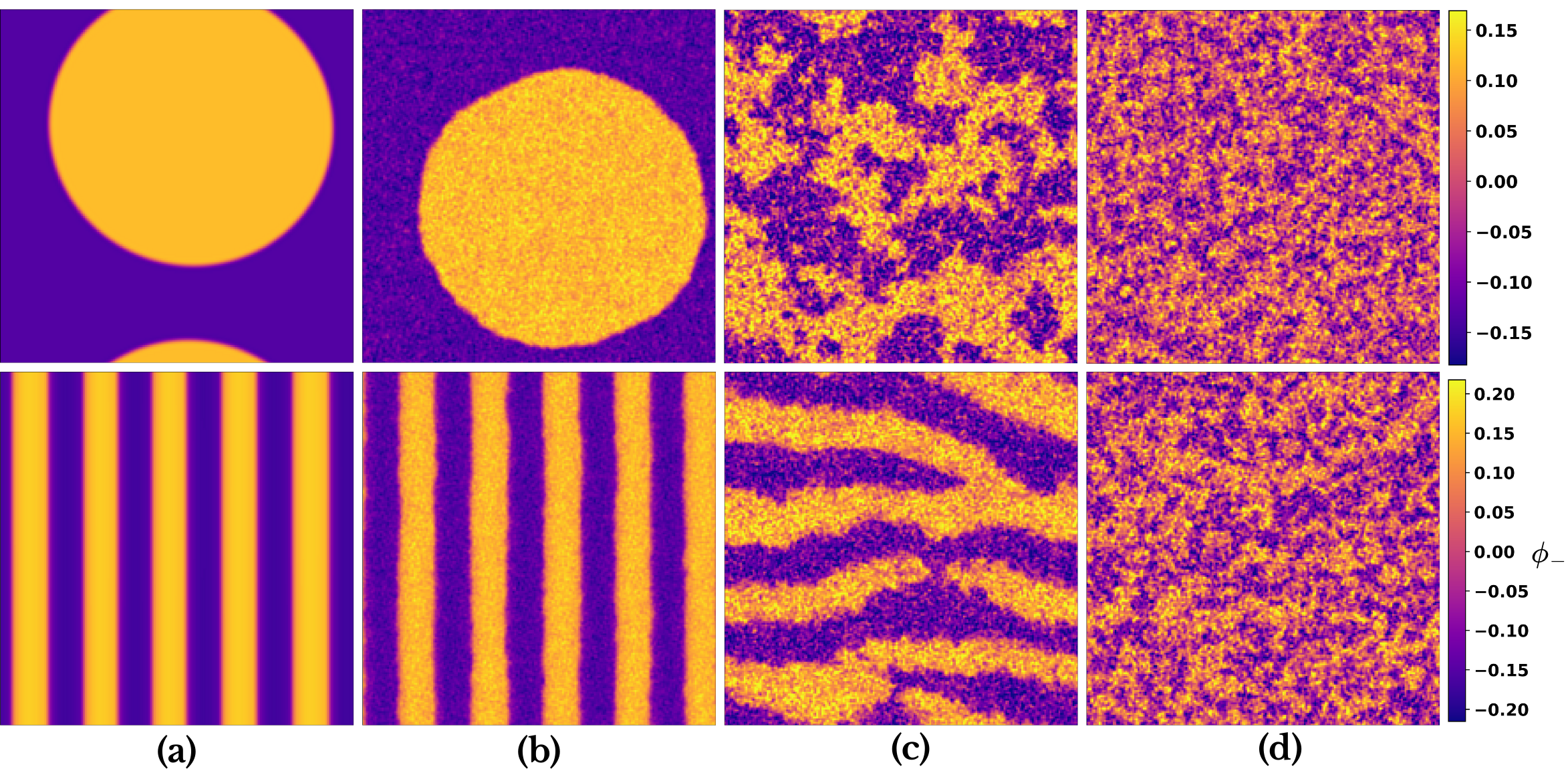} 
    \caption{Comparison of the charge fluctuation steady-state field configuration with and without the additive noise in Eq.~\eqref{eq:eom-final} for $L=128$, $\rho=0.6$ and $\lambda = 1.0$. From left to right, the effective noise intensities (Eq.~\eqref{eq:noise-in-eq}) are $\sigma =0$ (a), $0.01$ (b),  $0.03$ (c) and $0.04$ (d). Upper plots: $\delta = 0$. Lower plots: $\delta = 1, g_0 = 0, g_-= 0.5$. No appreciable quantitative differences in the results are found for (a) and (b), in agreement with results for Model A and B in \cite{Garcia-Sancho}. Raising the noise strength further makes the dynamics noise-dominated and promotes stripe reorientation.}
    \label{fig:comparison}
\end{figure}

Another simple case is the high-density equilibrium regime,    $\rho > \rho_c^{\mathrm{FT}}$ and $\delta = 0$. Here, all the drive-dependent couplings vanish and we simply set a positive value for $\lambda$. In this case, a steady-state amplitude is systematically found \footnote{This is supported by the weakly nonlinear stability analysis of Eq.~\eqref{eq:eom-final} \cite{thesisgui}, which can also be used to crudely estimate the amplitude values slightly above the onset, $\rho \gtrsim \rho_c^{\mathrm{FT}}$. On the other hand, for $\lambda < 0$, Eq.~\eqref{eq:eom-final} has no bounded solutions and the numerical integration is ill-behaved, thus making necessary the inclusion of higher-order derivative terms.} for the charge and density fluctuations. In all the runs performed, we consistently find two types of steady-state configurations: a round domain that resembles a bubble (see Fig.~\ref{fig:comparison}), or a single stripe whose orientation varies with realization. In either case, there is a single domain dominated by each species, which is consistent with the results of the FLM in the undriven regime. Furthermore, we verify that our equations of motion follow the Lifshitz-Slyozov law at high densities ($\rho>\rho_c^{\mathrm{FT}}$), with the growth of the SF peaks scaling as $t^{1/3}$, so that the characteristics of the coarsening dynamics do not depend on the particular value of $\lambda$. In fact, the cubic saturation terms due to $\lambda$ are more relevant to fix the magnitude of the steady-state amplitude and, as a result, their actual values may depend on the choice of $\lambda$. In the context of the Cahn-Hilliard equation or Model B \cite{Garcia-Sancho,Yoon,Kim2016}, these couplings are often set to unity, which corresponds to a rescaling of the field values. Therefore, for simplicity, we fix $\lambda=1.0$ for the remainder of this section.

The driven, high-density regime involves a large parameter space. Here, we cannot rely on the coupling values obtained in Ref.~\cite{DWRLG2}, as they are specialized for the low-density phase. However, we can reduce the parameter space by discarding the term $ g_+ \phi_+^2$ in Eq.~\eqref{eq:eom-final}; see the simplifying assumptions in Appendix~\ref{appx:coefficients}. We then proceed by fixing a reasonable value for the noise strengths, $\sigma$. Fig.~\ref{fig:comparison} compares typical runs with and without the conserved noises $\xi_{\pm}$ in \eqref{eq:eom-final} for $\rho=0.7$, $\delta =0$; and $\rho=0.7$, $\delta =1.0$, $g_-=0.5$, $g_0=0$. The linear coefficients are given by the expressions in \eqref{eq:linear-coefs}. We find that quantitative properties in the ordered phase are largely insensitive to the noise amplitude within a moderate range \footnote{This is not true, however, in the low-density regime and at interfaces, as we have verified for the microemulsion phase. In fact, for $\rho<\rho_c^{\mathrm{FT}}$, the fields will decay to zero, even if $\delta$ is maximal. Therefore, the noises act as source terms, which are crucial to the development of the microemulsion behavior. In another words, the microemulsion phase \textit{cannot} be observed without the noise terms in Eqs.~\eqref{eq:eom-final}.}, consistent with observations for Models A and B in \cite{Garcia-Sancho}. Accordingly, we fix the noise intensity to a moderate value, $\sigma = 0.01$. At very high noise levels, however, the system displays qualitatively new behaviors, most notably stripe reorientation [see Fig.~\ref{fig:comparison}(c,d)], suggesting that the high-noise limit of the model may contain additional phenomena worthy of future investigation.

\begin{figure}[ht]
    \centering
    \includegraphics[width=\columnwidth]{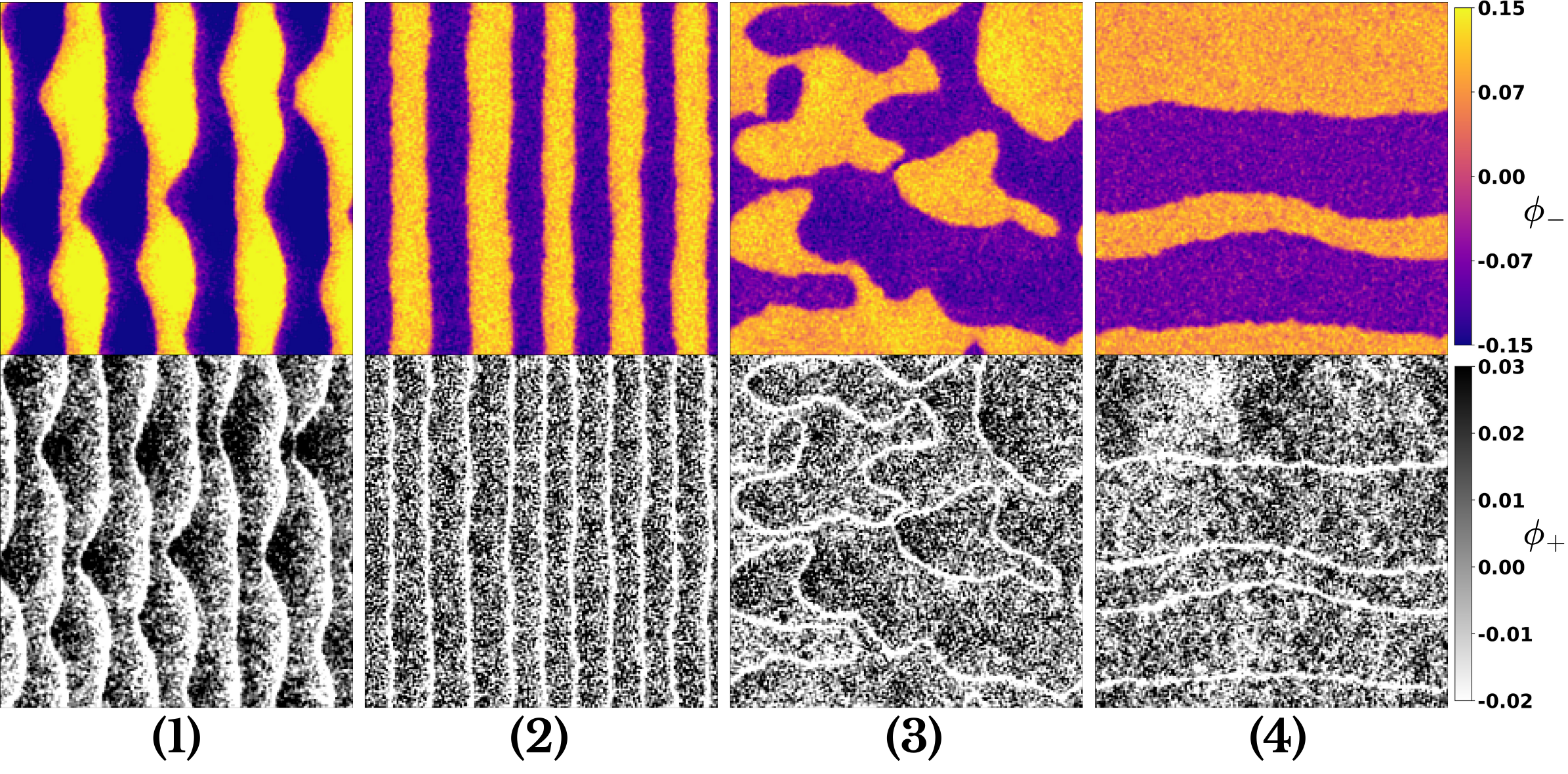} 
    \caption{Typical configurations of the field equations for regimes (1)-(4) mentioned in the text. Drive to the right. Upper (lower) plots are the charge fluctuation (density fluctuation) field configurations. We set $\rho = 0.7$, fixing all drive-independent linear coefficients, $D_\pm$, $\Gamma_\pm$ in Eqs.~\eqref{eq:eom-final} via the GE; see Eq.~\eqref{eq:linear-coefs}. In (1)-(3) we use $\delta = 1 $ and (1) $g_-=g_0 = 0$; (2) $g_-=0.5$ and $g_0 =0$; (3) $g_-=0.5$ and $g_0=0.05>g_0^*$. In (4), $\Delta v = 0.05<\Delta v^*$ (corresponding to $\delta = 0.24$), while $g_-=0.5$ and $g_0 = 0.0$. A movie related to this figure is available in \cite{movies}.}
    \label{fig:typconf-eom}
\end{figure}

Therefore, following these simplifications, we are left with two nonlinear couplings to consider: $g_0$ and $g_-$.  To sweep the parameter space, we shall limit ourselves to $g_0,g_- \in [0,1]$, which is a typical range indicated by the different coarse-graining prescriptions in the high-density regime. We also vary $\Delta v$, which depends on the drive and density as shown in  Eq.~\eqref{eq:linear-coefs}.  Figure~\ref{fig:typconf-eom} summarizes the four different regimes we encounter via numerical integration:
\begin{enumerate}
    \item Stripes perpendicular to the drive emerge already for $g_-=g_0=0$, solely due to the presence of the term proportional to the difference in the  charge and density field characteristic velocities: $\Delta v=v_--v_+$ [see Eq.~\eqref{eq:eom-final}]. However, in this case, the stripes are wavy or vertically modulated due to the absence of the nonlinear, drive-dependent saturation terms.
    \item For $g_0 \leq g_0^*$ and $g_- > 0$, the stripes retain their perpendicular alignment and the vertical modulation disappears. This suggests that the term proportional to $g_-$ in Eq.~\eqref{eq:eom-final} provides stripe stability. This regime has the best agreement between the field equations, the FLM, and the DWRLG phenomenology.
    \item  For $g_0 > g_0^*$ and $g_- \geq 0$, the perpendicular alignment is lost. In this regime, either horizontal stripes emerge or no stripe formation occurs at all, depending on the precise values of $g_0$ and $g_-$. Large values of $g_0$ destabilize vertical stripes, promoting alignment parallel to the drive.
    \item For $0<\Delta v < \Delta v^*$, vertical stripes are not formed in any simulation. This supports the conclusion that while $\Delta v > 0$ is a necessary condition for stripe formation with the correct alignment, it is not sufficient: $\Delta v$ must exceed a critical threshold $\Delta v^*$ to stabilize the vertical stripe pattern.
\end{enumerate} 
As an example, for $\rho = 0.7$, we can estimate  roughly that $\Delta v^* = 0.08(2)$ (associated, for this density, to $\delta \approx 0.38$) and $g_0^* = 0.02(1)$. Fig.~\ref{fig:typconf-eom} shows typical configurations for the previously mentioned regimes (1)-(4).

\begin{figure}[th]
\centering\includegraphics[width=\columnwidth]{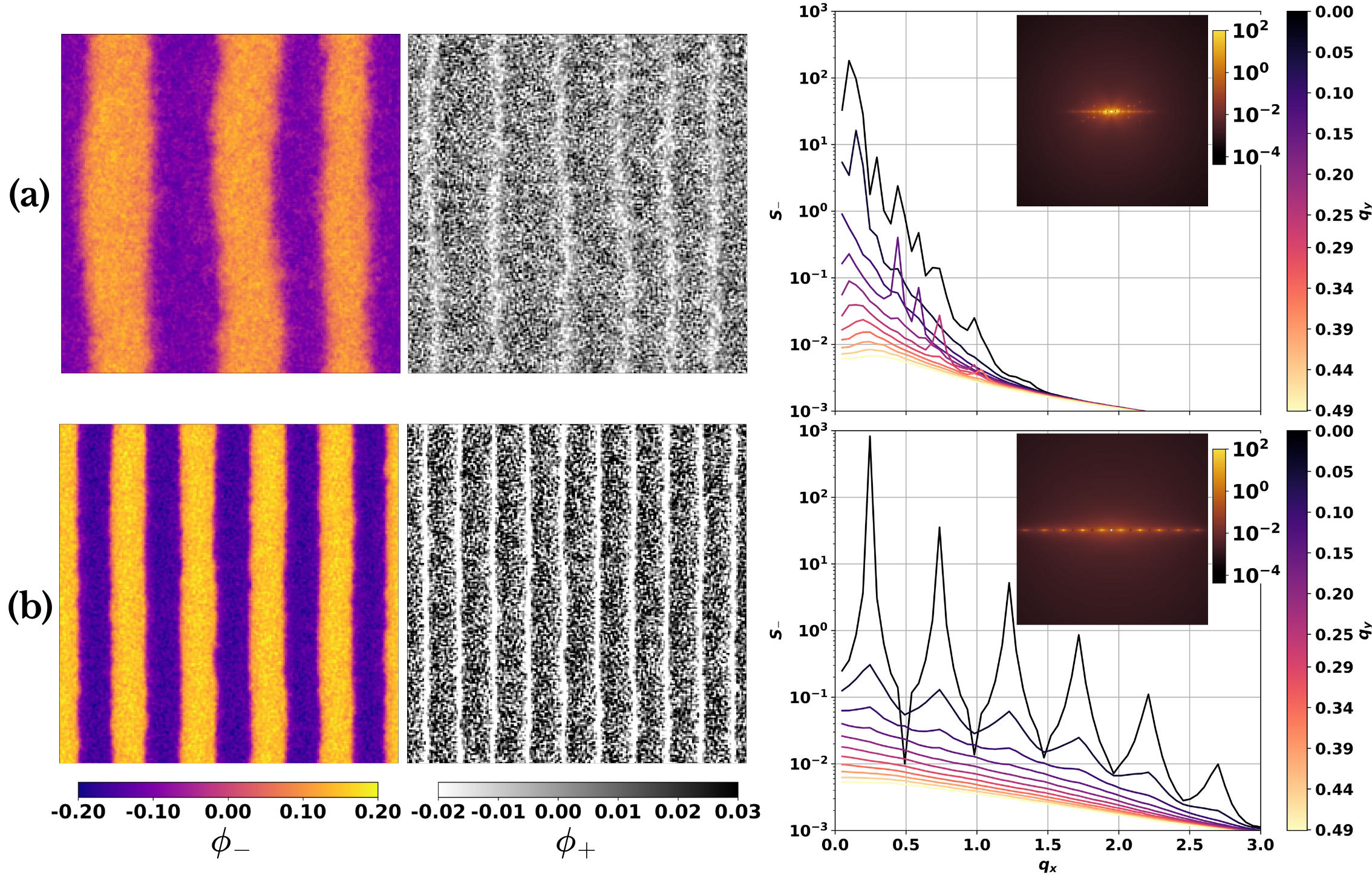}
\caption{Regimes of irregular stripes and regular stripes found by numerical integration of Eq.~\eqref{eq:eom-final}, analogous to the ordered regimes of the FLM illustrated in Fig.~\ref{fig:S_(qx,qy)}. We use $\delta = 1$, maintaining $\lambda = 1,\,g_0 = 0$ and $g_-=0.5$. (a) $\rho=0.4$. (b) $\rho=0.6$, with linear coefficients fixed by Eqs.~\eqref{eq:linear-coefs}. Charge fluctuation (left panels) and density fluctuation (center panels) of typical steady-state field configurations.  Right panels: charge fluctuation structure factor, $S_-(q_x,q_y)$, as a function of $q_{x}$ for $q_{y}\, L /2\pi=0,...,10$, in the two regimes. A movie related to this figure is available in \cite{movies}.}%
\label{fig:ft-regimes}%
\end{figure}

Our numerical results suggest that the term proportional to $g_-$ in Eq.~\eqref{eq:eom-final} is necessary to stabilize ordered stripes, analogous to the drive-independent stability term proportional to $\lambda$. Thus, like $\lambda$, we will fix the value of $g_-$ in the following. The choice to fix $g_-$ is further supported by the observation that, for finite $\Delta v> \Delta v^*$, we verified that keeping $0<g_-\lesssim 0.5$ does not affect the stripe orientation. Additionally, our simulations show that increasing $g_0$ consistently destabilizes perpendicular stripes, moving the system away from the phenomenology of the lattice models. Accordingly, we set $g_- = 0.5$ and $g_0 = 0$ in what follows \footnote{Note that it is not strictly necessary to drastically set $g_0 = 0$; any small value $g_0 \approx 0$ would yield similar results. However, since no particular value is preferred, we choose zero for simplicity and computational efficiency.}. In other words, we focus the remainder of our studies on Regime 2 shown in Fig.~\ref{fig:typconf-eom}. In this regime, we explore the effect of varying $\rho$ [which will vary the linear terms shown in Eq.~\eqref{eq:linear-coefs}] and $\Delta v$ (or equivalently $\delta$), and find that increasing either parameter also increases order along the drive direction. Conversely, when $\rho$ and $\Delta v$ are close to the thresholds $(\rho_c^{\mathrm{FT}},\Delta v^*)$, we find evidence of long-wavelength (Eckhaus-like) instabilities within the perpendicular stripe regime, which act as a mechanism to reduce the number of stripes, similarly to what occurs in the FLM. Based on this, we identify the large and near-threshold parameter regimes as the regular and irregular stripe regimes, respectively. The resemblance goes beyond visual inspection of the configurations: we also detect the clear signatures of the different regimes in the charge SF, with the irregular stripe regime [Fig.~\ref{fig:ft-regimes}(a)] showing a broad peak near the origin, whereas the regular stripe regime [Fig.~\ref{fig:ft-regimes}(b)] displays a sharp, Dirac delta-like peak accompanied by pronounced higher-order harmonics. However, \textit{unlike} the FLM irregular stripe regime, we observe weak harmonics around the peak and secondary maxima at $q_x>q^*$ for $q_y\neq0$, indicating stripes that are more ordered along $x$ yet weakly modulated along \footnote{In fact, this vertical modulation seems to be a feature of the equations that is already present if $g_{\pm,0}=0$; see, e.g., Fig.~\ref{fig:typconf-eom}\, (1).} $y$. Additionally, in the regular stripe regime, the continuum model further accentuates ordering: the SF is heavily localized at $q_y=0$, with modes at any $q_y\neq 0$ appearing almost as background noise; the resulting spectrum is strongly peaked and displays deep dips, resembling Bragg peaks in a diffraction grating that are consistent with a highly ordered stripe pattern.

For example, comparing with the results in Table~\ref{table:S_values}, the ratio of peak values $S_-(q^*,q_y)$ for $q_y L/2\pi = 0,1$ in the continuum model using $\delta = 1$ is $S_{-}\left(q^{\ast},0\right)/S_{-}\left(  q^{\ast}, 0.05\right)\simeq 52$ in the irregular stripe regime ($\rho=0.4$) and $S_{-}\left(q^{\ast},0\right)/S_{-}\left(  q^{\ast}, 0.05\right)\simeq 2700$ in the regular stripe regime ($\rho=0.6$) (nearly an order of magnitude larger than the FLM!). Table~\ref{table:FT_S_values} lists the peak values for the first three $q_y$ modes available in these two cases. In addition, we note that the density at which the regular stripe regime emerges in the continuum model appears to be shifted relative to the FLM: regular stripes are observed already at $\rho=0.6$ for $\delta=1$ (whereas in the FLM, $\rho_u \simeq 0.65$). However, note that $\rho$-dependence in the continuum model only refers to the GE results for the linear coefficients [see Eq.~\eqref{eq:linear-coefs}]. These assumptions are violated at high densities and we generally expect the location of the transition to be shifted due to the nonlinearities and noise.  
\begin{table}[htbp]
    \centering
    \begin{tabular}{|c|c|c|c|c|}
\hline
$\rho$ & $q^* $ & $S_{-}\left(q^{\ast},0\right)$  & $S_{-}\left(  q^{\ast}, 0.05\right)$  & $S_{-}\left(q^{\ast},0.1\right)$ \\
\hline
\hline
$0.4$ & $0.01$  & $181.04$  & $3.47$  & $0.56$\\
$0.6$ & $0.25$  & $826.19$  & $0.31$ & $0.072$\\
\hline
    \end{tabular}
    \caption{Comparison of the peak values of the charge fluctuation structure factor (third column) occurring at $q_{y}=0$ with peak values at $q_{y} \, L/2\pi=1,2$ in the irregular ($\rho=0.4, \delta = 1$) and the regular ($\rho=0.6, \delta = 1$) stripe regimes. We use $g_0 = 0$, $g_- = 0.5$ and $\lambda = 1$.}
    \label{table:FT_S_values}
\end{table}

These findings are summarized in Fig.~\ref{fig:ft-regimes}, which shows representative configurations with the charge fluctuation SF (averaged over $1000$ time points evenly spaced across the last $10^6$ time steps and $70$ realizations) for the irregular [Fig.~\ref{fig:ft-regimes}(a)] and regular [Fig.~\ref{fig:ft-regimes}(b)] stripe regimes. Together, they demonstrate that the continuum model not only reproduces the phenomenology of the FLM but also reveals subtle contrasts, such as the development of tiny harmonics in the intermediate regime \footnote{We can notice, however, a shoulder at higher $q_x$'s in Fig.~\ref{fig:S_(qx,qy)}(b)} and an amplified jump in SF peak values in the regular stripe regime.

\begin{figure}[ht]
    \centering
\includegraphics[width=\columnwidth]{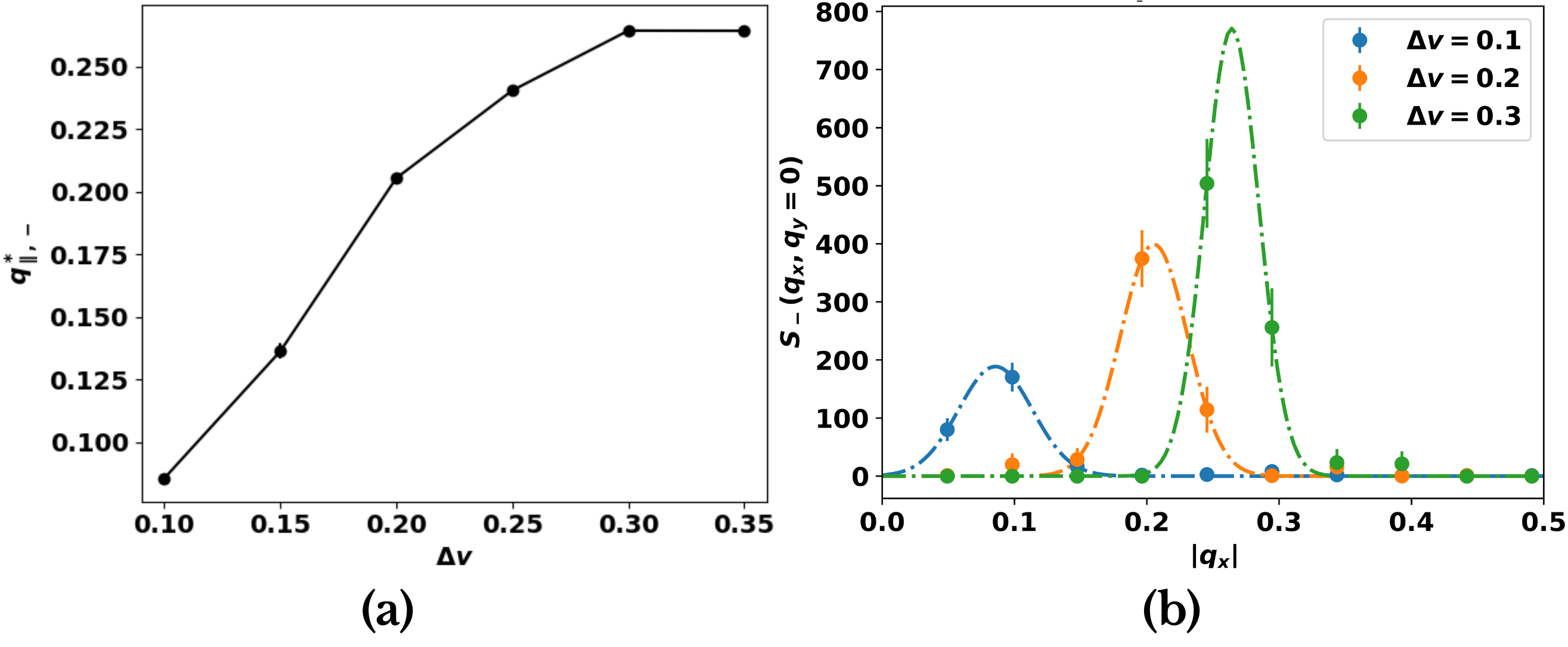} 
    \caption{Growth of order along the drive direction in the region where perpendicular stripes develop ($\Delta v>\Delta v^*$). We use $\rho = 0.7$, $\lambda = 1$, $g_0 = 0$ and $g_- = 0.5$ in Eq.~\eqref{eq:eom-final}. (a) Charge fluctuation  modulation characteristic wavevector, $q_{\parallel}^*$ (along the drive direction) as a function of $\Delta v$, the difference between the characteristic velocities of the charge and density fluctuation fields. Error bars are smaller than the symbols. (b) Charge fluctuation structure factor for three different values of $\Delta v$. Dashed lines are the best Gaussian fit around the peak region. Notice the peak signal becoming increasingly larger with increasing $\Delta v$.}
    \label{fig:deltav-q}
\end{figure}

As long-range order builds up along the drive direction, the stripes become not only better defined but also increase in number. We find that the steady-state characteristic modulation of the CF SF, $q^*$, increases with $\Delta v$ for $\Delta v > \Delta v^*$ at fixed $\rho = 0.7, g_-=0.5$ and $g_0 = 0$. Notably, this behavior is analogous to and consistent with the results reported in \cite{DWRLG2}, where $q^* \sim \Delta v$ was observed in the low-density regime. In Fig.~\ref{fig:deltav-q}(a), we plot the dependence of $q^*$ with $\Delta v$, in the regime where perpendicular stripes are well established, i.e., for $\Delta v > \Delta v^*$. Panel (b) shows that the CF SF has an increasingly prominent peak along the drive direction as $\Delta v$ grows, signaling the increase of order in that direction, as observed in the FLM. A more detailed analysis of how $q^*$ depends on $\Delta v$ is left for future investigation. 
 
Finally, by fixing $\rho$ and varying the drive-dependent parameters $\Delta v$, $g_0$, and $g_-$, we uncover a wide range of regimes in Eqs.~\eqref{eq:eom-final}, many of which have not been reported in the previous lattice-based studies. These regimes typically occur when $|g_0 \phi_+| \gg \Delta v$. To (partially) understand the origin of these new regimes, we reorganize the drive-dependent terms  proportional to $ \partial_x \phi_-$ in Eq.~\eqref{eq:eom-final} into the combination
\[\partial_x\left[\left(g_0 \phi_+ -\Delta v\right)\phi_- \right]. \label{eq:reorgdrive} \]
If $|g_0 \phi_+| \gg \Delta v$, we see from Eq.~\eqref{eq:reorgdrive} that the flow direction of the CF is strongly dependent on the sign of $\phi_+$, facilitating the formation of complex flow patterns.

\begin{figure}[ht]
 \centering
\includegraphics[width=0.45\columnwidth]{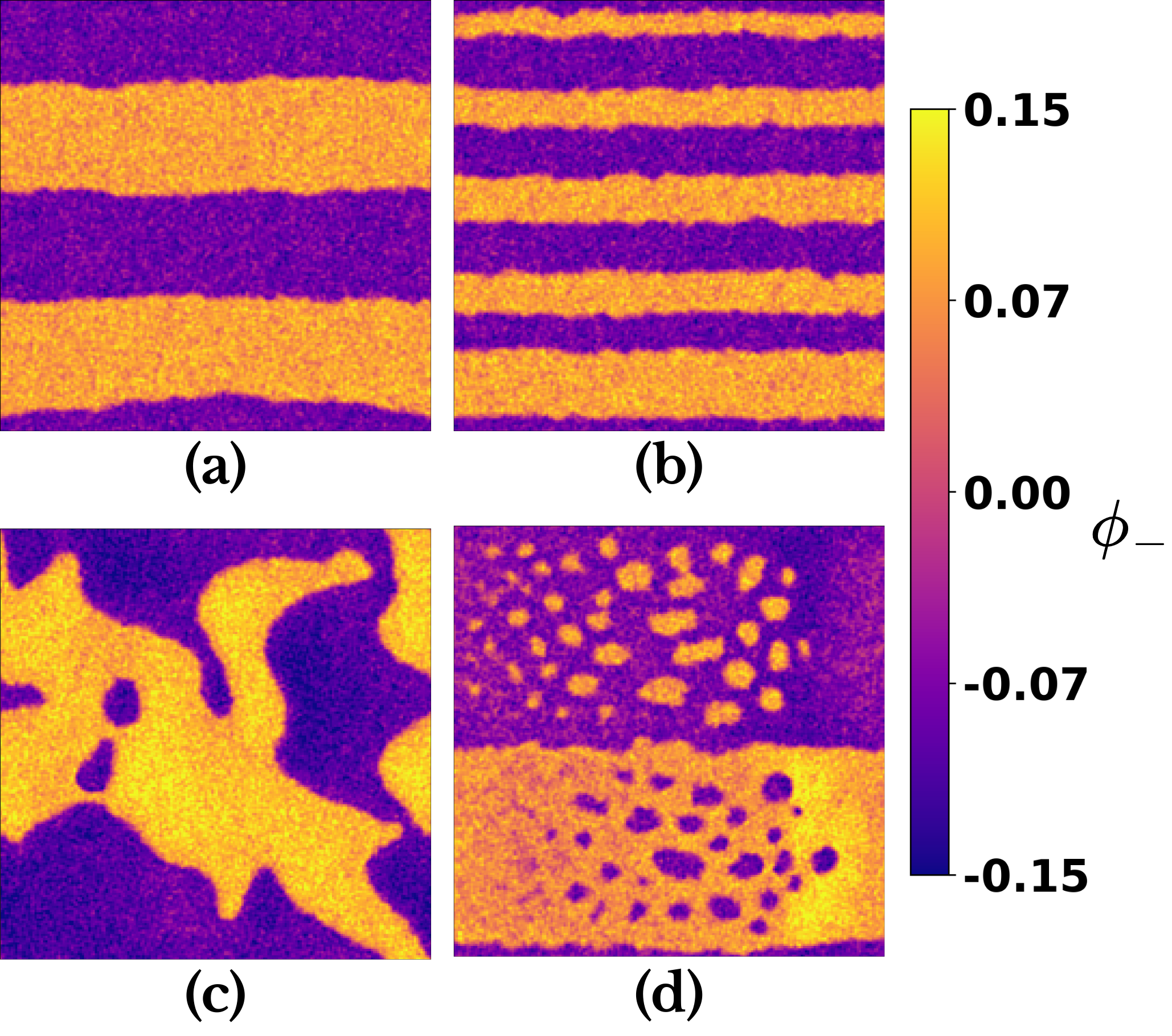} 
    \caption{Charge fluctuation $\phi_-$ snapshots at $t= 10^6$ of the different regimes found by numerical integration of Eq.~\eqref{eq:eom-final}. We use $\rho = 0.7$ [fixing the coefficients in Eq.~\eqref{eq:linear-coefs}], $\lambda=1$, and $\sigma = 0.01$ in all panels. (a) Regular and (b) irregular stripes aligned with the drive. (c) Spatio-temporal chaos. (d) Phase separation into bubbles modulated by shock-like envelopes. Parameters: (a) $\Delta v = 0.05$, $g_0=0.003$, $g_-=0.5$. (b) $\Delta v = 0.211$, $g_0 = 0.3$, $g_-=0.5$. (c) $\Delta v = 0.211$, $g_0 = 0.03$, $g_-=0.5$. (d) $\Delta v = 0.05$, $g_0=0.1$, $g_-=0.5$. A movie related to this figure is available in \cite{movies}.}
    \label{fig:mosaic}
\end{figure}

Fig.~\ref{fig:mosaic} shows snapshots taken at long times $t =  10^6$ that illustrate the variety of  new behaviors: regular [Fig.~\ref{fig:mosaic}(a)] and irregular [Fig.~\ref{fig:mosaic}(b)] stripes aligned \textit{with} the drive, spatio-temporal chaos [Fig.~\ref{fig:mosaic}(c)] and phase separation into bubbles modulated by shock-like envelopes [Fig.~\ref{fig:mosaic}(d)]. Because these regimes do not appear in either the FLM or the DWRLG, we have restricted our primary analysis to parameter regions that better reflect those models. Nevertheless, Fig.~\ref{fig:mosaic} clearly demonstrates that even a minimal set of continuum equations can generate a remarkably rich phenomenology. While a systematic exploration of how each term contributes to the various steady states would be valuable, such an endeavor is computationally demanding and falls beyond the scope of this study. It is worth noting that similar regimes have been observed in other contexts, such as parallel stripes in the KLS model \cite{Zia2010} and spatio-temporal chaos in the active non-reciprocal Cahn-Hilliard model \cite{Marchetti2024}.

\section{Discussion and conclusion\label{sec:conclusion}}

We investigate a model of a driven binary mixture (dubbed the ``FLM'') which effectively captures the behavior of the driven Widom-Rowlinson lattice gas in both the low- and high-density phases, but includes an additional intermediate regime with irregular stripes. In the high-density, ordered phase, earlier studies \cite{DWRLG1, DWRLG2} report standard coarsening dynamics in the absence of a drive, while the drive leads to the emergence of regular stripes oriented perpendicular to the drive. In Sec.~\ref{sec:Model}, we briefly characterized the FLM and observed striking similarities with the DWRLG phenomenology, suggesting that both models may belong to the same universality class within a properly chosen parameter space. Notably, we observe that perpendicular stripes emerge even when the system is initialized with parallel stripes \cite{thesisgui}, indicating a robust stripe-selection mechanism. This reorientation appears to proceed via an instability that is reminiscent of the Mullins-Sekerka instability of the interfaces \cite{thesisgui}.

Above the critical density, roughly estimated as $\rho_c \simeq 0.39$ for $\delta=0$ and $\rho_{\ell} = 0.40$ for $\delta=1$, we identify two additional transition densities in the FLM: (i) For $\delta > 0$, the density $\rho_u$ ($\simeq0.65$, for $\delta = 1$) marks a transition from short-range ordered (irregular) stripes to long-range ordered (regular) stripes [see, e.g., Figs.~\ref{fig:phase-diagram}(a,d)-~\ref{fig:S_(qx,qy)}(b,c)]. The irregular stripe region, $\rho_{\ell}< \rho < \rho_u$, is characterized by pronounced long-wavelength instabilities, which serve as a mechanism to reduce the number of stripes.  (ii) For $\delta \geq 0$, $\rho_g\simeq0.90$; marks the onset of a frozen regime for uniform initial conditions [Figs.~\ref{fig:phase-diagram}(a,b)], where coarsening arrests and glassy configurations appear. In Sec.~\ref{sec:Model} we offer a number of ways to characterize these two regimes.

In the driven low-density regime, the FLM shows indications of a peak in the CF structure factor [upper panel in Fig.~\ref{fig:lowd-figs-largeL}(a)] at an $L$-independent characteristic wavevector, $q^* = 0.101 \pm 0.001$. The DF structure factor [lower panel in Fig.~\ref{fig:lowd-figs-largeL}(a)] also shows a shoulder at $2q^*_-$, but remains peaked at $2\pi/L$, consistent with the results reported in Refs.~\cite{DWRLG1,DWRLG2}. Examining a snapshot  of a typical configuration in this regime [Fig.~\ref{fig:lowd-figs-largeL}(b)], we see that this microemulsion phase emerges from a tendency to form stripes perpendicular to the drive that is frustrated by the stronger effect of the drive in regions having low particle density.

We derive a continuum version of the FLM via a gradient expansion in Sec.~\ref{sec:Continuum}. The analysis yields a system of two (reciprocally) coupled nonlinear stochastic partial differential equations, Eq.~\eqref{eq:eom-final}. These equations are consistent with the field equations developed in Ref.~\cite{DWRLG2} for the low-density phase, but include additional terms that are important for describing the high-density regime, namely the terms proportional to $\lambda$ and $\Gamma_\pm$ in Eq.~\eqref{eq:eom-final}. Additionally, the charge fluctuation diffusion constant,\ $D_-$, changes sign at $\rho_c^{\mathrm{FT}} = 0.37\dots$, marking the onset of the transition from a disordered to an ordered regime. For comparison, the FLM estimates are $\rho_c^{(0)} \simeq0.39$ for the critical density at zero drive ($\delta=0$) and $\rho_{\ell} \simeq 0.40$ as the critical density for (irregular) stripe formation at $\delta=1$.

In Sec.~\ref{sec:numer-int}, we integrate the field equations numerically. We find that the PSSETD2H method is best suited for this task \footnote{Interestingly, we found no references in literature that describe these methods in detail, leading us to believe that such a description (and application) could be useful, particularly in the active matter context, since the non-reciprocal Cahn-Hilliard and Active Model B equations are analogous to Eq.~\eqref{eq:eom-final}}. Dealiasing techniques are straightforward to implement, effectively minimizing errors from spatial discretization, which is particularly important in the presence of noise \footnote{In fact, we find that it is necessary to use a much finer mesh in space and time than the one we use in Sec.~\ref{sec:numer-int} to numerically integrate an equation of type \eqref{eq:eom-final} with a noise of moderate intensity, $\sigma \approx 10^{-2}$, without recurring to dealiasing techniques.}. The remainder of Sec.~\ref{sec:numer-int} provides a qualitative overview of the continuum stochastic field theory, with minimal emphasis on quantitative analysis. A central aspect is that noise plays a crucial role in reproducing the features observed in previous studies. We observe not only a discontinuity in the structure factors $S_{\pm}(\mathbf{q})$ at the origin but also a peak in the charge fluctuation structure factor $S_-(\mathbf{q})$  at a nonzero wavevector along the drive direction, an indicator of the microemulsion phase.   

Turning to the high-density regime, the undriven ($\delta=0$) case displays standard coarsening dynamics consistent with the Lifshitz-Slyozov law in both the lattice model and the field theory. Our main findings for $\delta > 0$ are summarized in Figs.~\ref{fig:typconf-eom}-\ref{fig:ft-regimes}: We identify a regime in the field theory, in which the equations of motion best reproduce the features of the lattice models (see Fig.~\ref{fig:ft-regimes}\,(2)). In this regime, as both density and drive increase, the continuum model transitions from irregular- to regular-stripe configurations, similarly to the FLM. Additionally, we uncover several novel regimes not previously reported in the lattice studies, including stripes aligned parallel to the drive and turbulent-like behavior (see Fig.~\ref{fig:mosaic}). 

A key finding of Sec.~\ref{sec:numer-int} is that a nonzero difference in the characteristic velocities of the fields, $\Delta v$, is a \textit{necessary} condition for the formation of stripes aligned perpendicular to the drive, analogously to what is found for the microemulsion phase behavior in Ref.~\cite{DWRLG2}. When $\Delta v = 0$, such stripes do not form, even when the other drive-dependent couplings $g_0$ and $g_-$ are nonzero. Moreover, the modulation of the CF field (and, consequently, the DF field) along the drive direction, $q^*$, increases with the magnitude of $\Delta v$, in agreement with findings from the low-density regime in \cite{DWRLG2}; see Fig.~\ref{fig:deltav-q}. However, vertically aligned stripes emerge only when $\Delta v$ exceeds a finite threshold.

Finally, it is worth emphasizing that the stochastic continuum equations, Eq.~\eqref{eq:eom-final}, contain just the minimal set of drive-dependent couplings, which, at the mean field level, cannot produce the core phenomenology of the lattice-based models such as the discontinuities and anisotropies of the structure factors. Such phenomena would normally only be accessible through renormalization group analysis of the structure factor corrections due to the noise terms and nonlinearities. Nevertheless, by (numerically) seeking steady-state solutions of the stochastic equations, we effectively access these  corrections and find  behavior analogous to the nonequilibrium lattice models. Due to both analytical and computational limitations, our study has focused primarily on qualitative, comparative analyses. Nevertheless, we hope that this work provides a useful starting point for developing a more complete field-theoretical framework for this broader class of nonequilibrium systems.

\begin{acknowledgments}
We are grateful to Hugues Chaté and Alexandre Solon for many enlightening discussions, especially on the importance of dealiasing. 

G. F. O. and R. D. acknowledge support from CNPq, Brazil, under grant No. 161468/2021-7.

\end{acknowledgments}

\appendix
\section{Gradient expansion I -- equations of motion\label{appx:gradient-expansion}}

From the hopping rules stated in Section~\ref{sec:Model}, we can write the following equation for the evolution of $\rho _{A}\left( \mathbf{s},t\right) $ (and a similar one for $\rho _{B}$):
\begin{widetext}
\begin{equation}\label{eq:cont-eqn}
\Delta_t \rho _{A}\left( \mathbf{s},t\right) =-\bigg{\{}\Delta _{\mathrm{NN}}J_{A}^{\mathrm{NN}}\left( \mathbf{s},t\right) +\delta \Delta_{\mathrm{NN}}J_{A}^{\delta ,\mathrm{NN}}\left( \mathbf{s},t\right) +\Delta _{\mathrm{NNN}}J_{A}^{\mathrm{NNN}}\left( \mathbf{s},t\right) +\delta \Delta _{\mathrm{NNN}}J_{A}^{\delta ,\mathrm{NNN}}\left( \mathbf{s},t\right) \bigg{\}}
\end{equation}
\end{widetext}
where 
\begin{equation*}
\Delta _{t}\rho_A (\mathbf{s},t) \equiv \rho_A (\mathbf{s},t+\tau) - \rho_A(\mathbf{s},t)
\end{equation*}
is the change in the field at site $\mathbf{s}$ over a time increment $\tau$, and $\Delta_{\mathrm{NN}}J_{A}^{\mathrm{NN}}$ is the discrete divergence associated with NN hops, etc. Explicitly, the $\Delta _{\mathrm{NN}}J$'s (with $t$ suppressed) are:%
\begin{widetext}
\begin{align}\label{eq:divJs}
-\Delta _{\mathrm{NN}}J_{A}^{\mathrm{NN}}\left( i,j\right) &=\frac{1}{8}\sum_{\kappa =\pm 1}\bigg{\{}\rho _{A}(i-\kappa ,j)h(i,j)\slashed{\rho}^\leftrightarrow_B(i,j;\kappa )-\rho _{A}(i,j)h(i+\kappa ,j)\slashed{\rho}^\leftrightarrow_B(i+\kappa,j;\kappa )\bigg{\}}  \notag \\
&{}\ \ \ +\frac{1}{8}\sum_{\kappa =\pm 1}\bigg{\{}\rho _{A}(i ,j-\kappa)h(i,j)\slashed{\rho}^\updownarrow_B(i,j;\kappa )-\rho _{A}(i,j)h(i ,j+\kappa)\slashed{\rho}^\updownarrow_B(i,j+\kappa;\kappa )\bigg{\}}, \notag \\
-\Delta _{\mathrm{NN}}J_{A}^{\delta ,\mathrm{NN}}(i,j) &=\frac{1}{8}\sum_{\kappa =\pm 1}\kappa \bigg{\{}\rho _{A}(i-\kappa ,j)h(i,j)\slashed{\rho}^\leftrightarrow_B(i,j;\kappa )-\rho _{A}(i,j)h(i+\kappa ,j)\slashed{\rho}^\leftrightarrow_B(i+\kappa ,j;\kappa )\bigg{\}},
\end{align}
\end{widetext}
where 
\begin{align*}
\slashed{\rho}^\leftrightarrow_B(i,j;\kappa)=(1& -\rho _{B}(i+\kappa ,j)) \\
& (1-\rho _{B}(i,j+\kappa ))(1-\rho _{B}(i,j-\kappa )), \\
\slashed{\rho}^\updownarrow_B(i,j;\kappa)=(1& -\rho _{B}(i,j+\kappa)) \\
& (1-\rho _{B}(i+\kappa,j))(1-\rho _{B}(i-\kappa,j)) 
\end{align*}%
account for the NN $A$-$B$ exclusion for horizontal ($\leftrightarrow$) and vertical ($\updownarrow$) jumps, respectively, and $h(i,j) \equiv 1-\rho_A(i,j)-\rho_B(i,j)$ for the density of holes at site $\mathbf{s}=(i,j)$. There are analogous expressions for $\Delta _{\mathrm{NNN}}J$'s; see \cite{thesisgui} for details. From these, we can construct evolution equations for the deviations (from $\rho $ and $0$) of the density and charge fields (defined in Eq.~\eqref{eq:charge-dens}) $\phi _{\pm }\left( \mathbf{s},t\right) $. Let us write these symbolically as 
\begin{equation}\label{eq:phi-cont-eqn}
\Delta _{t}\phi _{\pm }=-\left\{ \Delta J_{\pm }+\delta \Delta J_{\pm
}^{\delta }\right\}  
\end{equation}

Since the $J$'s are polynomials in $\rho _{A,B}$ (up to sixth order), expressions like those in Eq.~\eqref{eq:divJs} are polynomial up to $O\left( \phi ^{6}\right) $. To arrive at the more familiar form of Eq.~\eqref{eq:continuity}, we must take the continuum limit and, to simplify the divergence terms, keep only the terms of lowest order in the gradients. The remainder of this Appendix provides one possible approach.

Consider our lattice to have spacing $\ell $ in continuous space ($\mathbf{r}=\left(x,y\right) $), so that the fields above are defined on $\mathbf{r}=\mathbf{s}\ell $. One path to a continuum limit is to let $\ell \rightarrow 0$ and assume that $\phi _{\pm }\left( x, y \right) $ is a smooth function, so that, e.g.,%
\begin{align*}
\phi_{\pm }&\left(i+1,j\right) =\phi _{\pm }\left( x +\ell ,y \right)\\
&=\phi _{\pm }\left( x, y \right) +\ell \partial _{x}\phi _{\pm}\left( x, y \right) +\frac{\ell ^{2}}{2!}\partial _{x }^{2}\phi_{\pm }\left( x, y \right) +...
\end{align*}%
Using the same approach for $\Delta _{t}\phi _{\pm }$ and letting the timestep ($\tau $) become infinitesimal, we arrive at%
\begin{equation*}
\Delta _{t}\phi _{\pm }=\tau \partial _{t}\phi _{\pm }+...
\end{equation*}%
Combining both sides and keeping only the relevant terms to fourth order in the derivatives and to second order in both fields, we obtain a set of equations similar to the deterministic parts of Eq.~\eqref{eq:eom-final}:%
{\small \begin{align}
\partial _{t}\phi _{+}& =D_{+}\nabla ^{2}\phi _{+}-\Gamma _{+}\nabla^{4}\phi _{+}-v_{+}\partial _{x}\phi _{+} \nonumber \\
& \qquad +\lambda _{+}\nabla ^{2}\phi _{+}^{2}+\lambda _{-}\nabla ^{2}\phi_{-}^{2}+g_{+}\partial _{x}\phi _{+}^{2}+g_{-}\partial _{x}\phi _{-}^{2}, \nonumber \\
\partial _{t}\phi _{-}& =D_{-}\nabla ^{2}\phi _{-}-\Gamma _{-}\nabla^{4}\phi _{-}-v_{-}\,\partial _{x}\phi _{-} \nonumber \\
& \qquad +\tilde{\lambda}_{+}\nabla (\phi _{-}\nabla \phi _{+})+\tilde{\lambda}_{-}\nabla (\phi _{+}\nabla \phi _{-})+g_{0}\partial _{x}(\phi_{+}\phi_{-}) \nonumber
\end{align}}
From the above equation, we see that the undriven currents appearing in Eq.~\eqref{eq:continuity}, $\mathbf{J}_{\pm}$, read
\begin{align} \label{eq:quadratic}
& \mathbf{J}_{+}= -\nabla\left[D_+\phi_+ - \Gamma_+\nabla^2\phi_+ +\lambda_+\phi^2_+ + \lambda_-\phi^2_-\right]\nonumber\\
& \mathbf{J}_{-}= -\nabla\left[D_-\phi_- - \Gamma_-\nabla^2 \phi_- \right] -\tilde{\lambda}_{+} \,\phi_{-}\nabla\phi _{+}-\tilde{\lambda}_{-}\,\phi _{+}\nabla \phi _{-},
\end{align} 
while the drive-dependent currents, $J_{\pm }^{\delta }$, are given by
\begin{align} \label{eq:quadratic-drive}
& J^\delta_{+}= -v_{+}\phi _{+}+g_{+}\phi _{+}^{2} + g_{-}\phi _{-}^{2}\nonumber\\
& J^\delta_{-}= -v_{-}\phi _{-}+g_{0}\phi _{+}\phi _{-}.
\end{align} 

Note that the form of the terms in Eqs.~\eqref{eq:quadratic}-\eqref{eq:quadratic-drive} is constrained by the $A\leftrightarrow B$ symmetry which implies that $\mathbf{J}_+$ ($\mathbf{J}_-$) cannot have odd (even) powers of $\phi_-$. Let us briefly highlight some characteristics of qualitative importance. First, based on this continuum limit of the FLM, we find that $D_{+}$ is positive for all $\rho $, which implies we can expect the DF to be stable, unless it is driven unstable through coupling to $\phi_{-}$. On the other hand, as expected of systems that display coarsening dynamics, $D_{-}$ changes sign when $\rho $ exceeds a critical value. 

Exploiting a Galilean transform, we will work in the co-moving frame of the DF, so that the $v_{+}\phi _{+}$ term is absent in\ $J_+^{\delta}$ and $-v_- \phi_-$ is replaced by $-(\Delta v) \phi_-$, where $\Delta v \equiv  v_- - v_+$. Note that both currents $J_{\pm}^{\delta}$ vanish when $\delta=0$, along with all their individual $v$ and $g$ terms.

Finally, we must add noise terms, $\xi_{\pm}$, which are difficult to derive from first principles, but we know they must conserve the total particle number and have zero mean. Taking the simplest possible approach, let $\xi_{A}$ and $\xi_{B}$ be the noise terms in the equations for $\rho_{A}$ and $\rho_{B}$. To first approximation, these are independent zero-mean conserved Gaussian noises, with variances $\sigma_A^2$ and $\sigma_B^2$. By symmetry under exchange of species $A$ and $B$, we must have $\sigma_A = \sigma_B$. In this case, 
\[\xi_+ = \xi_A + \xi_B, \qquad \xi_- = \xi_A - \xi_B, \nonumber\] 
have the same variance $2\sigma^2_A$, and are uncorrelated, $\langle\xi_+ \xi_-\rangle = 0$. Thus, it follows that $\xi_\pm$ are zero-mean conserved Gaussian noise terms, uncorrelated in time and with each other, having variance $\sigma^2 \equiv 2\sigma^2_A$, i.e., 
\begin{align}\label{eq:noise-in-eq-plusminus}
&\langle\xi_\pm(\mathbf{r},t)\rangle = 0 \nonumber \\
&\langle\xi_+(\mathbf{r},t)\xi_-(\mathbf{r}',t')\rangle = 0 \nonumber \\
&\langle\xi_\pm(\mathbf{r},t)\xi_\pm(\mathbf{r}',t')\rangle = -\sigma^2\nabla^2\delta(\mathbf{r}-\mathbf{r}')\delta(t-t').
\end{align}

Combining all the ingredients, we arrive at the SPDEs 
{\small\begin{align}
\partial _{t}\phi _{+}& =D_{+}\nabla ^{2}\phi _{+}-\Gamma _{+}\nabla^{4}\phi _{+} \nonumber \\
& +\lambda _{+}\nabla ^{2}\phi _{+}^{2}+\lambda _{-}\nabla ^{2}\phi_{-}^{2}+g_{+}\partial _{x}\phi _{+}^{2}+g_{-}\partial _{x}\phi _{-}^{2} + \xi_+, \nonumber \\
\partial _{t}\phi _{-}& =D_{-}\nabla ^{2}\phi _{-}-\Gamma _{-}\nabla^{4}\phi _{-}-\Delta v\,\partial _{x}\phi _{-} \nonumber \\
& +\tilde{\lambda}_{+}\nabla (\phi _{-}\nabla \phi _{+})+\tilde{\lambda}_{-}\nabla (\phi _{+}\nabla \phi _{-})+g_{0}\partial _{x}(\phi_{+}\phi_{-}) + \xi_- \label{eq:eom-prefinal}
\end{align} }
For convenience, we collected all the linear/quadratic terms on the first/second line. In Appendix \ref{appx:coefficients}, we show in some detail how to calculate the exact expressions of the linear coefficients. We also show how a set of assumptions is used to simplify the nonlinear terms in Eq.~\eqref{eq:eom-prefinal} to arrive at Eq.~\eqref{eq:eom-final}. 

\section{Gradient expansion II -- coefficients\label{appx:coefficients}}

To calculate the exact expression of the linear coefficients, we start by substituting 
 \begin{align*}
    & 2\rho_A = \phi_+ + \phi_- +\rho \nonumber \\
    & 2\rho_B = \phi_+ - \phi_- +\rho
\end{align*}
in Eq.~\eqref{eq:cont-eqn} to write the expressions in Eq.~\eqref{eq:divJs} (and the analogous expressions for species B) in terms of the density and charge fluctuations, but this time we only keep the resulting linear piece. Because the change of variables above involves the parameter $\rho$ (which is zeroth order in the fields), and since the $J$'s are sixth-order polynomials in $\rho_{A,B}$, the linear terms will have contributions coming from all possible orders in the fields. As a result, the corresponding linear coefficients are fifth-order polynomials in $\rho$.

The lattice update appearing in the different terms of Eq.~\eqref{eq:divJs} can be written in Fourier space in terms of the operator $e^{i \kappa \left(q_x+q_y\right)}$, such that for any field $f$ on the lattice,
\[\mathcal{F}^{-1} \{e^{i\kappa \left(q_x+q_y\right)} \widetilde{f}(q_x,q_y)\} = f(i+\kappa,j+\kappa),\nonumber\]
where $\mathcal{F}^{-1}$ is the inverse discrete Fourier transform. 

We take the discrete Fourier transform of the linear part of Eq.~\eqref{eq:divJs} and use the above property to express the lattice-updated fields using the exponentials. There will be analogous expressions for NNN hops. We keep working in Fourier space to substitute these expressions in Eq.~\eqref{eq:cont-eqn} and write
\begin{align*}
\Delta_t\phi_\pm(\qvec,t)= \mathcal{L}_{\pm}(\qvec) \phi_{\pm}(\qvec,t),
\end{align*}
where
\begin{widetext}
\begin{align}
    \mathcal{L}_{\pm}(\qvec) =& A_\pm + B_\pm \big{(}\cos\left(q_x\right) + \cos\left(q_y \right)\big{)} + C_\pm\big{(}\cos^2\left(q_x\right)+\cos^2\left(q_y\right)\big{)}  + E_\pm\cos\left(q_x\right)\cos\left(q_y\right) \nonumber \\ 
    & + G_\pm\big{(}\cos\left(q_x\right) +\cos\left(q_y\right)\big{)}\cos(q_x)\cos(q_y)\nonumber \\
    & +i\delta\big{(}H_\pm+d_1\cos\left(q_x\right) + K_\pm\cos\left(q_y\right) + d_2\cos\left(q_x\right)\cos\left(q_y\right) + d_3\cos^2\left(q_y\right)\big{)}\sin\left(q_x\right).
 \label{eq:full_op}
\end{align}
\end{widetext}
All coefficients are polynomials in $\rho$; their explicit forms are omitted for clarity and conciseness. Up to this point, no approximations have been introduced in the linear part: these are exact expressions, with $\mathcal{L}_\pm(\qvec)$ denoting the exact linear operators corresponding to the Fourier-space representation of the linear terms in the mass-transfer equation \eqref{eq:cont-eqn}. The gradient expansion (GE) of Eq.~\eqref{eq:full_op} yields, to lowest order,
\begin{align}
    \mathcal{L}_{\pm}(\qvec) \approx -D_\pm |\qvec|^2 - \Gamma_\pm|\qvec|^4 -  v_\pm\, i q_x .
 \label{eq:exp_op}
\end{align}
Comparing with expression \eqref{eq:full_op} and collecting similar terms, we find the following coefficients \footnote{For simplicity, we suppressed all the \textquotedblleft physical  units\textquotedblright\ for the coefficients, e.g., $D\propto \ell ^{2}\tau^{-1}$, etc. Their detailed forms play no role in the numerical studies.}:
\begin{alignat}{2}\label{eq:linear-coefs}
& D_{+}&& =\frac{2}{64}(2-\rho )^{2}\left( 3+4\rho +2\rho ^{3}-8\rho
^{2}\right) ,  \nonumber\\
& D_{-}&& =\frac{1}{64}(2-\rho )^{2}\left( 6-24\rho +23\rho ^{2}-5\rho
^{3}\right) ,  \nonumber \\
&\Gamma_+ &&= \frac{1}{384} (2- \rho)^2 \left(3 + 22\rho - 32\rho^2 +8 \rho ^3\right), \nonumber\\
&\Gamma_{-}&& = -\frac{1}{768}(2-\rho )^{2}\left( 6-60\rho +71\rho ^{2}-17\rho
^{3}\right) ,  \nonumber \\
&v_+ &&= -\frac{\delta}{64}  (2- \rho)^2 \left(-4 + 26\rho - 29\rho^2 + 8 \rho ^3\right), \nonumber \\
&v_- &&= \frac{3\, \delta }{32} (2- \rho)^2 \left(2-3\rho^2 + \rho ^3\right), \nonumber \\
&\Delta v&& \equiv v_- - v_+ = \frac{\delta }{64}(2-\rho )^{2}\left( 8+26\rho -47\rho
^{2}+14\rho ^{3}\right) .  
\end{alignat}

The remaining seven nonlinear coefficients $\lambda_\pm, \tilde{\lambda}_\pm$ and $g_{\pm,0}$ accompanying the quadratic terms in Eq.\eqref{eq:eom-prefinal} are still undetermined. Determining their dependence on $\rho$ through the GE is nontrivial, as the expression \eqref{eq:cont-eqn} includes contributions of all orders in the fields $\rho_{A,B}$, not just quadratic. For this reason, we choose to leave the quadratic coefficients as free parameters in the present study. Work is in progress to determine the exact functional dependence of the different nonlinear couplings on $\rho$ \cite{thesisgui}.

This is a large parameter space  (four $\lambda $'s, three $g$'s and $\sigma$), but a significant portion of it simply parametrizes the equilibrium $(\delta=0)$ behavior of the model. Therefore, we can introduce some simplifications to the equilibrium terms in order to better elucidate just the non-equilibrium features of interest. The specific simplifying approximations are 
\begin{enumerate}
    \item The GE analysis suggests $\tilde{\lambda}_{+}\gg \tilde{\lambda}_{-} \approx0$ for most realistic values of $\rho$, so we set $\tilde{\lambda}_-=0$.    \item Since stability for $\phi_{+}$ is always provided by $D_{+}>0$, we will drop the term $\lambda_+\phi^2_+$ from our equations \footnote{Note that such a term originates from a $\phi_+^3$ potential in the free-energy and, as such, is unstable. Therefore, one must include an additional $\gamma_+\, \phi^3_+$ (a $\phi_+^4$ potential) in Eq.~\eqref{eq:eom-prefinal} to ensure the stability of the theory. Since we avoid over-complicating our equations by seeking a minimal description and also considering that the DF is stable to linear-order in $\phi_+$, we can safely neglect the term proportional to $\lambda_+$.}. This choice is additionally supported by a weakly nonlinear stability analysis \footnote{Based on the approach of Cross \textit{et al.} \cite{CrossHohenberg, Cross2}, the weakly nonlinear stability analysis here consists of assuming that a dominant role in the instability of the homogeneous regime is played by a particular mode of $\phi _{-}$ with the Ansatz $\phi _{-}(\mathbf{r},t)=\alpha (t)\, \mathrm{Re}\, \{ e^{i\Psi (t)+i\mathbf{q} \cdot \mathbf{r}}\}$. As $A$-$B$ exclusion tends to favor vacancies between the charged regions, it is reasonable to assume  $\beta (t)\, \mathrm{Re}\, \{e^{i2\Psi (t)+i\mathbf{q} \cdot \mathbf{r}}\}$ for $\phi _{+}(\mathbf{r},t)$. In fact, the location of the peaks of the structure factors in the striped regime ($q_{-}=q^{\ast }=q_{+}/2$) displays this type of entraining of $\phi _{+}$ to $\phi _{-}$. Also, a perturbative scheme indicates $\beta \simeq \alpha ^{2}$. Details are found in \cite{thesisgui}.} near the critical $\rho^{\mathrm{FT}}_c$. We find that $\phi _{+}$ is of the same order as $\phi _{-}^{2}$, so we drop terms with powers of $\phi_+$ larger than 1 when we study the high-density regime.
    \item The GE suggests that the two remaining drive-independent nonlinear couplings, $\lambda_-$ and $\tilde{\lambda}_+$, are generally of the same order, but of opposite sign: $-\lambda_-\sim \tilde{\lambda}_+$, with $\tilde{\lambda}_+ >0 $. Thus, we reduce these couplings to a single effective $\lambda$, setting $-\lambda_-,\, \tilde{\lambda}_+= \lambda$. 
\end{enumerate}
 
It is also important to observe that, just as in mean-field theories of other statistical systems, different coarse-graining prescriptions (e.g., the present GE of the FLM or the MSRJD/Doi-Peliti developed in \cite{DWRLG2}) lead to different values for these coefficients, but they generally agree on their order of magnitude. Using the set of simplifying assumptions in Eq.~\eqref{eq:eom-prefinal} together with the linear coefficients in Eq.~\eqref{eq:linear-coefs}, we can write down the final form of the field equations used in this work: Eq.~\eqref{eq:eom-final}.

\section{Pseudospectral exponential time-differencing method \label{appx:pseudospectral}}

In Fourier space, Eq.~\eqref{eq:eom-final} for $\tilde{\Phi}\equiv \tilde{\Phi}(\mathbf{q},t) = (\fphi_-  (\mathbf{q},t),\fphi_+(\mathbf{q},t))^T$ can be written generally as 
\begin{equation}\label{eq:eom-ft2}
  \frac{d \widetilde{\Phi}}{dt}= \mathcal{L}(\mathbf{q})\widetilde{\Phi} + \mathrm{NL}[\widetilde{\Phi}] + \eta,
\end{equation}
 where $\mathrm{NL}[\widetilde{\Phi}]$ is the matrix of the nonlinear terms Fourier transform and  
 \begin{align*}
    &\mathcal{L}(\mathbf{q}) \coloneqq
    \begin{pmatrix}
     \mathcal{L}_-(\mathbf{q}) & 0 \\
     0 & \mathcal{L}_+(\mathbf{q})
     \end{pmatrix}
 \end{align*}
 is the matrix of  linear operators.

The noise $\eta \equiv \eta(\mathbf{q},t) $ is a vector with complex Gaussian random variable components, with zero mean and correlation
 \begin{align*}
        \langle \eta(\mathbf{q},t)\eta(\mathbf{q}',t')\rangle=4\pi^2
  q^2  \delta(\mathbf{q}+\mathbf{q}')\delta(t-t')    \begin{pmatrix}
     \sigma_-^2 & 0 \\
     0 & \sigma_+^2
     \end{pmatrix}.
 \end{align*}
Note that, since $\Phi(\mathbf{r})$ is a real field, then $\tilde{\Phi}(\mathbf{q})$, the Fourier transform, satisfies the condition $\tilde{\Phi}(-\mathbf{q})=\tilde{\Phi}^*$. This should be true for the noise as well.

The pseudospectral method is designed to numerically integrate partial differential equations, where we work directly with Eq.~\eqref{eq:eom-ft2} in Fourier space to solve Eq.~\eqref{eq:eom-final} by choosing an appropriate time-discretizing scheme \cite{Shen-Tang-Wang}. The word \textit{pseudo} refers to the fact that nonlinearities are calculated in direct space by means of the inverse Fourier transform, therefore we do not work fully in reciprocal, or Fourier, space. This method to calculate the nonlinearities introduces aliasing errors \cite{Roberts2011}, which must be dealt with using dealiasing techniques, see Appendix~\ref{appx:dealiasing}.

The coefficients of the fields Fourier transform are calculated using the FFTW fast Fourier transform (FFT) routine \cite{FFTW}. Many different conventions are used in literature, such that we advise checking the FFT routine documentation before any implementation. On the $L\times L$ periodic lattice with a $N\times N$ mesh, the physical wave modes $\mathbf{q}$ are related to the index modes by \[\mathbf{q}_{k} = 2\pi \mathbf{k}/L, \label{eq:physicalq}\] where the components $k_{x,y}$ of the index vector $\mathbf{k}$ are 
\[k_{x}, k_y \in \left\{-\frac{N}{2}, -\frac{N}{2}+1, \dots, \frac{N}{2} -1\right\}.\]

Using the FFT in a lattice transforms Eq.~\eqref{eq:eom-ft2} into a set of $N \times N$ coupled ordinary differential equations, where the Fourier amplitudes of the fields are stored as $N\times N$ complex arrays. We numerically integrate this system of equations in time by using the exponential time-diferencing (ETD) scheme, which is a class of methods that are well adapted to equations of type \eqref{eq:eom-ft2}  \cite{Cox}. 

The ETD scheme involves a specific change of variables that is similar to the interaction picture in quantum mechanics. Let
\[ \widetilde{\Phi}(\mathbf{q},t) = e^{\mathcal{L}(\mathbf{q}) t }\widetilde{U}(\mathbf{q},t) + e^{\mathcal{L}(\mathbf{q}) t }\int_0^t \mathrm{d}s\, e^{-\mathcal{L}(\mathbf{q}) s} \eta(\mathbf{q},s) ,\nonumber\]
where the exponential of the matrices are understood as the exponential of each entry (as they are diagonal in the charge/density basis). Then, the new variable $\widetilde{U} \equiv \widetilde{U}(\mathbf{q},t)$ satisfies the equation
\begin{equation}
   \frac{d \widetilde{U}}{dt}   = e^{ -\mathcal{L}(\mathbf{q}) t }\mathrm{NL}[\widetilde{\Phi}] .
\end{equation}

We are now in position to use any time-stepping method, such as Euler or Runge-Kutta, to integrate the new equation in time; each will result in a different ETD scheme. Here we use the stochastic second-order Heun (SH) method \cite{Toral}.
It is worth noting that the SH method achieves a balance between precision and simplicity. The deterministic part is approximated with an error of $O(\Delta t^2)$, while the stochastic part has an error of $O(\Delta t^{1/2})$. Although stochastic versions of the Runge-Kutta (RK) family exist, they are primarily limited to second-order (RK2) methods and are challenging to implement \cite{Rossler}.

Reversing the change of variables and denoting $\bullet_n$ as the fields at time $t = n \Delta t$, SH involves the calculation of the auxiliary field, $\tilde{A}_n \equiv \tilde{A}_n(\mathbf{q})$, given by
\[\widetilde{A}_n = e^{\mathcal{L}(\mathbf{q}) \Delta t }\left( \widetilde{\Phi}_n  + \Delta t\, \mathrm{NL}[\widetilde{\Phi}_n] \right) + \zeta_n , \nonumber \]
where $\zeta_n \equiv \zeta_n(\mathbf{q})$ are complex Gaussian random variables, with zero mean and correlation
{\small \begin{equation*}
           \langle \zeta_n(\mathbf{q})\zeta_{n'}(\mathbf{q}')\rangle=\frac{\left(e^{2\mathrm{Re}(\mathcal{L}(\mathbf{q}))\Delta t} - 1\right)}{2\mathrm{Re}(\mathcal{L}(\mathbf{q}))} 4\pi^2 q^2 \delta(\mathbf{q}+\mathbf{q}')\delta_{nn'}    \begin{pmatrix}
     \sigma_-^2 & 0 \\
     0 & \sigma_+^2
     \end{pmatrix}.
\end{equation*}}
For discrete $\mathbf{q}$, we further note the following correspondence \cite{Toral, Garcia-Sancho}
\[4\pi^2 \delta(\mathbf{q}+\mathbf{q}') \to \left(\frac{N}{\Delta x}\right)^2 \delta_{\mathbf{k+k}'}, \]
where $\Delta x = L/N$ is the spatial resolution. The fields at the next timestep, $t+\Delta t$, are calculated as
\begin{align}
&\widetilde{\Phi}_{n+1} = e^{\mathcal{L}(\mathbf{q}) \Delta t }\widetilde{\Phi}_n  \nonumber\\
&\qquad + \frac{\Delta t}{2}\left( e^{\mathcal{L}(\mathbf{q}) \Delta t }\mathrm{NL}[\widetilde{\Phi}_n] + \mathrm{NL}[\widetilde{A}_n]  \right)  + \zeta_n .
\end{align}
Observe that, at each iteration, we must repeat the procedure above for every value of $\mathbf{q}$, which is indexed by relation \eqref{eq:physicalq}.

\section{Dealiasing \label{appx:dealiasing}}

Although many authors have commented on the importance of dealiasing in pseudospectral methods, \textit{e.g.}, \cite{Orszag1971, Kravchenko1997, Wang2021}, such considerations are rarely mentioned in the physics literature. In particular, dealiasing is extremely important in fluid dynamics simulations, see e.g. \cite{Kravchenko1997, Hou, Margairaz2018, Robson2012}. Higher modes (or frequencies) which are not resolved, given our spatial resolution, are aliased to lower modes, generating spurious contributions to their amplitudes. This effect is referred to as ``aliasing error'' and is especially significant when we must compute nonlinear terms in the equation of interest, or in the case of stochastic equations, since the noise will constantly generate amplitudes for these higher modes. An equation such as \eqref{eq:eom-final} will then generally suffer from aliasing errors.

A general procedure to reduce aliasing error is to continuously set to zero amplitudes of modes beyond a certain threshold. Following \cite{Hou}, we define two dealiasing kernels
\begin{align*}
      \rho_{\mathrm{sharp}} (\kvec) =  
      \begin{dcases}
          1 \qquad \mathrm{if} \quad \frac{2k}{N} \leq \frac{2}{3}\\
          0 \qquad \mathrm{if} \quad \frac{2k}{N} > \frac{2}{3}
      \end{dcases}\\ 
      \rho_{\mathrm{smooth}}(\kvec) = \exp\left[-36\left(\frac{2k}{N}\right)^{36}\right] ,
\end{align*}
where $N$ is the number of points in the lattice and $k =|\mathbf{k}|$ is the modulus of the mode index in Fourier space (see Appendix~\ref{appx:pseudospectral}). The effect of the first kernel is simply to abruptly set to zero $1/3$ of the amplitudes of the higher modes, while the second kernel sets the amplitudes of the higher modes to zero smoothly. Using the smooth dealiasing kernel, we keep about $12\sim 15 \%$ more modes than the sharp dealiasing kernel. For $\rho_{\mathrm{smooth}}$, we use in the exponential the same factor of $36$ that is employed in \cite{Hou}.

In our numerical integrations, we study the effect of both dealiasing methods, but no appreciable differences in the simulations are found. Conversely, if dealiasing is not applied, spurious contributions from higher modes contaminate the small-scale dynamics, appearing either as artifacts in the results or as numerical instabilities. We choose to implement smooth dealiasing in the present work as it has several advantages in comparison with the $1/3$ dealiasing, as discussed in Ref.~\cite{Hou}. The smooth dealiasing is implemented by modifying  $\mathrm{NL}[\widetilde{\Phi}_n]$ when we compute  $\Phi_{n+1}$. Namely, we multiply these non-linear terms by the dealiasing kernel,
\[\mathrm{NL}[\widetilde{\Phi}_n] \rightarrow \rho_{\mathrm{smooth}}*\mathrm{NL}[\widetilde{\Phi}_n],\nonumber\]
 effectively setting to zero the large-$\mathbf{q}$ modes. 

If the time-discretizing scheme is a multi-staged method, such as RK2 or RK4, we must implement dealiasing at each stage. As an example, for RK2, two dealiasing operations are required, one for each stage. Therefore, dealiasing is computationally very costly. The method described here is an example of explicit dealiasing. It is worth mentioning that there are other dealiasing techniques, such as implicit \cite{Bowman2011}, random phase shift \cite{Sinhababu2021} dealiasing and, more recently, physics-informed machine learning dealiasing \cite{ChazoPaz2021}.

\bibliography{pfcddsdraft}

\begin{thebibliography}{79}%
\makeatletter
\providecommand \@ifxundefined [1]{%
 \@ifx{#1\undefined}
}%
\providecommand \@ifnum [1]{%
 \ifnum #1\expandafter \@firstoftwo
 \else \expandafter \@secondoftwo
 \fi
}%
\providecommand \@ifx [1]{%
 \ifx #1\expandafter \@firstoftwo
 \else \expandafter \@secondoftwo
 \fi
}%
\providecommand \natexlab [1]{#1}%
\providecommand \enquote  [1]{``#1''}%
\providecommand \bibnamefont  [1]{#1}%
\providecommand \bibfnamefont [1]{#1}%
\providecommand \citenamefont [1]{#1}%
\providecommand \href@noop [0]{\@secondoftwo}%
\providecommand \href [0]{\begingroup \@sanitize@url \@href}%
\providecommand \@href[1]{\@@startlink{#1}\@@href}%
\providecommand \@@href[1]{\endgroup#1\@@endlink}%
\providecommand \@sanitize@url [0]{\catcode `\\12\catcode `\$12\catcode
  `\&12\catcode `\#12\catcode `\^12\catcode `\_12\catcode `\%12\relax}%
\providecommand \@@startlink[1]{}%
\providecommand \@@endlink[0]{}%
\providecommand \url  [0]{\begingroup\@sanitize@url \@url }%
\providecommand \@url [1]{\endgroup\@href {#1}{\urlprefix }}%
\providecommand \urlprefix  [0]{URL }%
\providecommand \Eprint [0]{\href }%
\providecommand \doibase [0]{https://doi.org/}%
\providecommand \selectlanguage [0]{\@gobble}%
\providecommand \bibinfo  [0]{\@secondoftwo}%
\providecommand \bibfield  [0]{\@secondoftwo}%
\providecommand \translation [1]{[#1]}%
\providecommand \BibitemOpen [0]{}%
\providecommand \bibitemStop [0]{}%
\providecommand \bibitemNoStop [0]{.\EOS\space}%
\providecommand \EOS [0]{\spacefactor3000\relax}%
\providecommand \BibitemShut  [1]{\csname bibitem#1\endcsname}%
\let\auto@bib@innerbib\@empty
\bibitem [{\citenamefont {Dickman}\ and\ \citenamefont {Zia}(2018)}]{DWRLG1}%
  \BibitemOpen
  \bibfield  {author} {\bibinfo {author} {\bibfnamefont {R.}~\bibnamefont
  {Dickman}}\ and\ \bibinfo {author} {\bibfnamefont {R.~K.~P.}\ \bibnamefont
  {Zia}},\ }\bibfield  {title} {\bibinfo {title} {Driven {W}idom-{R}owlinson
  lattice gas},\ }\href@noop {} {\bibfield  {journal} {\bibinfo  {journal}
  {Phys. Rev. E}\ }\textbf {\bibinfo {volume} {97}},\ \bibinfo {pages} {062126}
  (\bibinfo {year} {2018})}\BibitemShut {NoStop}%
\bibitem [{\citenamefont {Lavrentovich}\ \emph {et~al.}(2021)\citenamefont
  {Lavrentovich}, \citenamefont {Dickman},\ and\ \citenamefont {Zia}}]{DWRLG2}%
  \BibitemOpen
  \bibfield  {author} {\bibinfo {author} {\bibfnamefont {M.~O.}\ \bibnamefont
  {Lavrentovich}}, \bibinfo {author} {\bibfnamefont {R.}~\bibnamefont
  {Dickman}},\ and\ \bibinfo {author} {\bibfnamefont {R.~K.~P.}\ \bibnamefont
  {Zia}},\ }\bibfield  {title} {\bibinfo {title} {Microemulsions in the driven
  {W}idom-{R}owlinson lattice gas},\ }\href@noop {} {\bibfield  {journal}
  {\bibinfo  {journal} {Phys. Rev. E}\ }\textbf {\bibinfo {volume} {104}},\
  \bibinfo {pages} {064135} (\bibinfo {year} {2021})}\BibitemShut {NoStop}%
\bibitem [{\citenamefont {Cross}\ and\ \citenamefont
  {Hohenberg}(1993)}]{CrossHohenberg}%
  \BibitemOpen
  \bibfield  {author} {\bibinfo {author} {\bibfnamefont {M.~C.}\ \bibnamefont
  {Cross}}\ and\ \bibinfo {author} {\bibfnamefont {P.~C.}\ \bibnamefont
  {Hohenberg}},\ }\bibfield  {title} {\bibinfo {title} {Pattern formation
  outside of equilibrium},\ }\href@noop {} {\bibfield  {journal} {\bibinfo
  {journal} {Rev. Mod. Phys.}\ }\textbf {\bibinfo {volume} {65}},\ \bibinfo
  {pages} {851} (\bibinfo {year} {1993})}\BibitemShut {NoStop}%
\bibitem [{\citenamefont {Hohenberg}\ and\ \citenamefont
  {Krekhov}(2015)}]{HohenbergKrekhov}%
  \BibitemOpen
  \bibfield  {author} {\bibinfo {author} {\bibfnamefont {P.}~\bibnamefont
  {Hohenberg}}\ and\ \bibinfo {author} {\bibfnamefont {A.}~\bibnamefont
  {Krekhov}},\ }\bibfield  {title} {\bibinfo {title} {An introduction to the
  ginzburg–landau theory of phase transitions and nonequilibrium patterns},\
  }\href {https://doi.org/10.1016/j.physrep.2015.01.001} {\bibfield  {journal}
  {\bibinfo  {journal} {Phys. Rep.}\ }\textbf {\bibinfo {volume} {572}},\
  \bibinfo {pages} {1} (\bibinfo {year} {2015})}\BibitemShut {NoStop}%
\bibitem [{\citenamefont {Cross}\ and\ \citenamefont
  {Greenside}(2009)}]{Cross2}%
  \BibitemOpen
  \bibfield  {author} {\bibinfo {author} {\bibfnamefont {M.}~\bibnamefont
  {Cross}}\ and\ \bibinfo {author} {\bibfnamefont {H.}~\bibnamefont
  {Greenside}},\ }\href {https://doi.org/10.1017/CBO9780511627200} {\emph
  {\bibinfo {title} {Pattern Formation and Dynamics in Nonequilibrium
  Systems}}}\ (\bibinfo  {publisher} {Cambridge University Press},\ \bibinfo
  {year} {2009})\BibitemShut {NoStop}%
\bibitem [{\citenamefont {Hohenberg}\ and\ \citenamefont
  {Halperin}(1977)}]{HohenbergHalperin}%
  \BibitemOpen
  \bibfield  {author} {\bibinfo {author} {\bibfnamefont {P.~C.}\ \bibnamefont
  {Hohenberg}}\ and\ \bibinfo {author} {\bibfnamefont {B.~I.}\ \bibnamefont
  {Halperin}},\ }\bibfield  {title} {\bibinfo {title} {Theory of dynamic
  critical phenomena},\ }\href@noop {} {\bibfield  {journal} {\bibinfo
  {journal} {Rev. Mod. Phys.}\ }\textbf {\bibinfo {volume} {49}},\ \bibinfo
  {pages} {435} (\bibinfo {year} {1977})}\BibitemShut {NoStop}%
\bibitem [{\citenamefont {Dickman}(2016)}]{Ron2016}%
  \BibitemOpen
  \bibfield  {author} {\bibinfo {author} {\bibfnamefont {R.}~\bibnamefont
  {Dickman}},\ }\bibfield  {title} {\bibinfo {title} {Phase coexistence far
  from equilibrium},\ }\href {https://doi.org/10.1088/1367-2630/18/4/043034}
  {\bibfield  {journal} {\bibinfo  {journal} {New J. Phys.}\ }\textbf {\bibinfo
  {volume} {18}},\ \bibinfo {pages} {043034} (\bibinfo {year}
  {2016})}\BibitemShut {NoStop}%
\bibitem [{\citenamefont {Calazans}\ and\ \citenamefont
  {Dickman}(2019)}]{Calazans2019}%
  \BibitemOpen
  \bibfield  {author} {\bibinfo {author} {\bibfnamefont {L.~F.}\ \bibnamefont
  {Calazans}}\ and\ \bibinfo {author} {\bibfnamefont {R.}~\bibnamefont
  {Dickman}},\ }\bibfield  {title} {\bibinfo {title} {Steady-state entropy: A
  proposal based on thermodynamic integration},\ }\href
  {https://doi.org/10.1103/PhysRevE.99.032137} {\bibfield  {journal} {\bibinfo
  {journal} {Phys. Rev. E}\ }\textbf {\bibinfo {volume} {99}},\ \bibinfo
  {pages} {032137} (\bibinfo {year} {2019})}\BibitemShut {NoStop}%
\bibitem [{\citenamefont {Katz}\ \emph {et~al.}(1983)\citenamefont {Katz},
  \citenamefont {Lebowitz},\ and\ \citenamefont {Spohn}}]{KLS2}%
  \BibitemOpen
  \bibfield  {author} {\bibinfo {author} {\bibfnamefont {S.}~\bibnamefont
  {Katz}}, \bibinfo {author} {\bibfnamefont {J.~L.}\ \bibnamefont {Lebowitz}},\
  and\ \bibinfo {author} {\bibfnamefont {H.}~\bibnamefont {Spohn}},\ }\bibfield
   {title} {\bibinfo {title} {Phase transitions in stationary nonequilibrium
  states of model lattice systems},\ }\href@noop {} {\bibfield  {journal}
  {\bibinfo  {journal} {Phys. Rev. B}\ }\textbf {\bibinfo {volume} {28}},\
  \bibinfo {pages} {1655(R)} (\bibinfo {year} {1983})}\BibitemShut {NoStop}%
\bibitem [{\citenamefont {Katz}\ \emph {et~al.}(1984)\citenamefont {Katz},
  \citenamefont {Lebowitz},\ and\ \citenamefont {Spohn}}]{KLS1}%
  \BibitemOpen
  \bibfield  {author} {\bibinfo {author} {\bibfnamefont {S.}~\bibnamefont
  {Katz}}, \bibinfo {author} {\bibfnamefont {J.~L.}\ \bibnamefont {Lebowitz}},\
  and\ \bibinfo {author} {\bibfnamefont {H.}~\bibnamefont {Spohn}},\ }\bibfield
   {title} {\bibinfo {title} {Nonequilibrium steady states of stochastic
  lattice gas models of fast ionic conductors},\ }\href@noop {} {\bibfield
  {journal} {\bibinfo  {journal} {J. Stat. Phys.}\ }\textbf {\bibinfo {volume}
  {34}},\ \bibinfo {pages} {497} (\bibinfo {year} {1984})}\BibitemShut
  {NoStop}%
\bibitem [{\citenamefont {Garrido}\ and\ \citenamefont
  {Marro}(1987)}]{Garrido1987}%
  \BibitemOpen
  \bibfield  {author} {\bibinfo {author} {\bibfnamefont {P.}~\bibnamefont
  {Garrido}}\ and\ \bibinfo {author} {\bibfnamefont {J.}~\bibnamefont
  {Marro}},\ }\bibfield  {title} {\bibinfo {title} {Ising models with
  anisotropic interactions: Stationary nonequilibrium states with a nonuniform
  temperature profile},\ }\href {https://doi.org/10.1016/0378-4371(87)90210-X}
  {\bibfield  {journal} {\bibinfo  {journal} {Physica A}\ }\textbf {\bibinfo
  {volume} {144}},\ \bibinfo {pages} {585} (\bibinfo {year}
  {1987})}\BibitemShut {NoStop}%
\bibitem [{\citenamefont {t.~Leung}\ \emph {et~al.}(1989)\citenamefont
  {t.~Leung}, \citenamefont {Schmittmann},\ and\ \citenamefont {Zia}}]{LSZ89}%
  \BibitemOpen
  \bibfield  {author} {\bibinfo {author} {\bibfnamefont {K.}~\bibnamefont
  {t.~Leung}}, \bibinfo {author} {\bibfnamefont {B.}~\bibnamefont
  {Schmittmann}},\ and\ \bibinfo {author} {\bibfnamefont {R.~K.~P.}\
  \bibnamefont {Zia}},\ }\bibfield  {title} {\bibinfo {title} {Phase
  transitions in a driven lattice gas with repulsive interactions},\
  }\href@noop {} {\bibfield  {journal} {\bibinfo  {journal} {Phys. Rev. Lett.}\
  }\textbf {\bibinfo {volume} {62}},\ \bibinfo {pages} {1772} (\bibinfo {year}
  {1989})}\BibitemShut {NoStop}%
\bibitem [{\citenamefont {Vall\'es}\ \emph {et~al.}(1989)\citenamefont
  {Vall\'es}, \citenamefont {t.~Leung},\ and\ \citenamefont {Zia}}]{VLZ89}%
  \BibitemOpen
  \bibfield  {author} {\bibinfo {author} {\bibfnamefont {J.~L.}\ \bibnamefont
  {Vall\'es}}, \bibinfo {author} {\bibfnamefont {K.}~\bibnamefont {t.~Leung}},\
  and\ \bibinfo {author} {\bibfnamefont {R.~K.~P.}\ \bibnamefont {Zia}},\
  }\bibfield  {title} {\bibinfo {title} {Driven nonequilibrium lattice systems
  with a shifted periodic boundary conditions},\ }\href@noop {} {\bibfield
  {journal} {\bibinfo  {journal} {J. Stat. Phys.}\ }\textbf {\bibinfo {volume}
  {56}},\ \bibinfo {pages} {43} (\bibinfo {year} {1989})}\BibitemShut {NoStop}%
\bibitem [{\citenamefont {Garrido}\ \emph {et~al.}(1990)\citenamefont
  {Garrido}, \citenamefont {Lebowitz}, \citenamefont {Maes},\ and\
  \citenamefont {Spohn}}]{GLMS90}%
  \BibitemOpen
  \bibfield  {author} {\bibinfo {author} {\bibfnamefont {P.~L.}\ \bibnamefont
  {Garrido}}, \bibinfo {author} {\bibfnamefont {J.~L.}\ \bibnamefont
  {Lebowitz}}, \bibinfo {author} {\bibfnamefont {C.}~\bibnamefont {Maes}},\
  and\ \bibinfo {author} {\bibfnamefont {H.}~\bibnamefont {Spohn}},\ }\bibfield
   {title} {\bibinfo {title} {Long-range correlations for conservative
  dynamics},\ }\href@noop {} {\bibfield  {journal} {\bibinfo  {journal} {Phys.
  Rev. A}\ }\textbf {\bibinfo {volume} {42}},\ \bibinfo {pages} {1954}
  (\bibinfo {year} {1990})}\BibitemShut {NoStop}%
\bibitem [{\citenamefont {Cheng}\ \emph {et~al.}(1991)\citenamefont {Cheng},
  \citenamefont {Garrido}, \citenamefont {Lebowitz},\ and\ \citenamefont
  {Vall{\'{e}}s}}]{CGLV91}%
  \BibitemOpen
  \bibfield  {author} {\bibinfo {author} {\bibfnamefont {Z.}~\bibnamefont
  {Cheng}}, \bibinfo {author} {\bibfnamefont {P.~L.}\ \bibnamefont {Garrido}},
  \bibinfo {author} {\bibfnamefont {J.~L.}\ \bibnamefont {Lebowitz}},\ and\
  \bibinfo {author} {\bibfnamefont {J.~L.}\ \bibnamefont {Vall{\'{e}}s}},\
  }\bibfield  {title} {\bibinfo {title} {Long-range correlations in stationary
  nonequilibrium systems with conservative anisotropic dynamics},\ }\href@noop
  {} {\bibfield  {journal} {\bibinfo  {journal} {Europhys. Lett.}\ }\textbf
  {\bibinfo {volume} {14}},\ \bibinfo {pages} {507} (\bibinfo {year}
  {1991})}\BibitemShut {NoStop}%
\bibitem [{\citenamefont {Schmittmann}\ and\ \citenamefont {Zia}(1991)}]{SZ91}%
  \BibitemOpen
  \bibfield  {author} {\bibinfo {author} {\bibfnamefont {B.}~\bibnamefont
  {Schmittmann}}\ and\ \bibinfo {author} {\bibfnamefont {R.~K.~P.}\
  \bibnamefont {Zia}},\ }\bibfield  {title} {\bibinfo {title} {Critical
  properties of a randomly driven diffusive system},\ }\href@noop {} {\bibfield
   {journal} {\bibinfo  {journal} {Phys. Rev. Lett.}\ }\textbf {\bibinfo
  {volume} {66}},\ \bibinfo {pages} {357} (\bibinfo {year} {1991})}\BibitemShut
  {NoStop}%
\bibitem [{\citenamefont {Schmittmann}\ \emph {et~al.}(1992)\citenamefont
  {Schmittmann}, \citenamefont {Hwang},\ and\ \citenamefont {Zia}}]{SHZ92}%
  \BibitemOpen
  \bibfield  {author} {\bibinfo {author} {\bibfnamefont {B.}~\bibnamefont
  {Schmittmann}}, \bibinfo {author} {\bibfnamefont {K.}~\bibnamefont {Hwang}},\
  and\ \bibinfo {author} {\bibfnamefont {R.~K.~P.}\ \bibnamefont {Zia}},\
  }\bibfield  {title} {\bibinfo {title} {Onset of spatial structures in biased
  diffusion of two species},\ }\href@noop {} {\bibfield  {journal} {\bibinfo
  {journal} {Europhys. Lett.}\ }\textbf {\bibinfo {volume} {19}},\ \bibinfo
  {pages} {19} (\bibinfo {year} {1992})}\BibitemShut {NoStop}%
\bibitem [{\citenamefont {Bassler}\ and\ \citenamefont
  {Zia}(1994)}]{Bassler1994}%
  \BibitemOpen
  \bibfield  {author} {\bibinfo {author} {\bibfnamefont {K.~E.}\ \bibnamefont
  {Bassler}}\ and\ \bibinfo {author} {\bibfnamefont {R.~K.~P.}\ \bibnamefont
  {Zia}},\ }\bibfield  {title} {\bibinfo {title} {Phase transitions in a
  nonequilibrium {P}otts model: A monte carlo study of critical behavior},\
  }\href {https://doi.org/10.1103/PhysRevE.49.5871} {\bibfield  {journal}
  {\bibinfo  {journal} {Phys. Rev. E}\ }\textbf {\bibinfo {volume} {49}},\
  \bibinfo {pages} {5871} (\bibinfo {year} {1994})}\BibitemShut {NoStop}%
\bibitem [{\citenamefont {Korniss}\ \emph {et~al.}(1997)\citenamefont
  {Korniss}, \citenamefont {Schmittmann},\ and\ \citenamefont {Zia}}]{KSZ97}%
  \BibitemOpen
  \bibfield  {author} {\bibinfo {author} {\bibfnamefont {G.}~\bibnamefont
  {Korniss}}, \bibinfo {author} {\bibfnamefont {B.}~\bibnamefont
  {Schmittmann}},\ and\ \bibinfo {author} {\bibfnamefont {R.~K.~P.}\
  \bibnamefont {Zia}},\ }\bibfield  {title} {\bibinfo {title} {Nonequilibrium
  phase transitions in a simple three-state lattice gas},\ }\href
  {https://doi.org/10.1007/BF02199117} {\bibfield  {journal} {\bibinfo
  {journal} {J. Stat. Phys.}\ }\textbf {\bibinfo {volume} {86}},\ \bibinfo
  {pages} {721} (\bibinfo {year} {1997})}\BibitemShut {NoStop}%
\bibitem [{\citenamefont {Schmittmann}\ and\ \citenamefont
  {Zia}(1995)}]{DDSbook}%
  \BibitemOpen
  \bibfield  {author} {\bibinfo {author} {\bibfnamefont {B.}~\bibnamefont
  {Schmittmann}}\ and\ \bibinfo {author} {\bibfnamefont {R.~K.~P.}\
  \bibnamefont {Zia}},\ }\href@noop {} {\emph {\bibinfo {title} {Statistical
  mechanics of driven diffusive systems}}}\ (\bibinfo  {publisher} {Academic
  Press},\ \bibinfo {address} {San Diego},\ \bibinfo {year} {1995})\BibitemShut
  {NoStop}%
\bibitem [{\citenamefont {Marro}\ and\ \citenamefont
  {Dickman}(1999)}]{MarroDickman99}%
  \BibitemOpen
  \bibfield  {author} {\bibinfo {author} {\bibfnamefont {J.}~\bibnamefont
  {Marro}}\ and\ \bibinfo {author} {\bibfnamefont {R.}~\bibnamefont
  {Dickman}},\ }\href {https://doi.org/10.1017/CBO9780511524288} {\emph
  {\bibinfo {title} {Nonequilibrium Phase Transitions in Lattice Models}}}\
  (\bibinfo  {publisher} {Cambridge University Press},\ \bibinfo {year}
  {1999})\BibitemShut {NoStop}%
\bibitem [{\citenamefont {Janssen}\ and\ \citenamefont
  {Schmittmann}(1986)}]{JS86}%
  \BibitemOpen
  \bibfield  {author} {\bibinfo {author} {\bibfnamefont {H.~K.}\ \bibnamefont
  {Janssen}}\ and\ \bibinfo {author} {\bibfnamefont {B.}~\bibnamefont
  {Schmittmann}},\ }\bibfield  {title} {\bibinfo {title} {Field theory of
  critical behaviour in driven diffusive systems},\ }\href
  {https://doi.org/10.1007/BF01312845} {\bibfield  {journal} {\bibinfo
  {journal} {Z. Phys. B}\ }\textbf {\bibinfo {volume} {64}},\ \bibinfo {pages}
  {503} (\bibinfo {year} {1986})}\BibitemShut {NoStop}%
\bibitem [{\citenamefont {tai Leung}\ and\ \citenamefont {Cardy}(1986)}]{LC86}%
  \BibitemOpen
  \bibfield  {author} {\bibinfo {author} {\bibfnamefont {K.}~\bibnamefont {tai
  Leung}}\ and\ \bibinfo {author} {\bibfnamefont {J.~L.}\ \bibnamefont
  {Cardy}},\ }\bibfield  {title} {\bibinfo {title} {Field theory of critical
  behavior in a driven diffusive system},\ }\href
  {https://doi.org/10.1007/BF01011310} {\bibfield  {journal} {\bibinfo
  {journal} {J. Stat. Phys.}\ }\textbf {\bibinfo {volume} {44}},\ \bibinfo
  {pages} {567} (\bibinfo {year} {1986})}\BibitemShut {NoStop}%
\bibitem [{\citenamefont {Dickman}\ and\ \citenamefont
  {Stell}(1995)}]{eqRWLG1}%
  \BibitemOpen
  \bibfield  {author} {\bibinfo {author} {\bibfnamefont {R.}~\bibnamefont
  {Dickman}}\ and\ \bibinfo {author} {\bibfnamefont {G.}~\bibnamefont
  {Stell}},\ }\bibfield  {title} {\bibinfo {title} {Critical behavior of the
  {W}idom-{R}owlinson lattice model},\ }\href@noop {} {\bibfield  {journal}
  {\bibinfo  {journal} {J. Chem. Phys.}\ }\textbf {\bibinfo {volume} {102}},\
  \bibinfo {pages} {8674} (\bibinfo {year} {1995})}\BibitemShut {NoStop}%
\bibitem [{\citenamefont {Brauns}\ and\ \citenamefont
  {Marchetti}(2024)}]{Marchetti2024}%
  \BibitemOpen
  \bibfield  {author} {\bibinfo {author} {\bibfnamefont {F.}~\bibnamefont
  {Brauns}}\ and\ \bibinfo {author} {\bibfnamefont {M.~C.}\ \bibnamefont
  {Marchetti}},\ }\bibfield  {title} {\bibinfo {title} {Nonreciprocal pattern
  formation of conserved fields},\ }\href@noop {} {\bibfield  {journal}
  {\bibinfo  {journal} {Phys. Rev. X}\ }\textbf {\bibinfo {volume} {14}},\
  \bibinfo {pages} {021014} (\bibinfo {year} {2024})}\BibitemShut {NoStop}%
\bibitem [{\citenamefont {Fausti}\ \emph {et~al.}(2021)\citenamefont {Fausti},
  \citenamefont {Tjhung}, \citenamefont {Cates},\ and\ \citenamefont
  {Nardini}}]{Fausti2021}%
  \BibitemOpen
  \bibfield  {author} {\bibinfo {author} {\bibfnamefont {G.}~\bibnamefont
  {Fausti}}, \bibinfo {author} {\bibfnamefont {E.}~\bibnamefont {Tjhung}},
  \bibinfo {author} {\bibfnamefont {M.~E.}\ \bibnamefont {Cates}},\ and\
  \bibinfo {author} {\bibfnamefont {C.}~\bibnamefont {Nardini}},\ }\bibfield
  {title} {\bibinfo {title} {Capillary interfacial tension in active phase
  separation},\ }\href {https://doi.org/10.1103/PhysRevLett.127.068001}
  {\bibfield  {journal} {\bibinfo  {journal} {Phys. Rev. Lett.}\ }\textbf
  {\bibinfo {volume} {127}},\ \bibinfo {pages} {068001} (\bibinfo {year}
  {2021})}\BibitemShut {NoStop}%
\bibitem [{\citenamefont {Fruchart}\ \emph {et~al.}(2021)\citenamefont
  {Fruchart}, \citenamefont {Hanai}, \citenamefont {Littlewood},\ and\
  \citenamefont {Vitelli}}]{Fruchart2021}%
  \BibitemOpen
  \bibfield  {author} {\bibinfo {author} {\bibfnamefont {M.}~\bibnamefont
  {Fruchart}}, \bibinfo {author} {\bibfnamefont {R.}~\bibnamefont {Hanai}},
  \bibinfo {author} {\bibfnamefont {P.~B.}\ \bibnamefont {Littlewood}},\ and\
  \bibinfo {author} {\bibfnamefont {V.}~\bibnamefont {Vitelli}},\ }\bibfield
  {title} {\bibinfo {title} {Non-reciprocal phase transitions},\ }\href
  {https://doi.org/10.1038/s41586-021-03375-9} {\bibfield  {journal} {\bibinfo
  {journal} {Nature}\ }\textbf {\bibinfo {volume} {592}},\ \bibinfo {pages}
  {363} (\bibinfo {year} {2021})}\BibitemShut {NoStop}%
\bibitem [{\citenamefont {Duan}\ \emph {et~al.}(2023)\citenamefont {Duan},
  \citenamefont {Agudo-Canalejo}, \citenamefont {Golestanian},\ and\
  \citenamefont {Mahault}}]{Golestanian2023}%
  \BibitemOpen
  \bibfield  {author} {\bibinfo {author} {\bibfnamefont {Y.}~\bibnamefont
  {Duan}}, \bibinfo {author} {\bibfnamefont {J.}~\bibnamefont
  {Agudo-Canalejo}}, \bibinfo {author} {\bibfnamefont {R.}~\bibnamefont
  {Golestanian}},\ and\ \bibinfo {author} {\bibfnamefont {B.}~\bibnamefont
  {Mahault}},\ }\bibfield  {title} {\bibinfo {title} {Dynamical pattern
  formation without self-attraction in quorum-sensing active matter: The
  interplay between nonreciprocity and motility},\ }\href
  {https://doi.org/10.1103/PhysRevLett.131.148301} {\bibfield  {journal}
  {\bibinfo  {journal} {Phys. Rev. Lett.}\ }\textbf {\bibinfo {volume} {131}},\
  \bibinfo {pages} {148301} (\bibinfo {year} {2023})}\BibitemShut {NoStop}%
\bibitem [{\citenamefont {Duan}\ \emph {et~al.}(2025)\citenamefont {Duan},
  \citenamefont {Agudo-Canalejo}, \citenamefont {Golestanian},\ and\
  \citenamefont {Mahault}}]{Golestanian2025}%
  \BibitemOpen
  \bibfield  {author} {\bibinfo {author} {\bibfnamefont {Y.}~\bibnamefont
  {Duan}}, \bibinfo {author} {\bibfnamefont {J.}~\bibnamefont
  {Agudo-Canalejo}}, \bibinfo {author} {\bibfnamefont {R.}~\bibnamefont
  {Golestanian}},\ and\ \bibinfo {author} {\bibfnamefont {B.}~\bibnamefont
  {Mahault}},\ }\bibfield  {title} {\bibinfo {title} {Phase coexistence in
  nonreciprocal quorum-sensing active matter},\ }\href
  {https://doi.org/10.1103/PhysRevResearch.7.013234} {\bibfield  {journal}
  {\bibinfo  {journal} {Phys. Rev. Res.}\ }\textbf {\bibinfo {volume} {7}},\
  \bibinfo {pages} {013234} (\bibinfo {year} {2025})}\BibitemShut {NoStop}%
\bibitem [{\citenamefont {Doi}(1976)}]{Doi}%
  \BibitemOpen
  \bibfield  {author} {\bibinfo {author} {\bibfnamefont {M.}~\bibnamefont
  {Doi}},\ }\bibfield  {title} {\bibinfo {title} {Second quantization
  representation for classical many-particle system},\ }\href@noop {}
  {\bibfield  {journal} {\bibinfo  {journal} {J. Phys. A: Math. Gen.}\ }\textbf
  {\bibinfo {volume} {9}},\ \bibinfo {pages} {1465} (\bibinfo {year}
  {1976})}\BibitemShut {NoStop}%
\bibitem [{\citenamefont {Peliti}(1985)}]{Peliti}%
  \BibitemOpen
  \bibfield  {author} {\bibinfo {author} {\bibfnamefont {L.}~\bibnamefont
  {Peliti}},\ }\bibfield  {title} {\bibinfo {title} {Path integral approach to
  birth-death processes on a lattice},\ }\href@noop {} {\bibfield  {journal}
  {\bibinfo  {journal} {J. Physique}\ }\textbf {\bibinfo {volume} {46}},\
  \bibinfo {pages} {1469} (\bibinfo {year} {1985})}\BibitemShut {NoStop}%
\bibitem [{\citenamefont {Skopenkov}(2023)}]{Skopenkov2023}%
  \BibitemOpen
  \bibfield  {author} {\bibinfo {author} {\bibfnamefont {M.}~\bibnamefont
  {Skopenkov}},\ }\bibfield  {title} {\bibinfo {title} {Discrete field theory:
  Symmetries and conservation laws},\ }\href
  {https://doi.org/10.1007/s11040-023-09459-4} {\bibfield  {journal} {\bibinfo
  {journal} {J. Math. Phys. Anal. Geom.}\ }\textbf {\bibinfo {volume} {26}},\
  \bibinfo {pages} {19} (\bibinfo {year} {2023})}\BibitemShut {NoStop}%
\bibitem [{\citenamefont {Mao}\ \emph {et~al.}(2019)\citenamefont {Mao},
  \citenamefont {Kuldinow}, \citenamefont {Haataja},\ and\ \citenamefont
  {Košmrlj}}]{Mao2019}%
  \BibitemOpen
  \bibfield  {author} {\bibinfo {author} {\bibfnamefont {S.}~\bibnamefont
  {Mao}}, \bibinfo {author} {\bibfnamefont {D.}~\bibnamefont {Kuldinow}},
  \bibinfo {author} {\bibfnamefont {M.~P.}\ \bibnamefont {Haataja}},\ and\
  \bibinfo {author} {\bibfnamefont {A.}~\bibnamefont {Košmrlj}},\ }\bibfield
  {title} {\bibinfo {title} {Phase behavior and morphology of multicomponent
  liquid mixtures},\ }\href {https://doi.org/10.1039/C8SM02045K} {\bibfield
  {journal} {\bibinfo  {journal} {Soft Matter}\ }\textbf {\bibinfo {volume}
  {15}},\ \bibinfo {pages} {1297} (\bibinfo {year} {2019})}\BibitemShut
  {NoStop}%
\bibitem [{\citenamefont {Saha}\ \emph {et~al.}(2020)\citenamefont {Saha},
  \citenamefont {Agudo-Canalejo},\ and\ \citenamefont
  {Golestanian}}]{Golestanian2020}%
  \BibitemOpen
  \bibfield  {author} {\bibinfo {author} {\bibfnamefont {S.}~\bibnamefont
  {Saha}}, \bibinfo {author} {\bibfnamefont {J.}~\bibnamefont
  {Agudo-Canalejo}},\ and\ \bibinfo {author} {\bibfnamefont {R.}~\bibnamefont
  {Golestanian}},\ }\bibfield  {title} {\bibinfo {title} {Scalar active
  mixtures: The nonreciprocal {C}ahn-{H}illiard model},\ }\href@noop {}
  {\bibfield  {journal} {\bibinfo  {journal} {Phys. Rev. X}\ }\textbf {\bibinfo
  {volume} {10}},\ \bibinfo {pages} {041009} (\bibinfo {year}
  {2020})}\BibitemShut {NoStop}%
\bibitem [{\citenamefont {Cox}\ and\ \citenamefont {Matthews}(2002)}]{Cox}%
  \BibitemOpen
  \bibfield  {author} {\bibinfo {author} {\bibfnamefont {S.}~\bibnamefont
  {Cox}}\ and\ \bibinfo {author} {\bibfnamefont {P.}~\bibnamefont {Matthews}},\
  }\bibfield  {title} {\bibinfo {title} {Exponential time differencing for
  stiff systems},\ }\href
  {https://doi.org/https://doi.org/10.1006/jcph.2002.6995} {\bibfield
  {journal} {\bibinfo  {journal} {J. Comput. Phys.}\ }\textbf {\bibinfo
  {volume} {176}},\ \bibinfo {pages} {430} (\bibinfo {year}
  {2002})}\BibitemShut {NoStop}%
\bibitem [{\citenamefont {Toral}\ and\ \citenamefont {Colet}(2014)}]{Toral}%
  \BibitemOpen
  \bibfield  {author} {\bibinfo {author} {\bibfnamefont {R.}~\bibnamefont
  {Toral}}\ and\ \bibinfo {author} {\bibfnamefont {P.}~\bibnamefont {Colet}},\
  }\href {https://doi.org/10.1002/9783527683147} {\emph {\bibinfo {title}
  {Stochastic Numerical Methods}}}\ (\bibinfo  {publisher} {Wiley},\ \bibinfo
  {year} {2014})\BibitemShut {NoStop}%
\bibitem [{\citenamefont {Hou}\ and\ \citenamefont {Li}(2007)}]{Hou}%
  \BibitemOpen
  \bibfield  {author} {\bibinfo {author} {\bibfnamefont {T.~Y.}\ \bibnamefont
  {Hou}}\ and\ \bibinfo {author} {\bibfnamefont {R.}~\bibnamefont {Li}},\
  }\bibfield  {title} {\bibinfo {title} {Computing nearly singular solutions
  using pseudo-spectral methods},\ }\href
  {https://doi.org/https://doi.org/10.1016/j.jcp.2007.04.014} {\bibfield
  {journal} {\bibinfo  {journal} {J. Comput. Phys.}\ }\textbf {\bibinfo
  {volume} {226}},\ \bibinfo {pages} {379} (\bibinfo {year}
  {2007})}\BibitemShut {NoStop}%
\bibitem [{\citenamefont {Widom}\ and\ \citenamefont {Rowlinson}(1970)}]{WR70}%
  \BibitemOpen
  \bibfield  {author} {\bibinfo {author} {\bibfnamefont {B.}~\bibnamefont
  {Widom}}\ and\ \bibinfo {author} {\bibfnamefont {J.~S.}\ \bibnamefont
  {Rowlinson}},\ }\bibfield  {title} {\bibinfo {title} {New model for the study
  of liquid–vapor phase transitions},\ }\href
  {https://doi.org/10.1063/1.1673203} {\bibfield  {journal} {\bibinfo
  {journal} {J. Chem. Phys.}\ }\textbf {\bibinfo {volume} {52}},\ \bibinfo
  {pages} {1670} (\bibinfo {year} {1970})}\BibitemShut {NoStop}%
\bibitem [{Note1()}]{Note1}%
  \BibitemOpen
  \bibinfo {note} {The rudiments of this model were introduced in \cite
  {DWRLG1}, where it played the role of an intermediary between simulations of
  the lattice gas and studies of a phenomenological field theory. In the
  present study, we investigate a full stochastic version of it.}\BibitemShut
  {Stop}%
\bibitem [{Note2()}]{Note2}%
  \BibitemOpen
  \bibinfo {note} {Generalizations to lattices in higher dimensions are
  straightforward.}\BibitemShut {Stop}%
\bibitem [{Note3()}]{Note3}%
  \BibitemOpen
  \bibinfo {note} {In our simulations, $\epsilon $ is chosen from a uniform
  distribution. From the results, $\epsilon $ serves only to redefine a
  diffusion length. Note that this multiplicative aspect of the noise may
  become relevant near the boundary values: $\rho _{A,B}=0,1$.}\BibitemShut
  {Stop}%
\bibitem [{Note4()}]{Note4}%
  \BibitemOpen
  \bibinfo {note} {The form of Eq.~\protect \eqref {eq:amount_moved} and these
  suppression factors are motivated in part by the remarkably good agreement
  between a conjectured formula for the current density and the observed one in
  DWRLG: Eq. (8) and Fig. 5 in \cite {DWRLG1}.}\BibitemShut {Stop}%
\bibitem [{\citenamefont {Oliveira}(2025)}]{thesisgui}%
  \BibitemOpen
  \bibfield  {author} {\bibinfo {author} {\bibfnamefont {G.~E.~F.}\
  \bibnamefont {Oliveira}},\ }\bibfield  {title} {\bibinfo {title} {Pattern
  formation in repulsive bicomponent driven diffusive systems}} (\bibinfo
  {year} {2025}),\ \bibinfo {note} {{P}hD thesis draft, Universidade Federal de
  Minas Gerais}\BibitemShut {NoStop}%
\bibitem [{Note5()}]{Note5}%
  \BibitemOpen
  \bibinfo {note} {Indeed, for $\rho $ as low as just $0.5$, certain
  configurations are \textquotedblleft frozen\textquotedblright . Examples
  include the checkerboard pattern (all white squares vacant, alternate black
  ones filled with $A$ or $B$) and similarly alternate rows or columns. Note
  that such configurations are also present in the WRLG, as $\rho =0,\pm 1$
  appears simply as $n=0,\pm 1$. As a result, the particular density $\rho _g$
  at which this transition occurs may depend on the initial
  condition.}\BibitemShut {Stop}%
\bibitem [{\citenamefont {Echeverr\'{\i}a-Alar}\ and\ \citenamefont
  {Clerc}(2020)}]{Alar-Clerc}%
  \BibitemOpen
  \bibfield  {author} {\bibinfo {author} {\bibfnamefont {S.}~\bibnamefont
  {Echeverr\'{\i}a-Alar}}\ and\ \bibinfo {author} {\bibfnamefont {M.~G.}\
  \bibnamefont {Clerc}},\ }\bibfield  {title} {\bibinfo {title} {Labyrinthine
  patterns transitions},\ }\href
  {https://doi.org/10.1103/PhysRevResearch.2.042036} {\bibfield  {journal}
  {\bibinfo  {journal} {Phys. Rev. Res.}\ }\textbf {\bibinfo {volume} {2}},\
  \bibinfo {pages} {042036} (\bibinfo {year} {2020})}\BibitemShut {NoStop}%
\bibitem [{\citenamefont {Freire~Oliveira}\ \emph {et~al.}(2025)\citenamefont
  {Freire~Oliveira}, \citenamefont {Dickman}, \citenamefont {Lavrentovich},\
  and\ \citenamefont {K.~P.~Zia}}]{movies}%
  \BibitemOpen
  \bibfield  {author} {\bibinfo {author} {\bibfnamefont {G.~E.}\ \bibnamefont
  {Freire~Oliveira}}, \bibinfo {author} {\bibfnamefont {R.}~\bibnamefont
  {Dickman}}, \bibinfo {author} {\bibfnamefont {M.}~\bibnamefont
  {Lavrentovich}},\ and\ \bibinfo {author} {\bibfnamefont {R.}~\bibnamefont
  {K.~P.~Zia}},\ }\href {https://doi.org/10.5281/zenodo.17704014} {\bibinfo
  {title} {Movie guide - pattern formation in a coupled driven diffusive
  system}} (\bibinfo {year} {2025})\BibitemShut {NoStop}%
\bibitem [{\citenamefont {Eckhaus}(1965)}]{Eckhaus}%
  \BibitemOpen
  \bibfield  {author} {\bibinfo {author} {\bibfnamefont {W.}~\bibnamefont
  {Eckhaus}},\ }\href@noop {} {\emph {\bibinfo {title} {Studies in Non-linear
  Stability Theory}}},\ Springer tracts in natural philosophy\ (\bibinfo
  {publisher} {Springer-Verlag},\ \bibinfo {year} {1965})\BibitemShut {NoStop}%
\bibitem [{Note6()}]{Note6}%
  \BibitemOpen
  \bibinfo {note} {Due to the symmetries of the systems we are considering, it
  can be shown that the cross correlation (associated with $S_{+-}$) vanishes,
  so that we investigated on these two SFs.}\BibitemShut {Stop}%
\bibitem [{Note7()}]{Note7}%
  \BibitemOpen
  \bibinfo {note} {Note that, more generally, the noises acting on $\phi _{\pm
  }(\protect \mathbf {r},t)$ originate from the noises in the underlying
  densities $\rho _{A,B}(\protect \mathbf {r},t)$, which are inherently
  multiplicative: i.e., $\xi _A$ vanishes in regions devoid of A particles, and
  analogously for $\xi _B$. If we consider small perturbations from a state
  with $\phi _+ \sim \rho $ and $\phi _- \sim 0$, the fluctuations are additive
  to a first approximation (as assumed here)}\BibitemShut {NoStop}%
\bibitem [{Note8()}]{Note8}%
  \BibitemOpen
  \bibinfo {note} {A similar feature occurs in the analysis of kinematic waves
  in drifting crystals \cite {driftingcrystal1}, where different characteristic
  velocities also play a central role in creating those periodic
  structures.}\BibitemShut {Stop}%
\bibitem [{Note9()}]{Note9}%
  \BibitemOpen
  \bibinfo {note} {As a reminder, a linear SPDE can be solved analytically. Not
  surprisingly, as they correspond to non-interacting fields, having isotropic
  parameters, the structure factors display only isotropic Ornstein-Zernike
  forms, despite the presence of $\Delta v$}\BibitemShut {NoStop}%
\bibitem [{Note10()}]{Note10}%
  \BibitemOpen
  \bibinfo {note} {This mesh size is chosen to balance numerical accuracy and
  computational cost. We have verified that further refining the mesh does not
  produce any appreciable changes in the results.}\BibitemShut {Stop}%
\bibitem [{Note11()}]{Note11}%
  \BibitemOpen
  \bibinfo {note} {This is supported by the weakly nonlinear stability analysis
  of Eq.~\protect \eqref {eq:eom-final} \cite {thesisgui}, which can also be
  used to crudely estimate the amplitude values slightly above the onset, $\rho
  \gtrsim \rho _c^{\protect \mathrm {FT}}$. On the other hand, for $\lambda <
  0$, Eq.~\protect \eqref {eq:eom-final} has no bounded solutions and the
  numerical integration is ill-behaved, thus making necessary the inclusion of
  higher-order derivative terms.}\BibitemShut {Stop}%
\bibitem [{\citenamefont {García-Ojalvo}\ and\ \citenamefont
  {Sancho}(1999)}]{Garcia-Sancho}%
  \BibitemOpen
  \bibfield  {author} {\bibinfo {author} {\bibfnamefont {J.}~\bibnamefont
  {García-Ojalvo}}\ and\ \bibinfo {author} {\bibfnamefont {J.~M.}\
  \bibnamefont {Sancho}},\ }\href {https://doi.org/10.1007/978-1-4612-1536-3}
  {\emph {\bibinfo {title} {Noise in Spatially Extended Systems}}}\ (\bibinfo
  {publisher} {Springer New York},\ \bibinfo {year} {1999})\BibitemShut
  {NoStop}%
\bibitem [{\citenamefont {Yoon}\ \emph {et~al.}(2020)\citenamefont {Yoon},
  \citenamefont {Jeong}, \citenamefont {Lee}, \citenamefont {Kim},
  \citenamefont {Kim}, \citenamefont {Lee},\ and\ \citenamefont {Kim}}]{Yoon}%
  \BibitemOpen
  \bibfield  {author} {\bibinfo {author} {\bibfnamefont {S.}~\bibnamefont
  {Yoon}}, \bibinfo {author} {\bibfnamefont {D.}~\bibnamefont {Jeong}},
  \bibinfo {author} {\bibfnamefont {C.}~\bibnamefont {Lee}}, \bibinfo {author}
  {\bibfnamefont {H.}~\bibnamefont {Kim}}, \bibinfo {author} {\bibfnamefont
  {S.}~\bibnamefont {Kim}}, \bibinfo {author} {\bibfnamefont {H.~G.}\
  \bibnamefont {Lee}},\ and\ \bibinfo {author} {\bibfnamefont {J.}~\bibnamefont
  {Kim}},\ }\bibfield  {title} {\bibinfo {title} {Fourier-spectral method for
  the phase-field equations},\ }\href {https://doi.org/10.3390/math8081385}
  {\bibfield  {journal} {\bibinfo  {journal} {Mathematics}\ }\textbf {\bibinfo
  {volume} {8}},\ \bibinfo {pages} {1385} (\bibinfo {year} {2020})}\BibitemShut
  {NoStop}%
\bibitem [{\citenamefont {Kim}\ \emph {et~al.}(2016)\citenamefont {Kim},
  \citenamefont {Lee}, \citenamefont {Choi}, \citenamefont {Lee},\ and\
  \citenamefont {Jeong}}]{Kim2016}%
  \BibitemOpen
  \bibfield  {author} {\bibinfo {author} {\bibfnamefont {J.}~\bibnamefont
  {Kim}}, \bibinfo {author} {\bibfnamefont {S.}~\bibnamefont {Lee}}, \bibinfo
  {author} {\bibfnamefont {Y.}~\bibnamefont {Choi}}, \bibinfo {author}
  {\bibfnamefont {S.-M.}\ \bibnamefont {Lee}},\ and\ \bibinfo {author}
  {\bibfnamefont {D.}~\bibnamefont {Jeong}},\ }\bibfield  {title} {\bibinfo
  {title} {Basic principles and practical applications of the cahn–hilliard
  equation},\ }\href {https://doi.org/10.1155/2016/9532608} {\bibfield
  {journal} {\bibinfo  {journal} {Math. Probl. Eng.}\ }\textbf {\bibinfo
  {volume} {2016}},\ \bibinfo {pages} {1} (\bibinfo {year} {2016})}\BibitemShut
  {NoStop}%
\bibitem [{Note12()}]{Note12}%
  \BibitemOpen
  \bibinfo {note} {This is not true, however, in the low-density regime and at
  interfaces, as we have verified for the microemulsion phase. In fact, for
  $\rho <\rho _c^{\protect \mathrm {FT}}$, the fields will decay to zero, even
  if $\delta $ is maximal. Therefore, the noises act as source terms, which are
  crucial to the development of the microemulsion behavior. In another words,
  the microemulsion phase \protect \textit {cannot} be observed without the
  noise terms in Eqs.~\protect \eqref {eq:eom-final}.}\BibitemShut {Stop}%
\bibitem [{Note13()}]{Note13}%
  \BibitemOpen
  \bibinfo {note} {Note that it is not strictly necessary to drastically set
  $g_0 = 0$; any small value $g_0 \approx 0$ would yield similar results.
  However, since no particular value is preferred, we choose zero for
  simplicity and computational efficiency.}\BibitemShut {Stop}%
\bibitem [{Note14()}]{Note14}%
  \BibitemOpen
  \bibinfo {note} {In fact, this vertical modulation seems to be a feature of
  the equations that is already present if $g_{\pm ,0}=0$; see, e.g., Fig.~\ref
  {fig:typconf-eom}\protect \, (1).}\BibitemShut {Stop}%
\bibitem [{Note15()}]{Note15}%
  \BibitemOpen
  \bibinfo {note} {We can notice, however, a shoulder at higher $q_x$'s in
  Fig.~\ref {fig:S_(qx,qy)}(b)}\BibitemShut {NoStop}%
\bibitem [{\citenamefont {Zia}(2010)}]{Zia2010}%
  \BibitemOpen
  \bibfield  {author} {\bibinfo {author} {\bibfnamefont {R.~K.~P.}\
  \bibnamefont {Zia}},\ }\bibfield  {title} {\bibinfo {title} {Twenty five
  years after {KLS}: A celebration of non-equilibrium statistical mechanics},\
  }\href {https://doi.org/10.1007/s10955-009-9884-0} {\bibfield  {journal}
  {\bibinfo  {journal} {J. Stat. Phys.}\ }\textbf {\bibinfo {volume} {138}},\
  \bibinfo {pages} {20} (\bibinfo {year} {2010})}\BibitemShut {NoStop}%
\bibitem [{Note16()}]{Note16}%
  \BibitemOpen
  \bibinfo {note} {Interestingly, we found no references in literature that
  describe these methods in detail, leading us to believe that such a
  description (and application) could be useful, particularly in the active
  matter context, since the non-reciprocal Cahn-Hilliard and Active Model B
  equations are analogous to Eq.~\protect \eqref {eq:eom-final}}\BibitemShut
  {NoStop}%
\bibitem [{Note17()}]{Note17}%
  \BibitemOpen
  \bibinfo {note} {In fact, we find that it is necessary to use a much finer
  mesh in space and time than the one we use in Sec.~\ref {sec:numer-int} to
  numerically integrate an equation of type \protect \eqref {eq:eom-final} with
  a noise of moderate intensity, $\sigma \approx 10^{-2}$, without recurring to
  dealiasing techniques.}\BibitemShut {Stop}%
\bibitem [{Note18()}]{Note18}%
  \BibitemOpen
  \bibinfo {note} {For simplicity, we suppressed all the \textquotedblleft
  physical units\textquotedblright \ for the coefficients, e.g., $D\propto \ell
  ^{2}\tau ^{-1}$, etc. Their detailed forms play no role in the numerical
  studies.}\BibitemShut {Stop}%
\bibitem [{Note19()}]{Note19}%
  \BibitemOpen
  \bibinfo {note} {Note that such a term originates from a $\phi _+^3$
  potential in the free-energy and, as such, is unstable. Therefore, one must
  include an additional $\gamma _+\protect \, \phi ^3_+$ (a $\phi _+^4$
  potential) in Eq.~\protect \eqref {eq:eom-prefinal} to ensure the stability
  of the theory. Since we avoid over-complicating our equations by seeking a
  minimal description and also considering that the DF is stable to
  linear-order in $\phi _+$, we can safely neglect the term proportional to
  $\lambda _+$.}\BibitemShut {Stop}%
\bibitem [{Note20()}]{Note20}%
  \BibitemOpen
  \bibinfo {note} {Based on the approach of Cross \protect \textit {et al.}
  \cite {CrossHohenberg, Cross2}, the weakly nonlinear stability analysis here
  consists of assuming that a dominant role in the instability of the
  homogeneous regime is played by a particular mode of $\phi _{-}$ with the
  ansatz $\phi _{-}(\protect \mathbf {r},t)=\alpha (t)\protect \, \protect
  \mathrm {Re}\protect \, \{ e^{i\Psi (t)+i\protect \mathbf {q} \cdot \protect
  \mathbf {r}}\}$. As $A$-$B$ exclusion tends to favor vacancies between the
  charged regions, it is reasonable to assume $\beta (t)\protect \, \protect
  \mathrm {Re}\protect \, \{e^{i2\Psi (t)+i\protect \mathbf {q} \cdot \protect
  \mathbf {r}}\}$ for $\phi _{+}(\protect \mathbf {r},t)$. In fact, the
  location of the peaks of the structure factors in the striped regime
  ($q_{-}=q^{\ast }=q_{+}/2$) displays this type of entraining of $\phi _{+}$
  to $\phi _{-}$. Also, a perturbative scheme indicates $\beta \simeq \alpha
  ^{2}$. Details are found in \cite {thesisgui}.}\BibitemShut {Stop}%
\bibitem [{\citenamefont {Shen}\ \emph {et~al.}(2011)\citenamefont {Shen},
  \citenamefont {Tang},\ and\ \citenamefont {Wang}}]{Shen-Tang-Wang}%
  \BibitemOpen
  \bibfield  {author} {\bibinfo {author} {\bibfnamefont {J.}~\bibnamefont
  {Shen}}, \bibinfo {author} {\bibfnamefont {T.}~\bibnamefont {Tang}},\ and\
  \bibinfo {author} {\bibfnamefont {L.-L.}\ \bibnamefont {Wang}},\ }\href
  {https://doi.org/10.1007/978-3-540-71041-7} {\emph {\bibinfo {title}
  {Spectral Methods}}},\ Vol.~\bibinfo {volume} {41}\ (\bibinfo  {publisher}
  {Springer Berlin Heidelberg},\ \bibinfo {year} {2011})\BibitemShut {NoStop}%
\bibitem [{\citenamefont {Roberts}\ and\ \citenamefont
  {Bowman}(2011)}]{Roberts2011}%
  \BibitemOpen
  \bibfield  {author} {\bibinfo {author} {\bibfnamefont {M.}~\bibnamefont
  {Roberts}}\ and\ \bibinfo {author} {\bibfnamefont {J.~C.}\ \bibnamefont
  {Bowman}},\ }\bibfield  {title} {\bibinfo {title} {Dealiased convolutions for
  pseudospectral simulations},\ }\href
  {https://doi.org/10.1088/1742-6596/318/7/072037} {\bibfield  {journal}
  {\bibinfo  {journal} {J. Phys. Conf. Ser.}\ }\textbf {\bibinfo {volume}
  {318}},\ \bibinfo {pages} {072037} (\bibinfo {year} {2011})}\BibitemShut
  {NoStop}%
\bibitem [{\citenamefont {Frigo}\ and\ \citenamefont {Johnson}(2005)}]{FFTW}%
  \BibitemOpen
  \bibfield  {author} {\bibinfo {author} {\bibfnamefont {M.}~\bibnamefont
  {Frigo}}\ and\ \bibinfo {author} {\bibfnamefont {S.}~\bibnamefont
  {Johnson}},\ }\bibfield  {title} {\bibinfo {title} {The design and
  implementation of {FFTW3}},\ }\href
  {https://doi.org/10.1109/JPROC.2004.840301} {\bibfield  {journal} {\bibinfo
  {journal} {Proc. IEEE}\ }\textbf {\bibinfo {volume} {93}},\ \bibinfo {pages}
  {216} (\bibinfo {year} {2005})}\BibitemShut {NoStop}%
\bibitem [{\citenamefont {Rößler}(2009)}]{Rossler}%
  \BibitemOpen
  \bibfield  {author} {\bibinfo {author} {\bibfnamefont {A.}~\bibnamefont
  {Rößler}},\ }\bibfield  {title} {\bibinfo {title} {Second order
  runge–kutta methods for itô stochastic differential equations},\ }\href
  {https://doi.org/10.1137/060673308} {\bibfield  {journal} {\bibinfo
  {journal} {SIAM J. Numer. Anal.}\ }\textbf {\bibinfo {volume} {47}},\
  \bibinfo {pages} {1713} (\bibinfo {year} {2009})}\BibitemShut {NoStop}%
\bibitem [{\citenamefont {Orszag}(1971)}]{Orszag1971}%
  \BibitemOpen
  \bibfield  {author} {\bibinfo {author} {\bibfnamefont {S.~A.}\ \bibnamefont
  {Orszag}},\ }\bibfield  {title} {\bibinfo {title} {On the elimination of
  aliasing in finite-difference schemes by filtering high-wavenumber
  components},\ }\href
  {https://doi.org/10.1175/1520-0469(1971)028<1074:OTEOAI>2.0.CO;2} {\bibfield
  {journal} {\bibinfo  {journal} {J. Atmos. Sci.}\ }\textbf {\bibinfo {volume}
  {28}},\ \bibinfo {pages} {1074} (\bibinfo {year} {1971})}\BibitemShut
  {NoStop}%
\bibitem [{\citenamefont {Kravchenko}\ and\ \citenamefont
  {Moin}(1997)}]{Kravchenko1997}%
  \BibitemOpen
  \bibfield  {author} {\bibinfo {author} {\bibfnamefont {A.}~\bibnamefont
  {Kravchenko}}\ and\ \bibinfo {author} {\bibfnamefont {P.}~\bibnamefont
  {Moin}},\ }\bibfield  {title} {\bibinfo {title} {On the effect of numerical
  errors in large eddy simulations of turbulent flows},\ }\href
  {https://doi.org/10.1006/jcph.1996.5597} {\bibfield  {journal} {\bibinfo
  {journal} {J. Comput. Phys.}\ }\textbf {\bibinfo {volume} {131}},\ \bibinfo
  {pages} {310} (\bibinfo {year} {1997})}\BibitemShut {NoStop}%
\bibitem [{\citenamefont {Wang}(2021)}]{Wang2021}%
  \BibitemOpen
  \bibfield  {author} {\bibinfo {author} {\bibfnamefont {C.}~\bibnamefont
  {Wang}},\ }\bibfield  {title} {\bibinfo {title} {Convergence analysis of
  fourier pseudo-spectral schemes for three-dimensional incompressible
  navier-stokes equations},\ }\href {https://doi.org/10.3934/era.2021019}
  {\bibfield  {journal} {\bibinfo  {journal} {Electron. Res. Arch.}\ }\textbf
  {\bibinfo {volume} {29}},\ \bibinfo {pages} {2915} (\bibinfo {year}
  {2021})}\BibitemShut {NoStop}%
\bibitem [{\citenamefont {Margairaz}\ \emph {et~al.}(2018)\citenamefont
  {Margairaz}, \citenamefont {Giometto}, \citenamefont {Parlange},\ and\
  \citenamefont {Calaf}}]{Margairaz2018}%
  \BibitemOpen
  \bibfield  {author} {\bibinfo {author} {\bibfnamefont {F.}~\bibnamefont
  {Margairaz}}, \bibinfo {author} {\bibfnamefont {M.~G.}\ \bibnamefont
  {Giometto}}, \bibinfo {author} {\bibfnamefont {M.~B.}\ \bibnamefont
  {Parlange}},\ and\ \bibinfo {author} {\bibfnamefont {M.}~\bibnamefont
  {Calaf}},\ }\bibfield  {title} {\bibinfo {title} {Comparison of dealiasing
  schemes in large-eddy simulation of neutrally stratified atmospheric flows},\
  }\href {https://doi.org/10.5194/gmd-11-4069-2018} {\bibfield  {journal}
  {\bibinfo  {journal} {Geosci. Model Dev.}\ }\textbf {\bibinfo {volume}
  {11}},\ \bibinfo {pages} {4069} (\bibinfo {year} {2018})}\BibitemShut
  {NoStop}%
\bibitem [{\citenamefont {Robson}\ \emph {et~al.}(2012)\citenamefont {Robson},
  \citenamefont {Nicoletopoulos}, \citenamefont {Hildebrandt},\ and\
  \citenamefont {White}}]{Robson2012}%
  \BibitemOpen
  \bibfield  {author} {\bibinfo {author} {\bibfnamefont {R.~E.}\ \bibnamefont
  {Robson}}, \bibinfo {author} {\bibfnamefont {P.}~\bibnamefont
  {Nicoletopoulos}}, \bibinfo {author} {\bibfnamefont {M.}~\bibnamefont
  {Hildebrandt}},\ and\ \bibinfo {author} {\bibfnamefont {R.~D.}\ \bibnamefont
  {White}},\ }\bibfield  {title} {\bibinfo {title} {Fundamental issues in fluid
  modeling: Direct substitution and aliasing methods},\ }\bibfield  {journal}
  {\bibinfo  {journal} {J. Chem. Phys.}\ }\textbf {\bibinfo {volume} {137}},\
  \href {https://doi.org/10.1063/1.4768421} {10.1063/1.4768421} (\bibinfo
  {year} {2012})\BibitemShut {NoStop}%
\bibitem [{\citenamefont {Bowman}\ and\ \citenamefont
  {Roberts}(2011)}]{Bowman2011}%
  \BibitemOpen
  \bibfield  {author} {\bibinfo {author} {\bibfnamefont {J.~C.}\ \bibnamefont
  {Bowman}}\ and\ \bibinfo {author} {\bibfnamefont {M.}~\bibnamefont
  {Roberts}},\ }\bibfield  {title} {\bibinfo {title} {Efficient dealiased
  convolutions without padding},\ }\href {https://doi.org/10.1137/100787933}
  {\bibfield  {journal} {\bibinfo  {journal} {SIAM J. Sci. Comput.}\ }\textbf
  {\bibinfo {volume} {33}},\ \bibinfo {pages} {386} (\bibinfo {year}
  {2011})}\BibitemShut {NoStop}%
\bibitem [{\citenamefont {Sinhababu}\ and\ \citenamefont
  {Ayyalasomayajula}(2021)}]{Sinhababu2021}%
  \BibitemOpen
  \bibfield  {author} {\bibinfo {author} {\bibfnamefont {A.}~\bibnamefont
  {Sinhababu}}\ and\ \bibinfo {author} {\bibfnamefont {S.}~\bibnamefont
  {Ayyalasomayajula}},\ }\bibfield  {title} {\bibinfo {title} {An improved
  dealiasing scheme for the fourth‐order {Runge‐Kutta} method: Formulation,
  accuracy and efficiency analysis},\ }\href {https://doi.org/10.1002/fld.4898}
  {\bibfield  {journal} {\bibinfo  {journal} {Int. J. Numer. Methods Fluids}\
  }\textbf {\bibinfo {volume} {93}},\ \bibinfo {pages} {559} (\bibinfo {year}
  {2021})}\BibitemShut {NoStop}%
\bibitem [{\citenamefont {{Chazo Paz}}\ \emph {et~al.}(2021)\citenamefont
  {{Chazo Paz}}, \citenamefont {{Flores}}, \citenamefont {{Martinez-Legazpi}},
  \citenamefont {{Nguyen}}, \citenamefont {{Santa Marta}}, \citenamefont
  {{Kahn}}, \citenamefont {{Bermejo}},\ and\ \citenamefont {{Del
  Alamo}}}]{ChazoPaz2021}%
  \BibitemOpen
  \bibfield  {author} {\bibinfo {author} {\bibfnamefont {C.}~\bibnamefont
  {{Chazo Paz}}}, \bibinfo {author} {\bibfnamefont {O.}~\bibnamefont
  {{Flores}}}, \bibinfo {author} {\bibfnamefont {P.}~\bibnamefont
  {{Martinez-Legazpi}}}, \bibinfo {author} {\bibfnamefont {C.}~\bibnamefont
  {{Nguyen}}}, \bibinfo {author} {\bibfnamefont {C.}~\bibnamefont {{Santa
  Marta}}}, \bibinfo {author} {\bibfnamefont {A.}~\bibnamefont {{Kahn}}},
  \bibinfo {author} {\bibfnamefont {J.}~\bibnamefont {{Bermejo}}},\ and\
  \bibinfo {author} {\bibfnamefont {J.~C.}\ \bibnamefont {{Del Alamo}}},\
  }\bibfield  {title} {\bibinfo {title} {{All-in-one, physics-informed
  dealiasing method to regularize cardiac 4D flow MRI data.}},\ }in\ \href@noop
  {} {\emph {\bibinfo {booktitle} {APS Division of Fluid Dynamics Meeting
  Abstracts}}},\ \bibinfo {series and number} {APS Meeting Abstracts}\
  (\bibinfo {year} {2021})\ p.\ \bibinfo {pages} {H14.009}\BibitemShut
  {NoStop}%
\bibitem [{\citenamefont {Das}\ \emph {et~al.}(2001)\citenamefont {Das},
  \citenamefont {Basu}, \citenamefont {Barma},\ and\ \citenamefont
  {Ramaswamy}}]{driftingcrystal1}%
  \BibitemOpen
  \bibfield  {author} {\bibinfo {author} {\bibfnamefont {D.}~\bibnamefont
  {Das}}, \bibinfo {author} {\bibfnamefont {A.}~\bibnamefont {Basu}}, \bibinfo
  {author} {\bibfnamefont {M.}~\bibnamefont {Barma}},\ and\ \bibinfo {author}
  {\bibfnamefont {S.}~\bibnamefont {Ramaswamy}},\ }\bibfield  {title} {\bibinfo
  {title} {Weak and strong dynamic scaling in a one-dimensional driven
  coupled-field model: Effects of kinematic waves},\ }\href@noop {} {\bibfield
  {journal} {\bibinfo  {journal} {Phys. Rev. E}\ }\textbf {\bibinfo {volume}
  {64}},\ \bibinfo {pages} {021402} (\bibinfo {year} {2001})}\BibitemShut
  {NoStop}%
\end{thebibliography}%

\end{document}